\documentclass[12pt]{article}
\usepackage[utf8]{inputenc}
\usepackage{hyperref}
\usepackage[margin=1in]{geometry}
\usepackage{setspace}
\usepackage{color}
\usepackage{titling}
\usepackage{graphicx}
\usepackage[bf,margin=20pt,font={small}]{caption}
\usepackage{subcaption}
\usepackage{tikz}
\usepackage{youngtab}
\usetikzlibrary{math}
\usepackage{amsmath,amssymb,dsfont}
\allowdisplaybreaks
\newcommand{\tr}{\operatorname{tr}}

\DeclareMathAlphabet{\mathpzc}{OT1}{pzc}{m}{it}

\usepackage{putex}

\definecolor{myblue}{rgb}{0.368417,0.506779,0.709798}
\definecolor{myorange}{rgb}{0.880722,0.611041l,0.142051}

\institution{SCGP}{Simons Center for Geometry and Physics, Stony Brook University, Stony Brook, NY 11794}
\institution{PU}{Department of Physics, Princeton University, Princeton, NJ 08544}
\institution{PCTS}{Princeton Center for Theoretical Science, Princeton University, Princeton, NJ 08544}

\title{RG Limit Cycles and 
Unconventional Fixed Points in Perturbative QFT}
\authors{Christian B. Jepsen,\worksat{\SCGP} Igor R.~Klebanov\worksat{\PU,\PCTS} and Fedor K.~Popov\worksat{\PU}
}

\abstract{
We study quantum field theories with sextic interactions in $3-\epsilon$ dimensions, where the scalar fields $\phi^{ab}$ form irreducible representations under the $O(N)^2$ or $O(N)$ global symmetry group. We calculate the beta functions up to four-loop order and find the Renormalization Group fixed points. 
In an example of large $N$ equivalence, the parent $O(N)^2$ theory and its anti-symmetric projection exhibit identical large $N$ beta functions which possess real fixed points. 
However, for projection to the symmetric traceless representation of $O(N)$, the large $N$ equivalence is violated by
the appearance of an additional double-trace operator not inherited from the parent theory. Among the large $N$ fixed points of this daughter theory we find complex CFTs. 
The symmetric traceless $O(N)$ model also exhibits very interesting phenomena 
when it is analytically continued to small non-integer values of $N$. 
Here we find unconventional fixed points, which we call ``spooky." They are located at real values of the coupling constants $g^i$, but two eigenvalues of the Jacobian matrix $\partial \beta^i/\partial g^j$ are complex. When these complex conjugate eigenvalues cross the imaginary axis, a Hopf bifurcation occurs, giving rise to RG limit cycles. This crossing occurs for 
$N_{\rm crit} \approx 4.475$, and 
for a small range of $N$ above this value we find
RG flows which lead to limit cycles. }

\date{}

\begin{document}
\maketitle

\tableofcontents

\section{Introduction and Summary}
The Renormalization Group (RG) is among the deepest ideas in modern theoretical physics. There is a variety of possible RG behaviors, and limit cycles are among
the most exotic and mysterious. Their possibility was mentioned in the classic review \cite{Wilson:1973jj} in the context of connections between RG and dynamical systems 
(for a recent discussion of these connections, see \cite{Gukov:2016tnp}).
 However, there has been relatively little research on RG limit cycles.
They have appeared in quantum mechanical systems \cite{Glazek:2002hq,Braaten_2004,Gorsky:2013yba,Dawid:2017ahd}, in particular, in a description of the Efimov bound states \cite{efimov1970energy} (for a review, see \cite{Bulycheva:2014twa}). The status of RG limit cycles in QFT is less clear.
They have been searched for in unitary 4-dimensional QFT \cite{Fortin:2012cq}, but turned out to be impossible \cite{Fortin:2012hn,Luty:2012ww}, essentially due to the constraints imposed by the $a$-theorem \cite{Cardy:1988cwa,Jack:1990eb,Komargodski:2011vj}.\footnote{See, however, \cite{Morozov:2003ik,Curtright:2011qg}, where it is argued that QFTs may exhibit multi-valued $c$ or $a$-functions that do not rule out limit cycles.}
 
In this paper we report some progress on RG limit cycles in the context of perturbative QFT.  
We demonstrate their existence in a simple $O(N)$ symmetric model of scalar fields with sextic interactions in $3-\epsilon$ dimensions.
As expected, the limit cycles appear when the theory is continued to a range of parameters where it is non-unitary. 
The scalar fields form a symmetric traceless $N\times N$ matrix, and imposition of the $O(N)$ symmetry restricts the number of sextic operators to $4$.   
When we consider an analytic continuation of this model to non-integer real values of $N$
(a mathematical framework for such a continuation was presented in \cite{Binder:2019zqc}), we find a surprise.
In the range $4.465 < N < 4.534$, as well as in three other small ranges of $N$, there are unconventional RG fixed points which we call ``spooky." These fixed points are located at real values of the sextic couplings $g^i$, but only two of the eigenvalues of the Jacobian matrix $\partial \beta_i/\partial g_j$ are real; the other two are complex conjugates of each other. 
This means that a pair of nearly marginal operators at the spooky fixed points have complex scaling dimensions.\footnote{These special complex dimensions appear in addition to the complex dimensions of certain evanescent operators that are typically present in $\epsilon$ expansions \cite{Hogervorst:2015akt}. The latter dimensions have large real parts and are easily distinguished from our nearly marginal operators. Some of the operators with complex dimensions we observe resemble evanescent operators in that they interpolate to vanishing operators at integer values of $N$; this is discussed in section \ref{RGcycle}.}
At the critical value $N_{\rm crit}\approx 4.475$, the two complex eigenvalues of the Jacobian become purely imaginary. As a result, for $N$ slightly bigger than  $N_{\rm crit}$, where the real part of the complex eigenvalues becomes negative, there are RG flows which lead to limit cycles. In the theory of dynamical systems this phenomenon is called a Hopf 
(or Poincar\` e-Andronov-Hopf) bifurcation \cite{hopf1942bifurcation}. 
The possibility of RG limit cycles appearing via a Hopf bifurcation
was generally raised in \cite{Gukov:2016tnp}, but no specific examples were provided.  As we demonstrate in section \ref{RGcycle}, the symmetric traceless $O(N)$ model in $3-\epsilon$ dimensions provides a simple perturbative example of this phenomenon.

We show that there is no conflict between the limit cycles we have found and the $F$-theorem
\cite{Myers:2010xs, Jafferis:2011zi, Klebanov:2011gs, Casini:2012ei, Giombi:2014xxa,Fei:2015oha,Jack:2015tka,Jack:2016utw}. 
This is because the analytic continuation to non-integer values of $N$ below $5$ violates the unitarity of the symmetric traceless $O(N)$ model, so that the $F$-function is not monotonic.
We feel that the simple perturbative realization of limit cycles we have found is interesting, and we hope that there are analogous phenomena in other models and dimensions.

Our paper also sheds new light on the large $N$ behavior of the matrix models in $3-\epsilon$ dimensions.
Among the fascinating features of various large $N$ limits (for a recent brief overview, see \cite{Klebanov:2018fzb}) are the ``large $N$ equivalences," which relate models that are certainly different at finite $N$. An incomplete list of the conjectured large $N$ equivalences includes 
\cite{Eguchi:1982nm,Kachru:1998ys,Lawrence:1998ja,Bershadsky:1998mb,Bershadsky:1998cb,Armoni:2003gp,Kovtun:2004bz,Unsal:2006pj}.
Some of them appear to be valid, even non-perturbatively, while others are known to break down dynamically. For example, in the non-supersymmetric orbifolds of the ${\cal N}=4$ supersymmetric Yang-Mills theory \cite{Douglas:1996sw,Kachru:1998ys,Lawrence:1998ja,Bershadsky:1998mb,Bershadsky:1998cb}, there are perturbative instabilities in the large $N$ limit due to the beta functions for certain double-trace couplings having no real zeros \cite{Tseytlin:1999ii,Dymarsky:2005uh,Dymarsky:2005nc,Pomoni:2008de}.

In section \ref{Sexticmat} we study the RG flows of three scalar theories in $3-\epsilon$ dimensions with sextic interactions:
the parent $O(N)^2$ symmetric model of $N\times N$ matrices $\phi^{ab}$, and its two daughter theories which have $O(N)$ symmetry.
For each model, we list all sextic operators marginal in three dimensions, compute the associated beta functions up to 4 loops, and determine the fixed points. 
One of our motivations for this study is to investigate the large $N$ orbifold equivalence and its violation in the simple context of purely scalar theories. 
We observe evidence of large $N$ equivalence between the parent $O(N)^2$ theory and the daughter $O(N)$ theory of antisymmetric matrices: 
both theories have 3 invariant operators, and the large $N$ beta functions are identical. However, the large $N$ equivalence of the parent theory with the daughter $O(N)$ theory of symmetric traceless matrices is violated by the appearance of an additional invariant operator in the latter. 
The large $N$ fixed points in this theory occur at a complex value of the 
coefficient of this operator. As a result, instead of the conventional CFT in the parent theory, we find a ``complex CFT" \cite{Gorbenko:2018ncu,Gorbenko:2018dtm} (see also
\cite{Kaplan:2009kr}) in the daughter theory. As discussed above, analytical continuation of this model to small non-integer $N$ leads to the appearance of the spooky fixed points and limit cycles.

\section{The Beta Function Master Formula}
\label{masterSec}

In a general sextic scalar theory with potential
\begin{align}
V(\phi)=
\frac{\lambda_{iklmnp}}{6!}
\phi_i
\phi_k
\phi_l
\phi_m
\phi_n
\phi_p
\end{align}
the beta function receives a two-loop contribution from the Feynman diagram
\begin{align*}
\begin{matrix}
\text{
\includegraphics{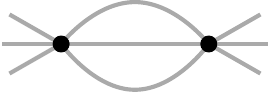}
}
\end{matrix}
\end{align*}
In \cite{Jack:2015tka,Jack:2016utw,Osborn:2017ucf} one can find explicit formulas for the corresponding two-loop beta function in $d=3-\epsilon$ dimensions. Equation (6.1) of the latter reference reads
\begin{align}
\beta_V(\phi)
=
-2\epsilon\,V(\phi)
+\frac{1}{3(8\pi)^2}
V_{ijk}(\phi)
V_{ijk}(\phi)\,, \label{6.2}
\end{align}
where $V_{i...j}(\phi)\equiv \frac{\partial}{\partial\phi^i}...\frac{\partial}{\partial\phi^j}V(\phi)$. By taking the indices to stand for doublets of sub-indices, this formula can be used to compute the beta functions of matrix tensor models. In order to apply the formula to models of symmetric or anti-symmetric matrices, however, we need to slightly modify it. Letting $i$ and $j$ stand for doublets of indices, we define the object $C^{ij}$ via the momentum space propagator:
\begin{align}
\big<\widetilde\phi^i(k)\widetilde\phi^j(-k)\big>_0
=\frac{C^{ij}}{k^2}\,.
\end{align}
With this definition in hand, equation \eqref{6.2} straightforwardly generalizes to
\begin{align}
\beta_V(\phi)
=
-2\epsilon\,V(\phi)
+\frac{C^{ii'}C^{jj'}C^{kk'}}{3(8\pi)^2}
V_{ijk}(\phi)
V_{i'j'k'}(\phi)\,.
\end{align}
At four-loops the following four kinds of Feynman diagrams contribute to the beta function:
\begin{align*}
\begin{matrix}
\text{
\includegraphics{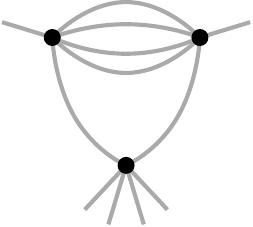}
}
\end{matrix}
\hspace{5mm}
\begin{matrix}
\text{
\includegraphics{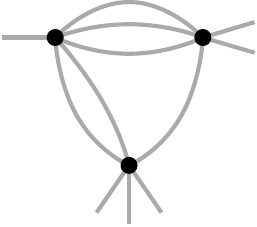}
}
\end{matrix}
\hspace{5mm}
\begin{matrix}
\text{
\includegraphics{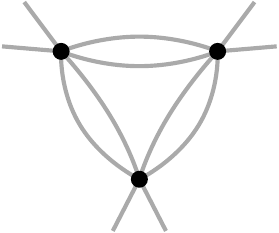}
}
\end{matrix}
\hspace{5mm}
\begin{matrix}
\text{
\includegraphics{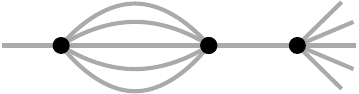}
}
\end{matrix}
\end{align*}
The resulting four-loop beta function can be read off from equation (6.2) of \cite{Osborn:2017ucf}:
\begin{gather}
	\beta^{(4)}_V = \frac{1}{(8\pi)^4}\bigg(\frac16 V_{ij} V_{iklmn} V_{jklmn} - \frac43 V_{ijk} V_{ilmn} V_{jklmn} -\frac{\pi^2}{12} V_{ijkl} V_{klmn}+\bigg) + \phi_i \gamma^{\phi}_{ij} V_j\,,
\end{gather}
where the anomalous dimension $\gamma^{\phi}_{ij}$ is given by
\begin{gather}
\gamma^{\phi}_{ij} = \frac{1}{90(8\pi)^4} \lambda_{iklmnp} \lambda_{jklmnp}\,.
\end{gather}
The above two equations also admit of straightforward generalizations by contracting indices through the $C^{ij}$ matrix.

Before proceeding to matrix models, we can review the beta function obtained by the above formulas in the case of a sextic $O(M)$ vector model described by the action 
\begin{align}
S=\int d^{3-\epsilon}x\,
\bigg(
\frac{1}{2} \left(\partial_\mu  \phi^j\right)^2 
+
\frac{g}{6!}\,\big(\phi^i\phi^i\big)^3
\bigg)\,,
\end{align}
where the field $\phi^i$ is an $M$-component vector. The four-loop beta function of this vector model is given by \cite{Hager:2002uq, Osborn:2017ucf}
\begin{gather}
\beta_g=-2\epsilon g+\frac{192(3M+22)}{6!(8\pi)^2}g^2
\\
-\frac{1}{(6!)^2(8\pi)^4}\Big(9216(53M^2+858M+3304)+1152\pi^2(M^3+34M^2+620M+2720)\Big)g^3
 \nonumber \,.
\end{gather}
This equation provides a means of checking the beta functions of the matrix models, which reduce to the vector model when all couplings are set to zero except for the coupling, denoted $g_3$ below, associated with the triple trace operator.

\section{Sextic Matrix Models}
\label{Sexticmat}

\begin{figure}
\label{MatricOperators}
\begin{subfigure}{1\textwidth}
\begin{align*}
\begin{matrix}
\text{
\scalebox{0.9}{
\includegraphics{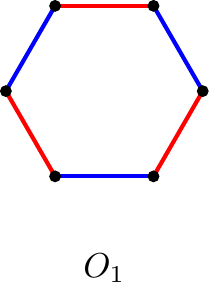}
}}
\end{matrix}
\hspace{15mm}
\begin{matrix}
\text{
\scalebox{0.9}{
\includegraphics{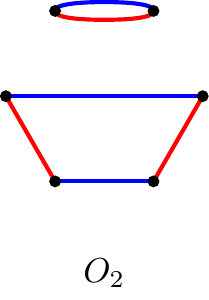}
}}
\end{matrix}
\hspace{15mm}
\begin{matrix}
\text{
\scalebox{0.9}{
\includegraphics{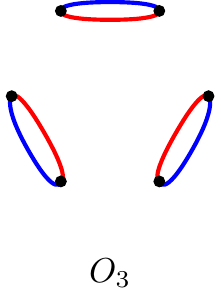}
}}
\end{matrix}
\hspace{15mm}
\begin{matrix}
\text{
\scalebox{0.9}{
\includegraphics{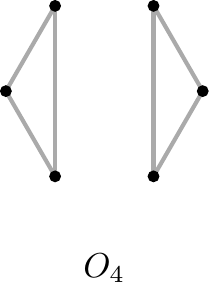}
}}
\end{matrix}
\end{align*}
\end{subfigure}
\caption{The sextic operators in matrix models. The double trace operator $O_4$ exists only in the theory of symmetric matrices.}
\label{matrixOps}
\end{figure}

We now turn to matrix models in $d=3-\epsilon$ dimensions.  The parent theory we consider has the Lagrangian given by
\begin{align}
	S = \int d^{3-\epsilon}x \left[\frac{1}{2}\left( \partial_\mu  \phi^{ab} \right)^2  +\frac{1}{6!}
\Big(
g_1
O_1(x)
+
g_2
O_2(x)
+
g_3
O_3(x)
\Big) \right],\label{matrixPot}
\end{align}
where the dynamical degrees of freedom are scalar matrices $\phi^{ab}$ which transform under the action of a global $O(N)\times O(N)$ symmetry.
The three operators in the potential are
\begin{align}
O_1=&\phi^{a_1b_1}
\phi^{a_2b_1}
\phi^{a_2b_2}
\phi^{a_3b_2}
\phi^{a_3b_3}
\phi^{a_1b_3}= {\rm tr} \left[\phi \phi^T \right]^3 \nonumber
\\
O_2=&\phi^{ab}
\phi^{ab}
\phi^{a_1b_1}
\phi^{a_2b_1}
\phi^{a_2b_2}
\phi^{a_1b_2} = {\rm tr} \left[\phi \phi^T\right] {\rm tr} \left[\phi \phi^T\right]^2
\\
O_3=&(\phi^{ab}
\phi^{ab})^3 = \left({\rm tr}\left[ \phi \phi^T\right]\right)^3 \nonumber.
\end{align}
They make up all sextic operators that are invariant under the global symmetry. Later we will also study projections of the parent theory that have only a global $O(N)$ symmetry that rotates first and second indices at the same time. In such models it becomes possible to construct singlets via contractions between first and second indices, and therefore there is an additional sextic scalar:
\begin{align}
O_4=\big(\phi^{a_1a_2}\phi^{a_2a_3}\phi^{a_3a_1}\big)^2=\left( {\rm tr} \left[ \phi^3\right] \right)^2 \,.
\end{align}
The sextic operators are depicted diagrammatically in fig. \ref{matrixOps}. We could also introduce an operator containing $\tr\left[\phi\right]$, but since the orbifolds we will study are models of symmetric traceless and anti-symmetric matrices, the trace is identically zero. In the anti-symmetric model, the operator $O_4$ vanishes, but it is non-vanishing in the symmetric orbifold, and so in this case we will introduce this additional marginal operator to the Lagrangian and take the potential to be given by
\begin{align}
V(x)=
\frac{1}{6!}
\Big(
g_1
O_1(x)
+
g_2
O_2(x)
+
g_3
O_3(x)
+g_4O_4(x)
\Big)\,.
\end{align}

To study the large $N$ behavior of these matrix models, we introduce rescaled coupling constants $\lambda_1$, $\lambda_2$, $\lambda_3$, $\lambda_4$. To simplify expressions, it will be convenient to also rescale the coupling constants by a numerical prefactor. We therefore define the rescaled couplings by
\begin{align}
\label{eqMatrixScalings}
g_1=6!(8\pi)^2\frac{\lambda_1}{N^2}
\hspace{10mm}
g_2=6!(8\pi)^2\frac{\lambda_2}{N^3}
\hspace{10mm}
g_3=6!(8\pi)^2\frac{\lambda_3}{N^4}
\hspace{10mm}
g_4=6!(8\pi)^2\frac{\lambda_4}{N^3}\,.
\end{align}
To justify these powers of $N$, let us perform a scaling $\phi^{ab} \rightarrow \sqrt{N}\phi^{ab}$. Then the coefficient of each $q$-trace term in the action scales as $N^{2-q}$. This is the standard scaling in the 't Hooft limit, which ensures that each term in the action is of order $N^2$.

\subsection{The $O(N)^2$ parent theory}

For the matrix model parent theory, the momentum space propagator is given by
\begin{align}
\big<\widetilde\phi^{ab}(k)\widetilde\phi^{a'b'}(-k)\big>_0
=\frac{\delta^{aa'}\delta^{bb'}}{k^2}\,.
\end{align}
Computing the four-loop beta functions and taking the large $N$ limit with scalings \eqref{eqMatrixScalings}, we find that, up to $\mathcal{O}(\frac{1}{N})$ corrections,
\begin{align}
\beta_{\lambda_1}=&-2\lambda_1\epsilon
+72 \lambda_1^2 -288 (17 + \pi^2) \lambda_1^3
\nonumber
\\
\beta_{\lambda_2}=&-2\lambda_2\epsilon
+432 \lambda_1^2 + 96 \lambda_1 \lambda_2-864  (90 + 7 \pi^2)\lambda_1^3 -864  (10 + \pi^2) \lambda_1^2 \lambda_2
\label{parentBeta}
\\ \nonumber
\beta_{\lambda_3}=&-2\lambda_3\epsilon
+168 \lambda_1^2+ 192 \lambda_1 \lambda_2+ 32\lambda_2^2 -432  (210 + 23 \pi^2)\lambda_1^3  -1152  (39 + 4 \pi^2)\lambda_1^2 \lambda_2
\\ &
+ 4608 \lambda_1^2 \lambda_3  -768  (6 + \pi^2)\lambda_1 \lambda_2^2  - \frac{128}{3}\pi^2 \lambda_2^3
\nonumber
\end{align}
These beta functions have two non-trivial fixed points, which are both real. But one of these fixed points, which comes from balancing the 2-loop and 4-loop contributions, is not perturbatively reliable in an $\epsilon$ expansion around $\epsilon=0$ because all the couplings at this fixed points contain terms of order $\mathcal{O}(\epsilon^0)$. The other fixed point is given by
\begin{align}
\lambda_1=&\,\frac{\epsilon}{36}+\frac{17+\pi^2}{324}\epsilon^2,
\quad
\lambda_2=\, -\frac{\epsilon}{2}-\frac{22+7\pi^2}{36}\epsilon^2,
\quad
\lambda_3=\,\frac{295}{108}\epsilon+\frac{4714+6301\pi^2}{1944}\epsilon^2\,.
\label{parentFixed}
\end{align}
At this fixed point the matrix $\frac{\partial\beta_{\lambda_i}}{\partial\lambda_j}$ has eigenvalues 
\begin{align}
\left\{-2\epsilon+\frac{32}{9}\epsilon^2,\frac{2\epsilon}{3}-\frac{44+10\pi^2}{27}\epsilon^2,2\epsilon-\frac{34+2\pi^2}{9}\epsilon^2\right\}.
\end{align}
Each eigenvalue $m_i$ corresponds to a nearly marginal operator with scaling dimension 
\begin{equation}
\Delta_i= d+m_i= 3-\epsilon+m_i\ .
\label{scaldim}
\end{equation} 
Thus, negative eigenvalues correspond to slightly relevant operators, which cause an instability of the fixed point.
The only unstable direction, corresponding to eigenvalue $-2\epsilon+\frac{32}{9}\epsilon^2$, is
\begin{align}
\left(\frac{245}{3}+\frac{4225\pi^2-4188}{36}\epsilon\right)\lambda_1+
\left(10+\frac{67\pi^2-28}{6}\right)\lambda_2+
\lambda_3.
\end{align}
The above comments relate to the $O(N)^2$ matrix model at $N=\infty$. We can also study the model at finite $N$. One interesting quantity is $N_{\rm min}$, the smallest value of $N$ at which the fixed point that interpolates to the large $N$ solution \eqref{parentFixed} appears as a solution to the beta functions. This fixed point emerges along with another fixed point, and right at $N_{\rm min}$ these solutions to the beta functions are identical, so that the matrix $\left(\frac{\partial \beta_i}{\partial g_j}\right)$ is degenerate. So we arrive at the following system of equations
\begin{gather}
	\beta_i (\lambda_i, N) = 0, \hspace{10mm} \det\left( \frac{\partial \beta_i}{\partial \lambda_j}\right) (\lambda_i, N) = 0. \label{eqSys}
\end{gather}
This system of equations can easily be solved numerically to zeroth order in $\epsilon$, and with a zeroth order solution in hand the first order solution can be obtained by linearizing the system of equations. We find that $N_{\rm min} = 23.2541 - 577.350\epsilon$, which nicely fits the results of a numerical study where we compute $N_{\rm min}$ at different values of $\epsilon$:
\begin{center}
\begin{tabular}{|c|c|c|c|c|c|c|}
		\hline
		$\epsilon$ &   0 & 0.001 & 0.002 & 0.003 & 0.004 & 0.005 \\
		\hline
		$N_{\rm min}$ & 23.255 & 22.682 & 22.124 & 21.576  & 21.039 & 20.511\\
		\hline
	\end{tabular}
\end{center}
These values result in a numerical fit $N_{\rm min}(\epsilon) = 23.255 - 553.7 \epsilon$, which coincides with the result stated above.

\begin{figure}
\begin{subfigure}{1\textwidth}
\begin{center}
\scalebox{0.7}{
\includegraphics{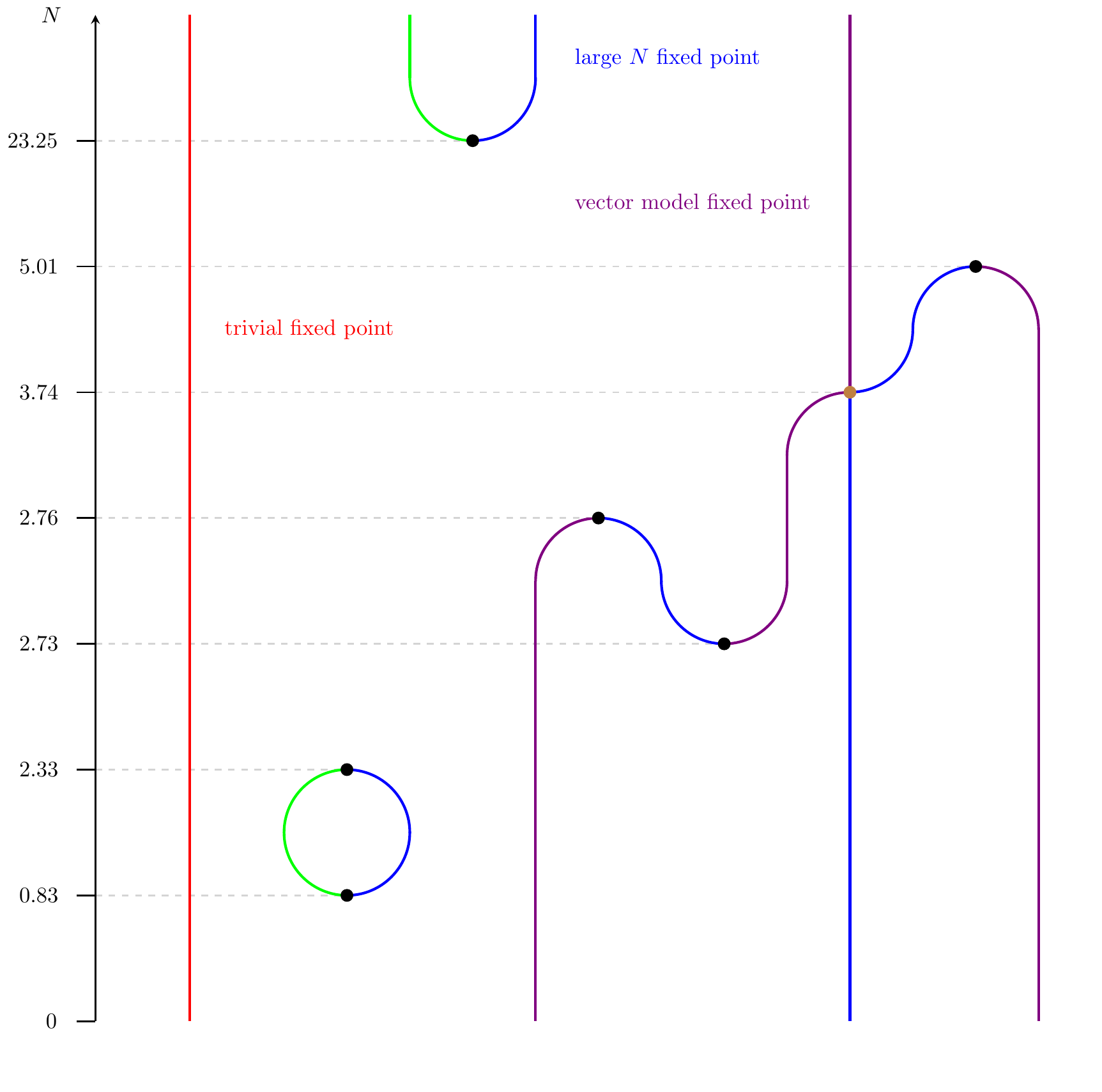}
}\end{center}
\end{subfigure}
\vspace{10mm}
\begin{subfigure}{1\textwidth}
\begin{center}
\scalebox{0.8}{
\begin{tabular}{ |c|c|c|c| } 
 \hline
 $N$ & $g_1/\epsilon$ & $g_2/\epsilon$ & $g_3/\epsilon$ \\ 
 \hline
$23.2541-577.350 \epsilon$ 
& $20.3055+1085.34\epsilon$ 
& $-10.2467-671.121 \epsilon$
& $2.64544+226.967 \epsilon$ \\ 
 \hline
$5.01072 +14.4537\epsilon$ 
& $18.4283 +56.2132 \epsilon$
& $37.3192 +141.611 \epsilon$
& $22.5095 +65.4233 \epsilon$ \\
  \hline
$\sqrt{14}+ \mathcal{O}(\epsilon)$
& $\mathcal{O}(\epsilon^2)$
& undetermined $\mathcal{O}(\epsilon)$
& $15\pi^2/2+\mathcal{O}(\epsilon)$ \\
  \hline
$2.75605 -0.0161858\epsilon$ 
& $477.273 +5099.17 \epsilon$
& $-829.732 -8328.37\epsilon$
& $382.831 +3255.35 \epsilon$\\
  \hline
$2.72717 -0.757475\epsilon$ 
& $210.819 +1081.1\epsilon$
& $-428.594 -2397.37\epsilon$
& $270.026 +1676.65\epsilon$\\
  \hline
  $2.33265 -0.316279\epsilon$ 
& $755.558 +5809.01\epsilon$
& $-1059.23 -8206.69\epsilon$
& $438.184 +3265.96\epsilon$\\
  \hline
$0.827007 +8.10374\epsilon$ 
& $237.478 +3365.73\epsilon$
& $-261.049 -4508.85\epsilon$
& $220.926 +2109.71\epsilon$\\
  \hline
\end{tabular}
}
\end{center}
\end{subfigure}
\caption{The real perturbative fixed points of the $O(N)^2$ matrix model parent theory, the intersection point (marked in \textcolor{brown}{brown}), and the critical points at which they merge and disappear (marked in \textbf{black}) as a function of $N$ for small $\epsilon$. Fixed points that are IR-unstable in all three directions are drawn in \textcolor{red}{red}, those unstable in two directions are drawn in \textcolor{violet}{violet}, those unstable in one direction are drawn in \textcolor{blue}{blue}, and those that are stable in all three directions are drawn in \textcolor{green}{green}. The four-loop corrections to the third point on the list, where two fixed lines intersect, are undetermined for any $\mathcal{O}(\epsilon^2)$ value of $\lambda_2$.
}
\label{MatFlow}
\end{figure}

\begin{figure}
	\centering
\includegraphics[width=0.4\textwidth]{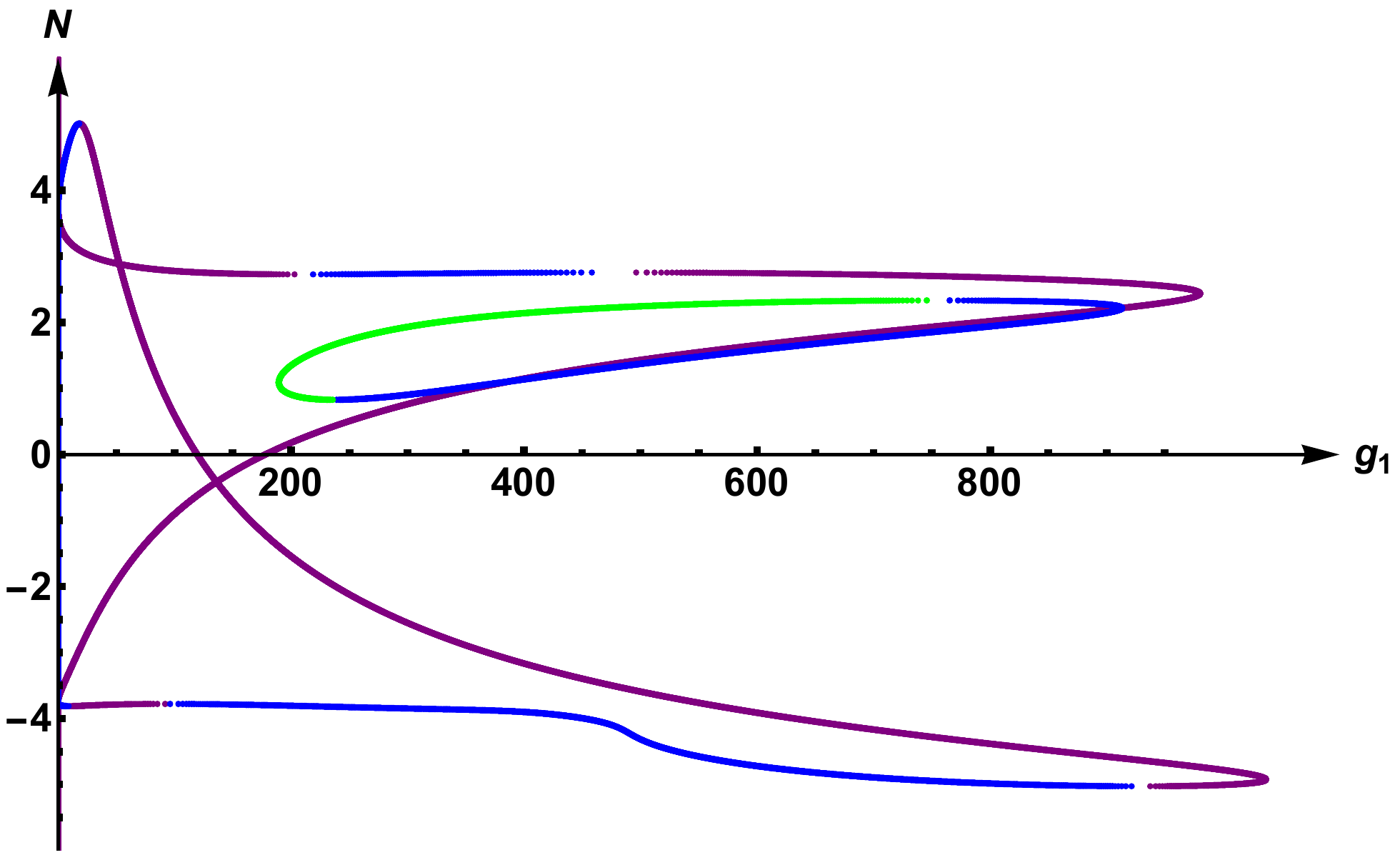}%
\quad\quad  \includegraphics[width=0.4\textwidth]{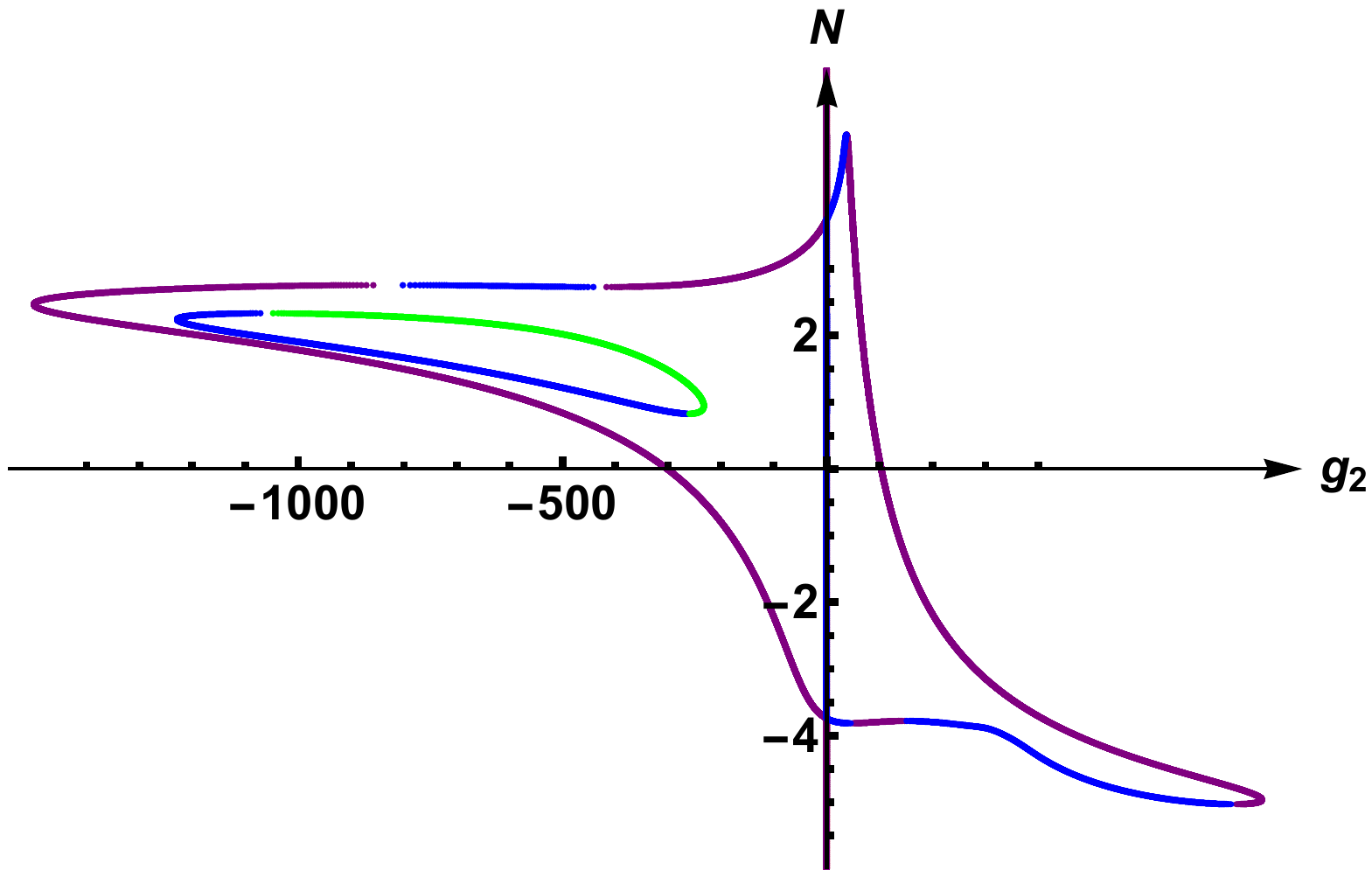}
\includegraphics[width=0.45\textwidth]{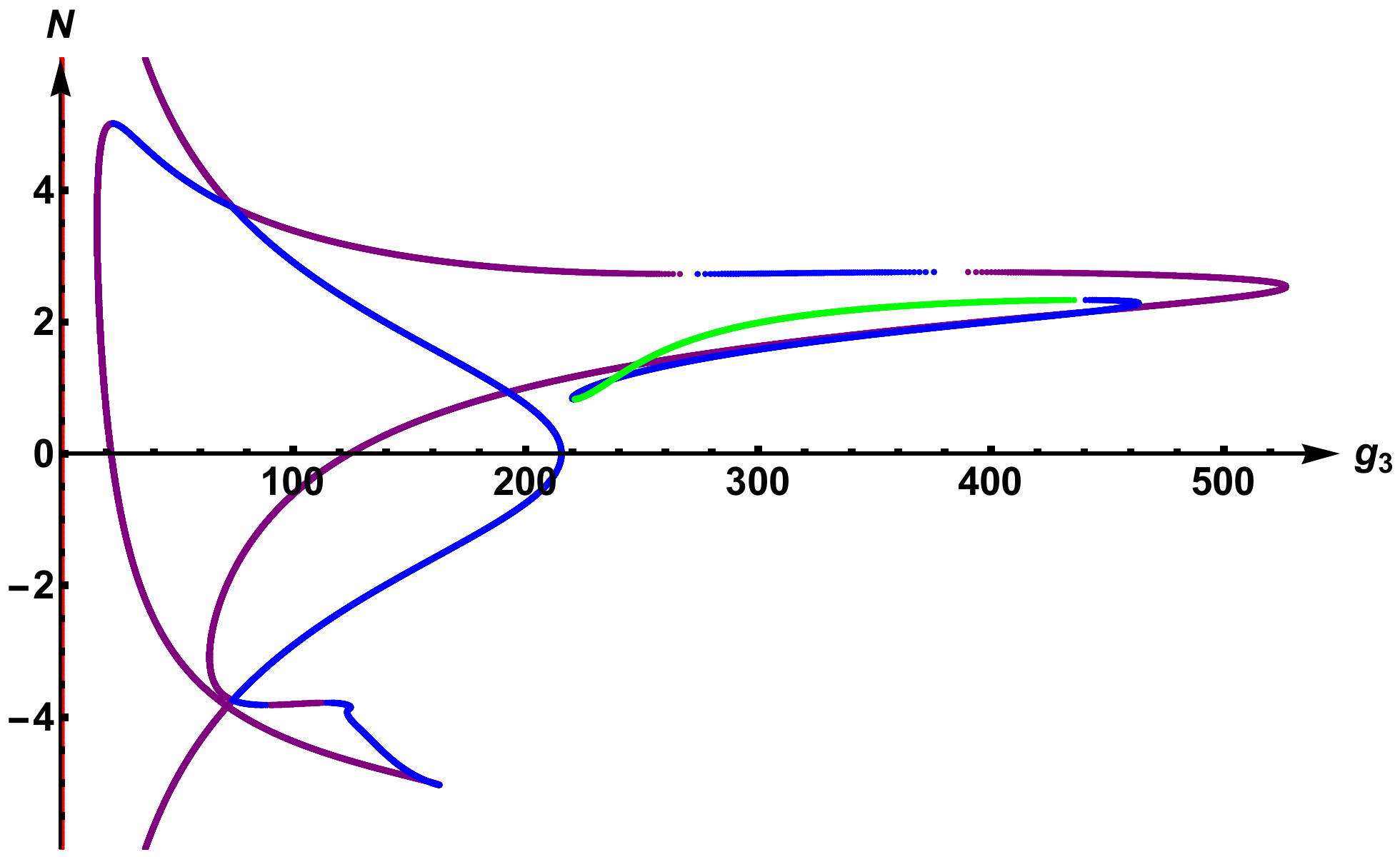}
\caption{The locations of the real perturbative fixed points of the $O(N)^2$ matrix model in the space of coupling constants as a function of $N$ for small $\epsilon$. The colors indicate the number of stable directions associated with a given fixed point as in figure \ref{MatFlow}.}
\label{actualMatFlow}
\end{figure}

If we take $N$ to be finite and $\epsilon \ll \frac{1}{N^2}$, we can provide some more details about the number and stability of fixed points for different values of $N$. For $N> 23.2541 - 577.350\epsilon$ there are three non-trivial, real, perturbatively accessible fixed points, which in the large $N$ limit , to leading order in $\epsilon$, scale with $N$ as 
\begin{align}
& \hspace{3cm} g_1 = g_2 = 0, \hspace{20mm} g_3 = \frac{6!(8\pi)^2}{288}\frac{\epsilon}{N^2},
\nonumber
\\
&g_1 =  \frac{6!(8\pi)^2}{36}\frac{\epsilon}{N^2} ,
\hspace{10mm} 
g_2 = -\frac{10}{36}\cdot 6!(8\pi)^2\frac{\epsilon}{N^3},
 \hspace{10mm} 
 g_3 = \frac{6!(8\pi)^2}{288}\frac{\epsilon}{N^2},
 \label{eq:threeFix}
 \\
& g_1 =  \frac{6!(8\pi)^2}{36}\frac{\epsilon}{N^2} ,
\hspace{10mm} 
g_2 = -\frac{1}{2} \cdot 6!(8\pi)^2\frac{\epsilon}{N^3},
 \hspace{10mm} 
 g_3 = \frac{295}{108}\cdot 6!(8\pi)^2\frac{\epsilon}{N^4}.
 \nonumber
\end{align}
The first of these three fixed points is identical to the vector model fixed point; that is to say, the symmetry is enhanced from $O(N)^2$ to $O(N^2)$. This fixed point extends to all $N$ in the small $\epsilon$ regime we are considering:
\begin{align}
g_1 = g_2 = 0, \hspace{20mm} g_3 = \frac{6!(8\pi)^2}{96(22+3N^2)}\epsilon.
\end{align}
The third fixed point in \eqref{eq:threeFix} extends to the regime where $N^2>\frac{1}{\epsilon}$ and becomes the large $N$ solution discussed above. This fixed point merges with the second fixed point in \eqref{eq:threeFix} at a critical point situated at $N(\epsilon) = 23.2541 - 577.350\epsilon$ And so at intermediate values of $N$, only the vector model fixed point exists. But as we keep decreasing $N$ we encounter another critical point at $N(\epsilon)=5.01072+14.4537\epsilon$, from which two new solutions to the vanishing beta functions emerge. As $N$ further decreases past the value $N(\epsilon)=2.75605-0.0161858\epsilon$, another pair of fixed points appear, but then at $N(\epsilon)=2.72717-0.757475\epsilon$ two of the fixed points merge and become complex. Then at $N(\epsilon)=2.33265-0.316279 \epsilon$ two new fixed points appear, but these disappear again at $N(\epsilon)=0.827007+8.10374\epsilon$, so that for $N$ below this value there are a total of three real non-trivial fixed points. The behaviour of the various fixed points as a function of $N$ is summarized in more detail in figures \ref{MatFlow} and \ref{actualMatFlow}.

\subsection{The $O(N)$ 
model of antisymmetric matrices}

For the theory of antisymmetric matrices $\phi^T = - \phi$ the momentum space propagator is given by
\begin{align}
\big<\widetilde\phi^{ab}(k)\widetilde\phi^{a'b'}(-k)\big>_0
=\frac{1}{2k^2}
\big(
\delta^{aa'}\delta^{bb'}
-\delta^{ab'}\delta^{ba'}
\big)\,.
\end{align}
Performing the large $N$ expansion using the scalings \eqref{eqMatrixScalings}  we get the large $N$ beta functions
\begin{align}
\beta_{\lambda_1}=&
- 2 \lambda_1 \epsilon
+18 \lambda_1^2 - 18(17+\pi^2) \lambda_1^3
\nonumber
\\
\beta_{\lambda_2}=&
- 2 \lambda_2 \epsilon
+108 \lambda_1^2+ 24 \lambda_1 \lambda_2  - 54(90+7\pi^2) \lambda_1^3 - 54(10+\pi^2)\lambda_1^2\lambda_2 
\label{antiBeta}
\\ \nonumber
\beta_{\lambda_3}=&
-
 2 \lambda_3 \epsilon+
42 \lambda_1^2+ 48 \lambda_1 \lambda_2+ 8 \lambda_2^2
- 27(210+23\pi^2) \lambda_1^3
-72(39+4\pi^2)\lambda_1^2\lambda_2
\nonumber
\\ 
&
+ 288 \lambda_1^2 \lambda_3
-  48(6+\pi^2) \lambda_1 \lambda_2^2 - \frac{8}{3}\pi^2 \lambda_2^3\,.
\nonumber
\end{align}
These beta-functions are equivalent to \eqref{parentBeta} up to a redefinition of the rescaled couplings by a factor of four, which is compatible with this daughter theory being equivalent in the large $N$ limit to the parent theory studied in the previous section. 

\begin{figure}
\begin{subfigure}{1\textwidth}
\begin{center}
\scalebox{0.8}{
\includegraphics{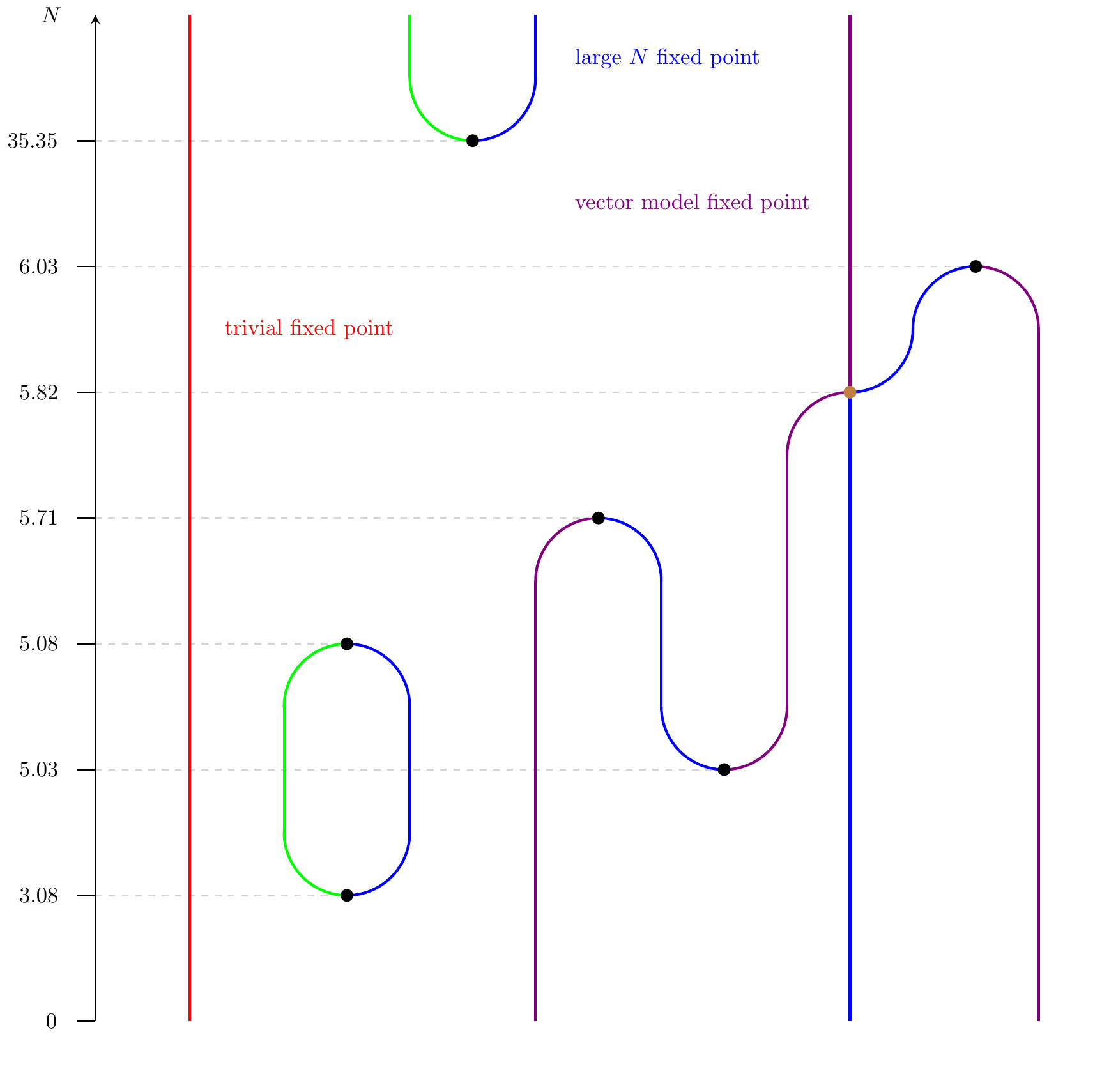}
}\end{center}
\end{subfigure}
\vspace{10mm}
\begin{subfigure}{1\textwidth}
\begin{center}
\scalebox{0.9}{
\begin{tabular}{ |c|c|c|c| } 
 \hline
 $N$ & $\lambda_1/\epsilon$ & $\lambda_2/\epsilon$ & $\lambda_3/\epsilon$ \\ 
 \hline
$35.3546-673.428 \epsilon$ 
& $49.5253+2344.67\epsilon$ 
& $-14.7886-819.812 \epsilon$
& $2.27483+172.497 \epsilon$ \\ 
 \hline
$6.02669 +7.37013\epsilon$ 
& $13.2186 +135.952 \epsilon$
& $46.5606 +358.588 \epsilon$
& $52.3442 +184.725 \epsilon$ \\
  \hline
$(1+\sqrt{113})/2+ \mathcal{O}(\epsilon)$
& $\mathcal{O}(\epsilon^2)$
& undetermined $\mathcal{O}(\epsilon)$
& $15\pi^2/2+\mathcal{O}(\epsilon)$ \\
  \hline
$5.70601 +0.540694\epsilon$ 
& $1835.96 +12199.7 \epsilon$
& $-1514.42 -9969.85\epsilon$
& $315.529 +1975.47\epsilon$\\
  \hline
$5.07531 -0.0278896\epsilon$ 
& $1742.93 +14681.9\epsilon$
& $-1228.95 -10464.7\epsilon$
& $275.926 +2170.35\epsilon$\\
  \hline
$5.03275 -0.586724\epsilon$ 
& $350.124 +3001.15\epsilon$
& $-404.283 -3356.64\epsilon$
& $180.867 +1310.49\epsilon$\\
  \hline
$3.08122 +8.26176\epsilon$ 
& $666.939 +7903.77\epsilon$
& $-373.592 -5369.46\epsilon$
& $170.179  +1403.34\epsilon$ \\
 \hline
\end{tabular}
}
\end{center}
\end{subfigure}
\caption{The real perturbative fixed points of the antisymmetric matrix model, their intersection point (marked in \textcolor{brown}{brown}), and the critical points at which they merge and disappear (marked in \textbf{black}) as a function of $N$ for small $\epsilon$. Fixed points that are IR-unstable in all three directions are drawn in \textcolor{red}{red}, those unstable in two directions are drawn in \textcolor{violet}{violet}, those unstable in one direction are drawn in \textcolor{blue}{blue}, and those that are stable in all three directions are drawn in \textcolor{green}{green}. 
}
\label{antiMatFlow}
\end{figure}

\begin{figure}
	\centering
\includegraphics[width=0.4\textwidth]{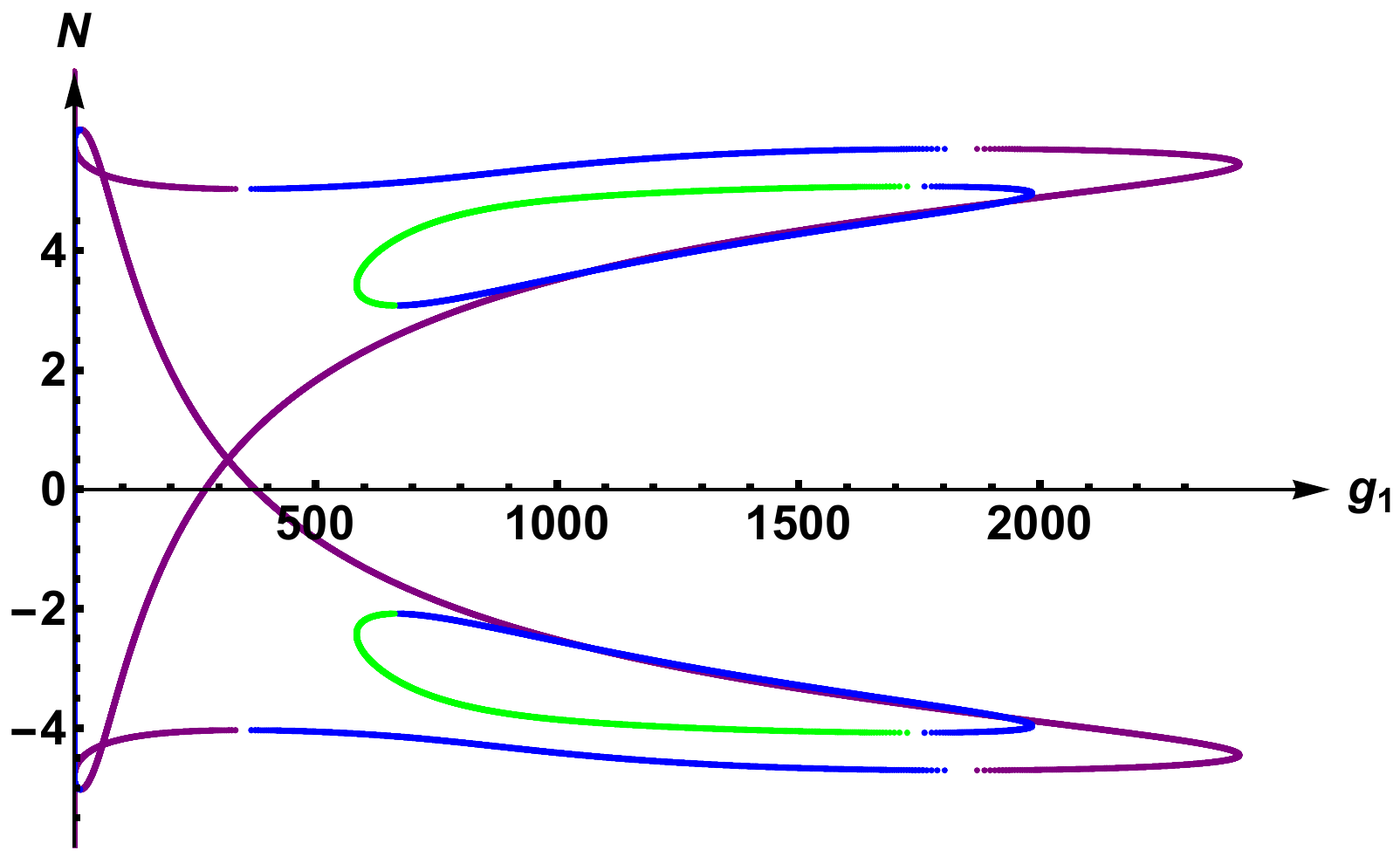}%
\quad\quad \includegraphics[width=0.4\textwidth]{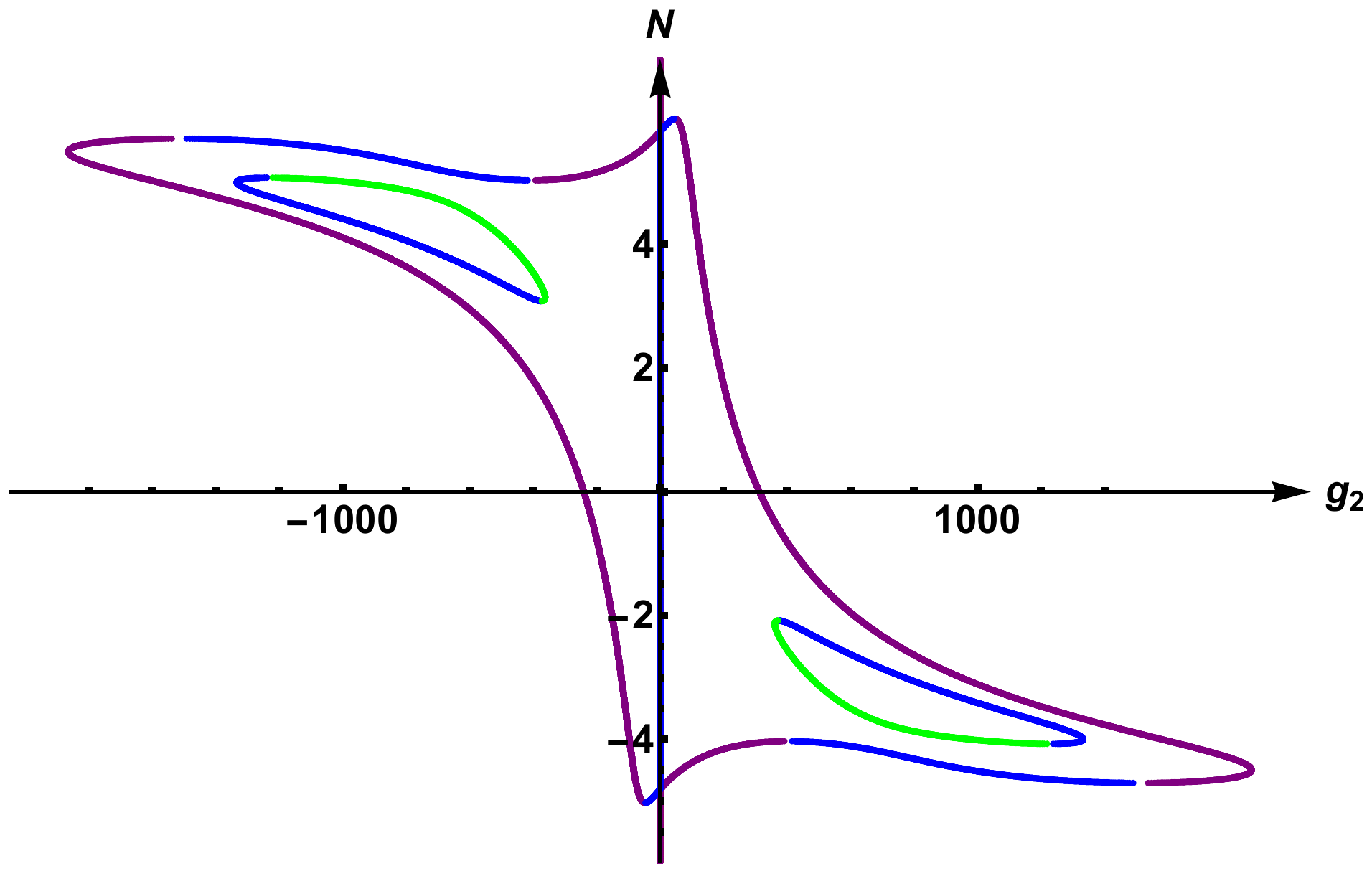}\\
\includegraphics[width=0.4\textwidth]{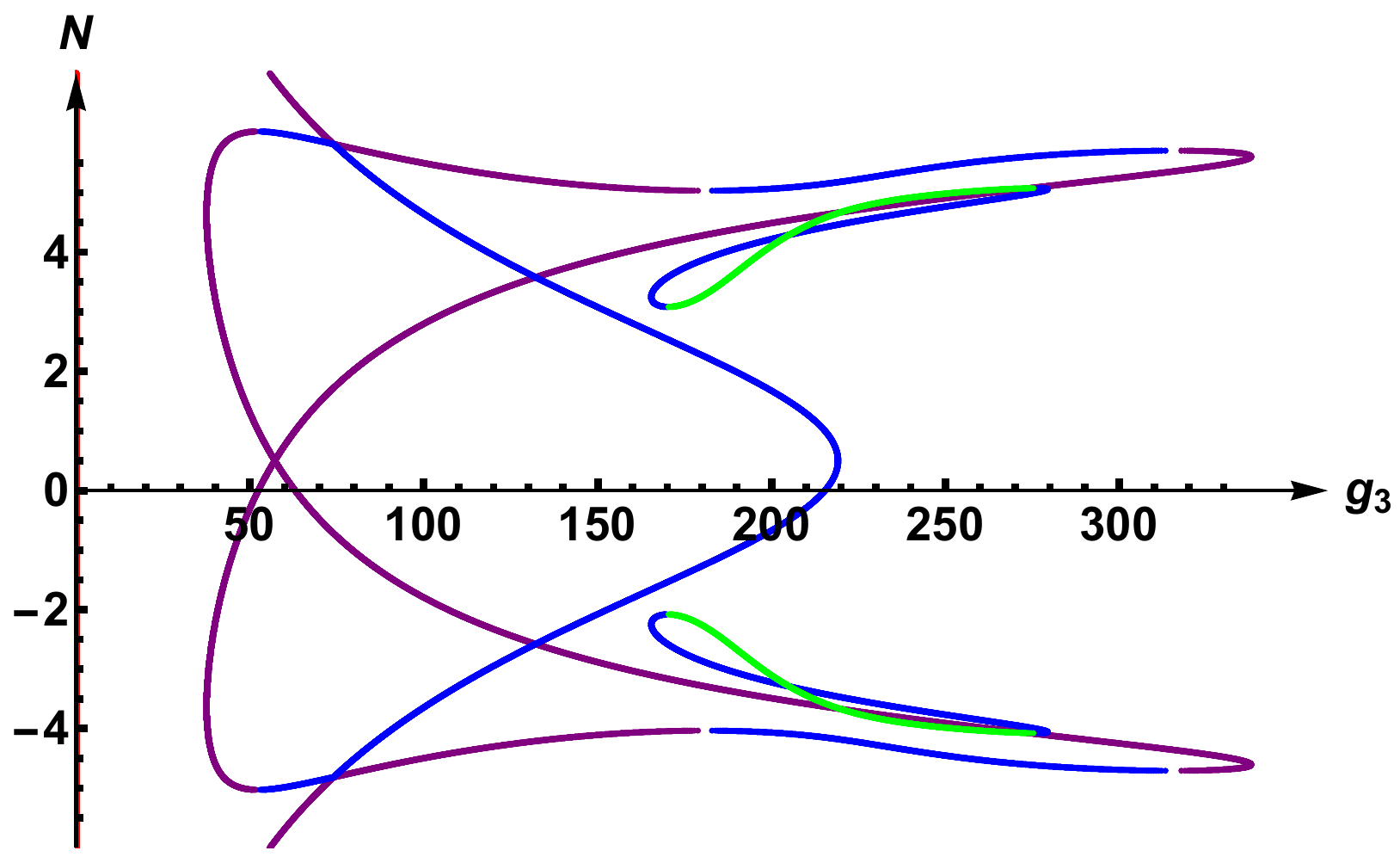}
\caption{The locations of the real perturbative fixed points of the anti-symmetric matrix model in the space of coupling constants as a function of $N$ for small $\epsilon$. The colors indicate the number of stable directions associated with a given fixed point as in figure \ref{antiMatFlow}.}
\label{actualAntiMatFlow}
\end{figure}

We can also study the behaviour of this model for finite $N$ and $\epsilon\ll 1$. For $N> 35.3546 - 673.428\, \epsilon$ there are three (real, perturbatively accessible) fixed points, which in the large $N$ limit (keeping $\epsilon \ll  \frac{1}{N^2}$) to leading order in $\epsilon$ scale with $N$ as 
\begin{align}
&\hspace{30mm} g_1 = g_2 = 0, \hspace{20mm} g_3 = \frac{6!(8\pi)^2}{144}\frac{\epsilon}{N^2},
\nonumber
\\
&g_1 =  \frac{6!(8\pi)^2}{9}\frac{\epsilon}{N^2}, 
\hspace{10mm} 
g_2 = -\frac{10}{9}\cdot6!(8\pi)^2\frac{\epsilon}{N^3},
 \hspace{10mm} 
 g_3 = \frac{6!(8\pi)^2}{144}\frac{\epsilon}{N^2},
 \label{eq:athreeFix}
 \\
& g_1 =  \frac{6!(8\pi)^2}{9}\frac{\epsilon}{N^2},
\hspace{10mm} 
g_2 = -2\cdot 6!(8\pi)^2\frac{\epsilon}{N^3},
 \hspace{10mm} 
 g_3 = \frac{295}{27}\cdot 6!(8\pi)^2\frac{\epsilon}{N^4}.
 \nonumber
\end{align}
The first of these three fixed points is the vector model fixed point, and it is present more generally in the small $\epsilon$ regime we are considering:
\begin{align}
g_1 = g_2 = 0, \hspace{20mm} g_3 = \frac{6!(8\pi)^2}{48(44-3N+3N^2)}\epsilon.
\end{align}
The third fixed point in \eqref{eq:athreeFix} extends to the regime where $N^2>\frac{1}{\epsilon}$ and becomes the large $N$ solution discussed above. This fixed point merges with the second fixed point in \eqref{eq:athreeFix} at a critical point situated at $N(\epsilon) = 35.3546 - 673.428\, \epsilon$ And so at intermediate values of $N$, only the vector model fixed point exists. But as we keep decreasing $N$ we encounter another critical point at $N(\epsilon)=6.02669+7.37013\epsilon$, from which two new solutions to the vanishing beta functions emerge. As $N$ further decreases past the value $N(\epsilon)=5.70601+0.540694\epsilon$, another pair of fixed points appear, and past $N(\epsilon)=5.075310-0.0278896\epsilon$ yet another pair of fixed point appear (in this range of $N$, all seven non-trivial solutions to the vanishing beta functions are real). But already below $N(\epsilon)=5.03275-0.586724\epsilon$, two of the fixed points become complex, and below $N(\epsilon)=3.08122+8.26176\epsilon$ two more fixed points become complex, so that for $N$ below this value there are a total of three real non-trivial fixed points. The behaviour of the various fixed points as a function of $N$ is summarized in more detail in figures \ref{antiMatFlow} and \ref{actualAntiMatFlow}.

\subsection{Symmetric traceless matrices and violation of large $N$ equivalence}

There is a projection of the parent theory of general real matrices $\phi^{ab}$ which restricts them to symmetric matrices $\phi=\phi^T$. In order to have an irreducible representation of $O(N)$ we should also require them to be traceless $\tr\phi=0$. Then the propagator is given by
\begin{align}
\big<\widetilde\phi^{ab}(k)\widetilde\phi^{a'b'}(-k)\big>_0
=\frac{1}{2k^2}
\Big(
\delta^{aa'}\delta^{bb'}
+\delta^{ab'}\delta^{ba'}
-\frac{2}{N}\delta^{ab}\delta^{a'b'}
\Big)\,.
\end{align}
The operators $O_{1,2,3,4}$ are actually independent for $N>5$, while for $N=2,3,4,5$ there are linear relations between them:
\begin{itemize}
	\item $N=2: O_4=0, \quad O_3 = 2 O_2 = 4 O_1$,
	\item $N=3: O_3 =2O_2,\quad  2O_4 = 3O_3+6O_1$,
	\item $N=4,5: 18O_2 + 8O_4 = 24O_1+3O_3$.
\end{itemize}
We will see that the existence of these relations for small integer values of $N$ has interesting implications for the analytic continuation of the theory from $N>5$ to $N<5$. 

Let us first discuss the large $N$ theory.
For the rescaled couplings $\lambda_1$, $\lambda_2$, and $\lambda_3$, the large $N$ beta functions are the same as \eqref{antiBeta} for the anti-symmetric model.
But now there is an additional coupling constant, whose large $N$ beta function is given by
\begin{align}
\beta_{\lambda_4}=&-2\epsilon \lambda_4
+72 \lambda_1^2  + 36 \lambda_1 \lambda_4  + 6 \lambda_4^2 - 738 \lambda_1^2 \lambda_4
-18 (180 + 11 \pi^2)\lambda_1^3
\,.
\end{align}
Consequently, the RG flow now has five non-trivial fixed points, two of which are real fixed points but with coupling constants containing $\mathcal{O}(\epsilon^0)$ terms. Another pair of fixed points is given by
\begin{gather}
\lambda_1=\frac{\epsilon}{9}
+\frac{17+\pi^2}{81}\epsilon^2,
\quad
\lambda_2=-2\epsilon-\frac{22+7\pi^2}{9}\epsilon^2,
\quad
\lambda_3=\frac{295}{27}\epsilon
+\frac{4714+6301\pi^2}{486}\epsilon^2,
\nonumber
\\
\lambda_4=
\frac{-3\pm i\sqrt{39}}{18}\epsilon
+\frac{273-78\pi^2\pm i\sqrt{39}(67+12\pi^2)}{2106}\epsilon^2\,.
\end{gather}
The first three coupling constants assume the same value as for the anti-symmetric model, a rescaled version of \eqref{parentFixed} of the parent theory, but the additional coupling constant assumes a complex value, thus breaking large $N$ equivalence and suggesting that the fixed point is unstable and described by a complex CFT \cite{Gorbenko:2018ncu,Gorbenko:2018dtm}.

We find that the eigenvalues of $\frac{\partial\beta_{\lambda_i}}{\partial\lambda_j}$ at this complex fixed point are
\begin{align}
\big\{-2\epsilon+\frac{32}{9}\epsilon^2,\,
\mp 2i\sqrt{\frac{13}{3}}\epsilon\pm 2i\frac{67+12\pi^2}{9\sqrt{39}}\epsilon^2,\,
\frac{2}{3}\epsilon-2\frac{22+5\pi^2}{27}\epsilon^2,\,
2\epsilon-2\frac{17+\pi^2}{9}\epsilon^2
\big\}
\end{align}
where the imaginary eigenvalue is associated with a complex linear combination of $\lambda_1$ and $\lambda_4$. 
 Thus, there is actually a pair of complex large $N$ fixed points:
at one of them there is an operator of complex dimension $d+ iA=3-\epsilon + i A$, while at the other it has dimension $d- i A$,\footnote{As $N$ is reduced, the two complex conjugate fixed points 
persist down to arbitrarily small $N$. For finite $N$, however, the complex scaling dimensions are no longer of the form $d\pm i A$: the real part deviates from $d$, which is consistent with 
the behavior of general complex CFTs \cite{Gorbenko:2018ncu,Gorbenko:2018dtm}.} where
$A=2\sqrt{\frac{13}{3}}\epsilon - 2\frac{67+12\pi^2}{9\sqrt{39}}\epsilon^2$. 
Thus, this pair of complex fixed points satisfy the criteria to be identified as complex CFTs \cite{Gorbenko:2018ncu,Gorbenko:2018dtm}. 
In our large $N$ theory, the scaling dimensions $d\pm i A$ correspond to the double-trace operator $O_4$, so that the single-trace operator $\tr \phi^3$
should have scaling dimension $\frac{1}{2}(d\pm i A)$. Indeed, we find that its two-loop anomalous dimension is, for large $N$, 
\begin{equation}
	\gamma_{\tr \phi^3} = 6\left(3\lambda_1 + \lambda_2\right)  = \epsilon \pm i \sqrt{\frac{13}{3}}\epsilon \ .
\end{equation}
Therefore,
\begin{equation}
 \Delta_{\tr\phi^3} =3 \left ( \frac{d}{2}-1 \right ) + \gamma_{\tr \phi^3}=  \frac{3-\epsilon}{2} \pm i \sqrt{\frac{13}{3}}\epsilon = \frac{d \pm i A}{2}\ .
\end{equation}
Scaling dimensions of this form are ubiquitous in large $N$ complex CFTs 
\cite{Pomoni:2008de,Kaplan:2009kr,Giombi:2017dtl,Giombi:2018qgp}. 
In the dual AdS description they correspond to fields violating the Breitenlohner-Freedman stability bound.

\begin{figure}
\centering
\scalebox{0.6}{
\includegraphics{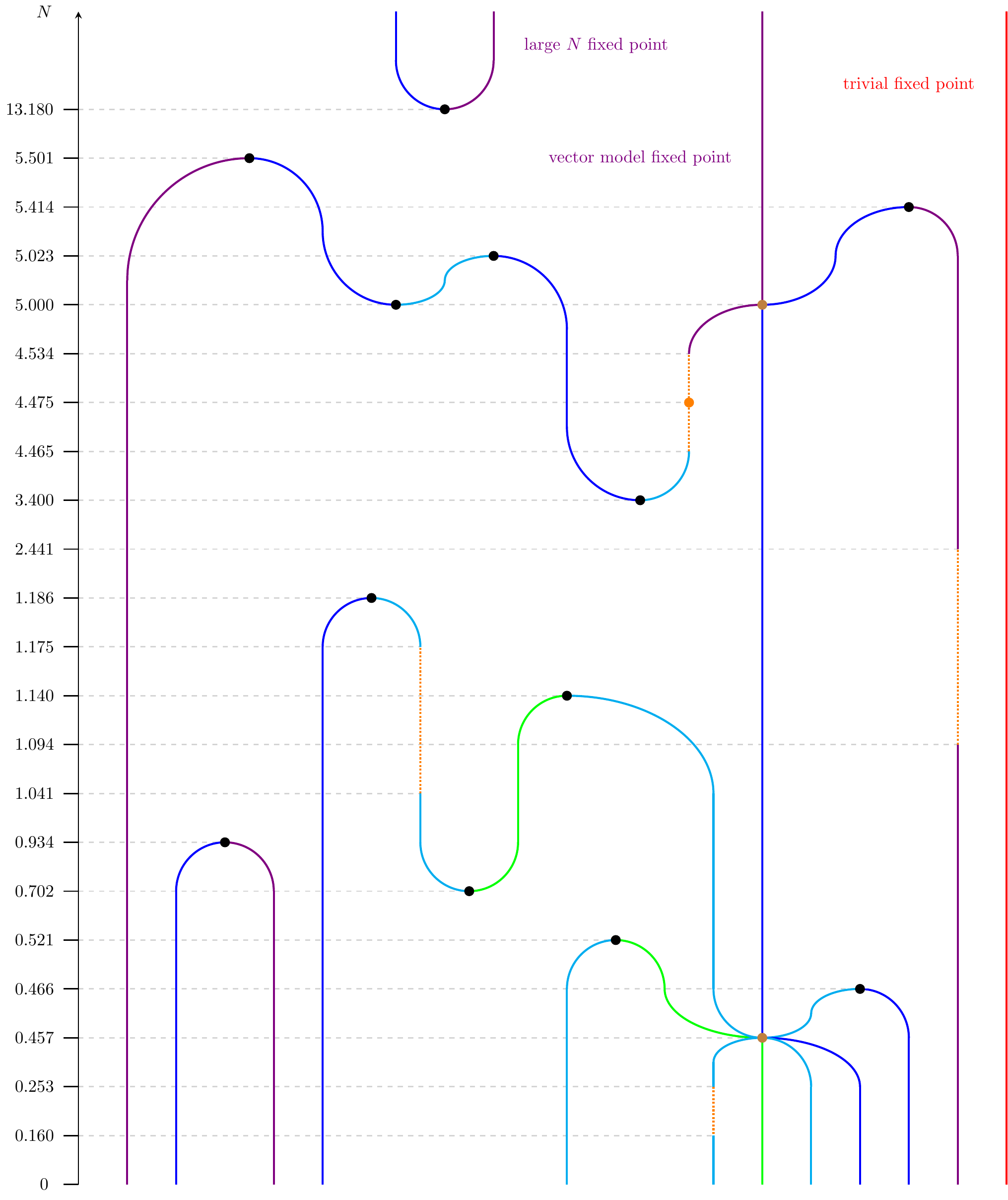}
}
			\scalebox{0.6}{
				\begin{tabular}{ |c|c|c|c|c| } 
					\hline
					$N$ & $g_1/\epsilon$ 
					& $g_2/\epsilon$ 
					& $g_3/\epsilon$ 
					& $g_4/\epsilon$ \\ 
					\hline
					$13.1802-57.5808 \epsilon$ 
					& $37.9805+498.738\epsilon$ 
					& $13.7692+157.614 \epsilon$
					& $0.774624+9.43200 \epsilon$ 
					& $21.5178+155.312 \epsilon$ \\ 
					\hline
					$5.50104 -0.966432\epsilon$ 
					& $1424.22 +11076.8 \epsilon$
					& $-1176.03 -9116.73 \epsilon$
					& $247.515 +1873.61 \epsilon$ 
					& $-454.872-3511.98 \epsilon$ \\
					\hline
					$5.41410 +13.7204\epsilon$ 
					& $24.4748 +360.178\epsilon$
					& $57.2276 +450.992 \epsilon$
					& $39.8006 -29.6552 \epsilon$ 
					& $-2.62055-19.2614 \epsilon$ \\
					\hline
					$5.02251+ 0.314146\epsilon$
					& $1132.14+13268.0\epsilon$
					& $-775.767-9368.16\epsilon$
					& $185.009+1864.18\epsilon$ 
					& $-372.446-4364.10\epsilon$ \\
					\hline
					$5 +\mathcal{O}(\epsilon)$ 
					& $\mathcal{O}(\epsilon^2)$
					& undetermined $\mathcal{O}(\epsilon)$
					& $15\pi^2/2+\mathcal{O}(\epsilon)$
					& $\mathcal{O}(\epsilon^2)$ \\
					\hline
					$5 $ 
					& $868.525 +8195.57 \epsilon$
					& $-651.394 -6497.79\epsilon$
					& $182.588 +1618.14 \epsilon$
					& $-289.508-2731.86 \epsilon$ \\
					\hline
					$3.39974+5.04412\epsilon$ 
					& $308.575 +3818.19\epsilon$
					& $-149.500 -2394.44\epsilon$
					& $113.071 +818.926\epsilon$
					& $-100.935-1242.36 \epsilon$ \\
					\hline
					$1.18613 -1.96911\epsilon$ 
					& $113.631+136.626\epsilon$
					& $-445.062 -3310.43\epsilon$
					& $475.932 +3758.3\epsilon$
					& $573.101+3747.7 \epsilon$ \\
					\hline
					$1.139999 -0.0564804\epsilon$ 
					& $7.14941 +103.455\epsilon$
					& $-121.617 -1749.67\epsilon$
					& $281.382 +2487.82\epsilon$
					& $113.505+1635.81\epsilon$ \\
					\hline
					$0.934072 -0.0890231\epsilon$ 
					& $0.0911386 +344.846\epsilon$
					& $-2777.40 -9338.97\epsilon$
					& $1172.45 +4559.95\epsilon$
					& $2333.04+8376.93 \epsilon$ \\
					\hline
					$0.701527 +10.3604\epsilon$ 
					& $12.8934 +848.994\epsilon$
					& $-57.8652 -4059.74\epsilon$
					& $279.112 +3827.54\epsilon$
					& $67.4704+4336.08 \epsilon$ \\
					\hline
					$0.521281 -14.4794\epsilon$ 
					& $3.96346 -441.552\epsilon$
					& $-16.5232 +1957.63\epsilon$
					& $257.847+606.789\epsilon$
					& $22.3424-2270.44 \epsilon$ \\
					\hline
					$0.465602 -6.81219\epsilon$ 
					& $1.79072 -162.063\epsilon$
					& $24.3958 -1503.26\epsilon$
					& $228.454 +2430.09\epsilon$
					& $-15.2518+919.203 \epsilon$ \\
					\hline
					$(\sqrt{33}-3)/6 +\mathcal{O}(\epsilon)$ 
					& undetermined $\mathcal{O}(\epsilon)$ 
					& undetermined $\mathcal{O}(\epsilon)$ 
					& $24\pi^2 +\mathcal{O}(\epsilon)$
					& $\mathcal{O}(\epsilon)$  \\
					\hline
				\end{tabular}
			}
	\caption{The perturbative real fixed points of the symmetric matrix model, the intersection points (marked in \textcolor{brown}{brown}), and the critical points at which they merge and disappear (marked in \textbf{black}) as a function of $N$ for small $\epsilon$. Fixed points that are IR-unstable in all four directions are drawn in \textcolor{red}{red}, those unstable in three directions are drawn in \textcolor{violet}{violet}, those unstable in two direction are drawn in \textcolor{blue}{blue}, those unstable in one direction are drawn in \textcolor{cyan}{cyan}, and those that are stable in all four directions are drawn in \textcolor{green}{green}. The \textcolor{orange}{orange} dotted lines denote the segments of ``spooky" fixed points, where two eigenvalues of $\frac{\partial \beta_i}{\partial g_j}$ are complex, and at the orange vertex those eigenvalues are purely imaginary. 
} 
	\label{MatFlowSym}
\end{figure}

Let us also note that the symmetric orbifold has a fixed point where only the twisted sector coupling is non-vanishing:
\begin{align}
	\lambda_{1,2,3}=0,
\quad
\lambda_4=\frac{\epsilon}{3}\,.
\end{align}
It could be connected to the fact that in the large $N$ limit of the parent theory the $O_4$ could not contribute to the beta functions of the other operators and therefore we can safely set $\lambda_{1,2,3}=0$ without setting $\lambda_4\neq 0$. 

\begin{figure}
	\centering
	\includegraphics[width=0.45\textwidth]{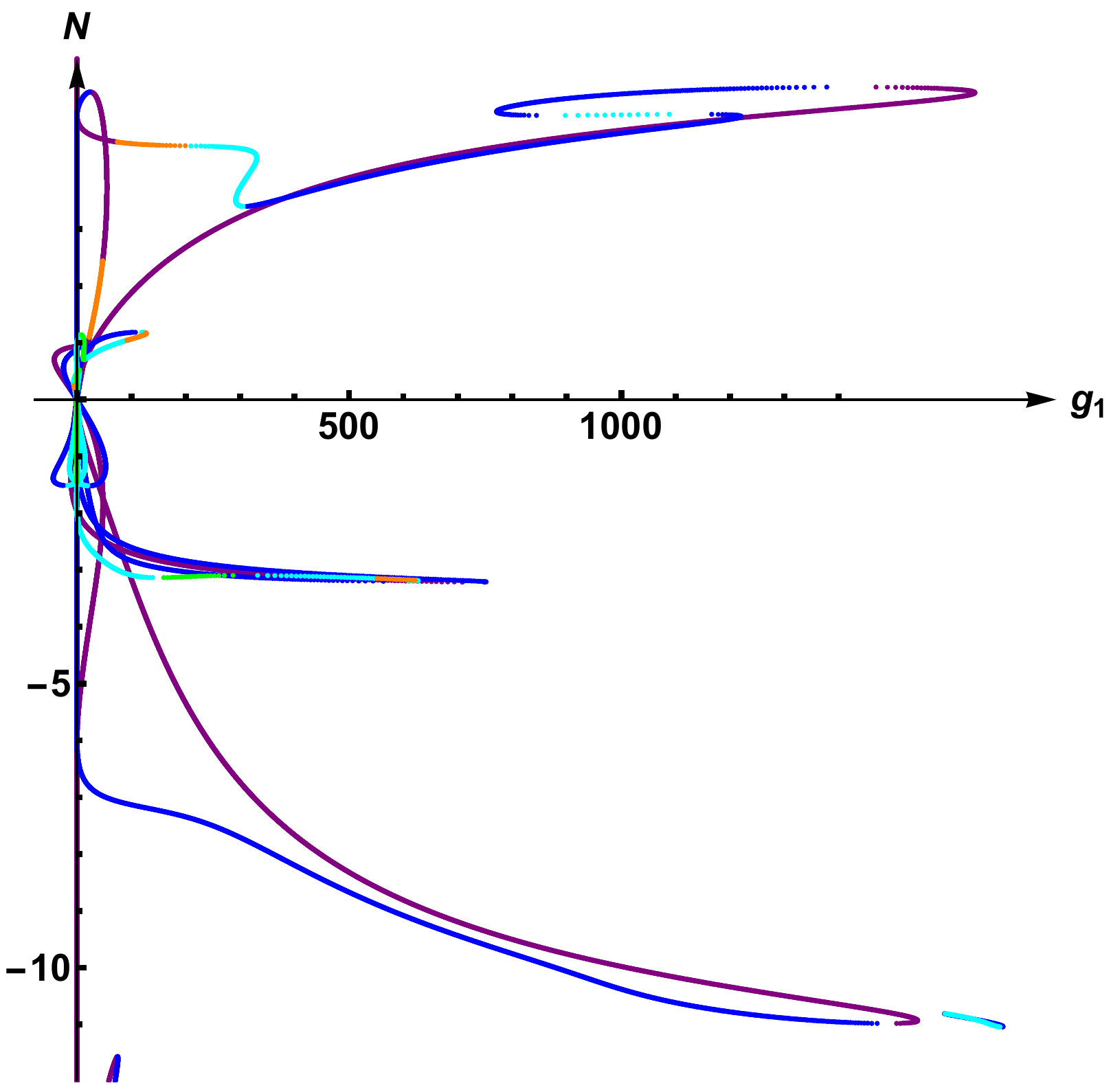}%
	\quad\quad \includegraphics[width=0.45\textwidth]{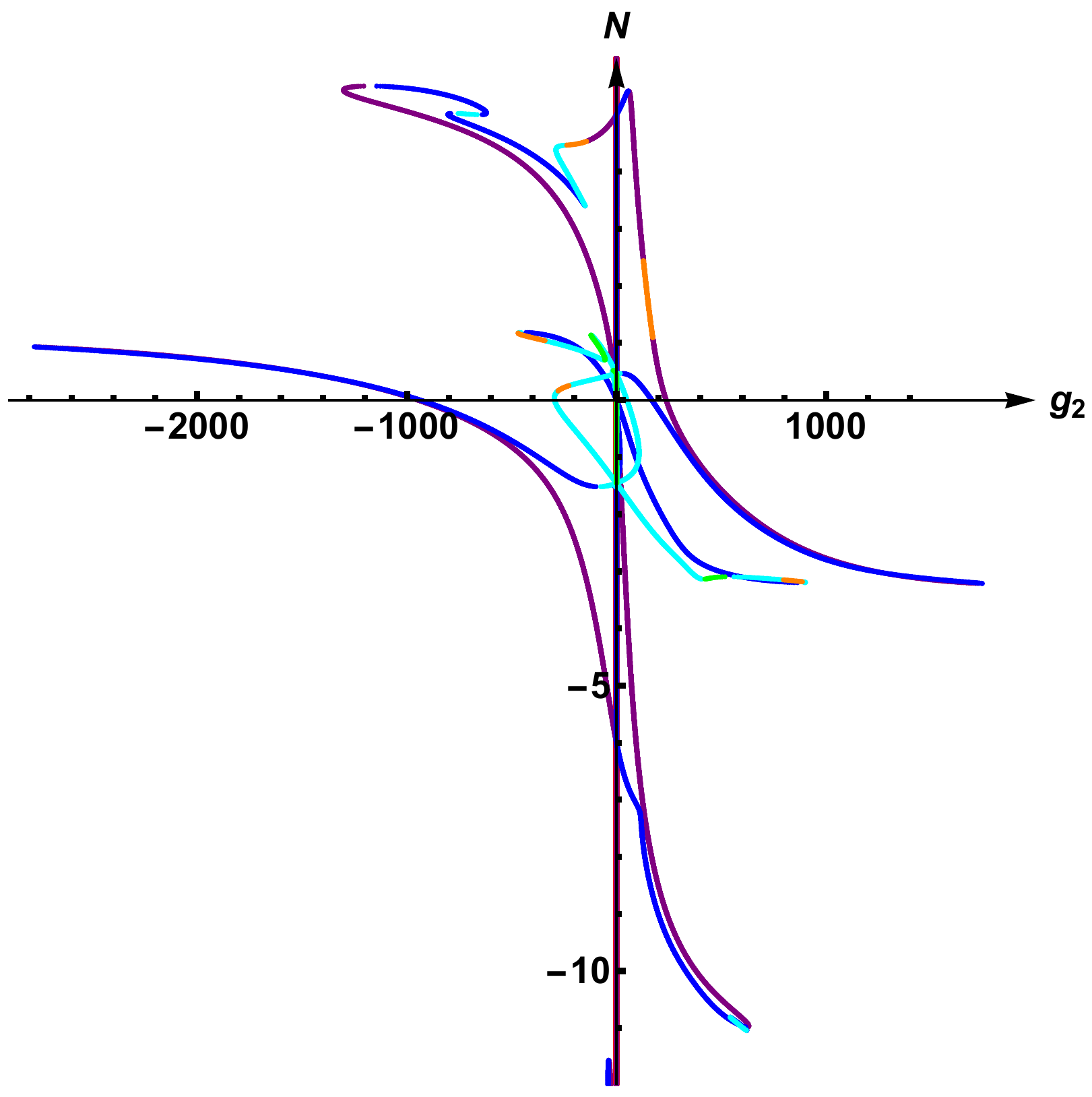}\\
	\includegraphics[width=0.45\textwidth]{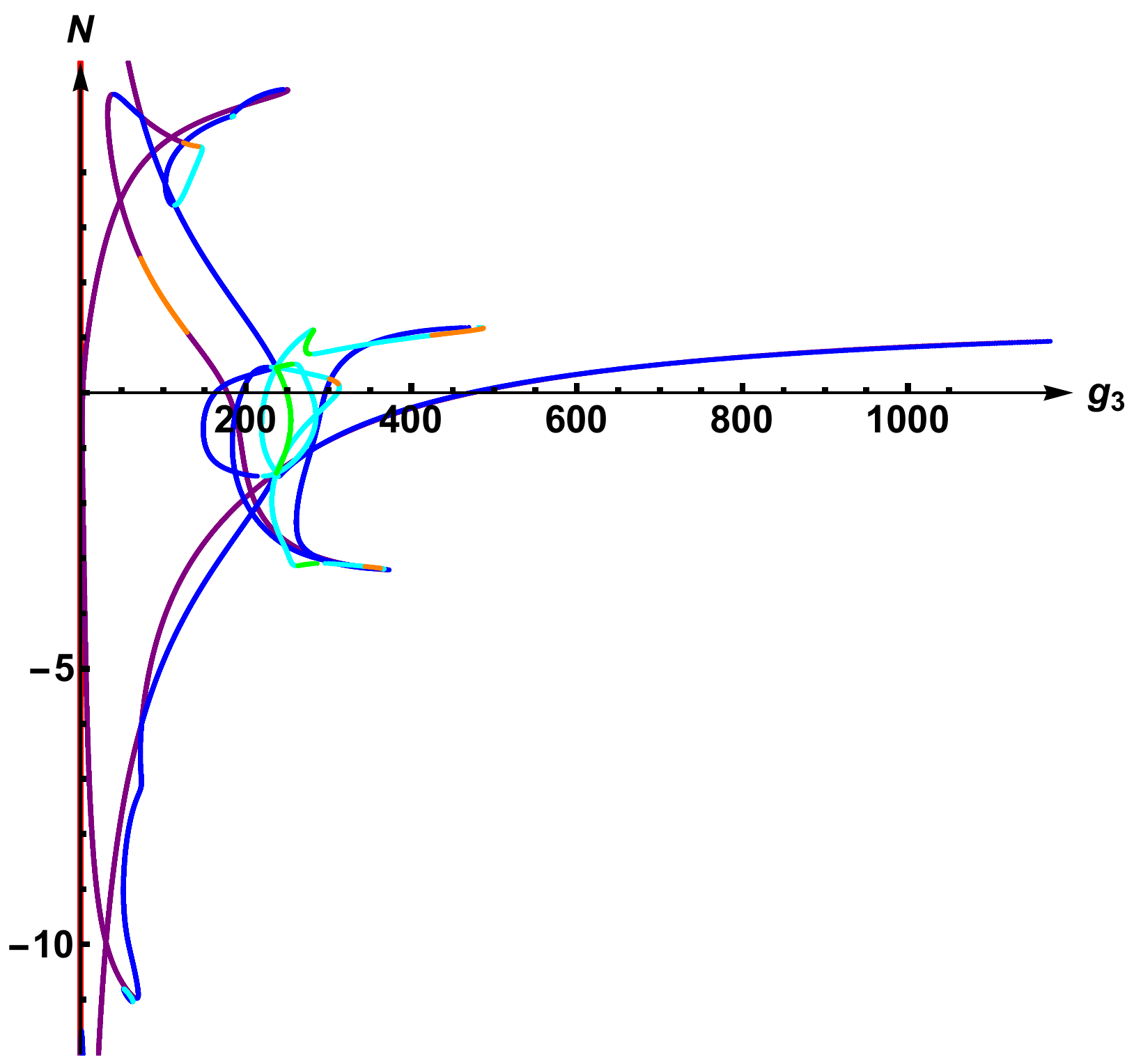}%
	\quad \quad \includegraphics[width=0.45\textwidth]{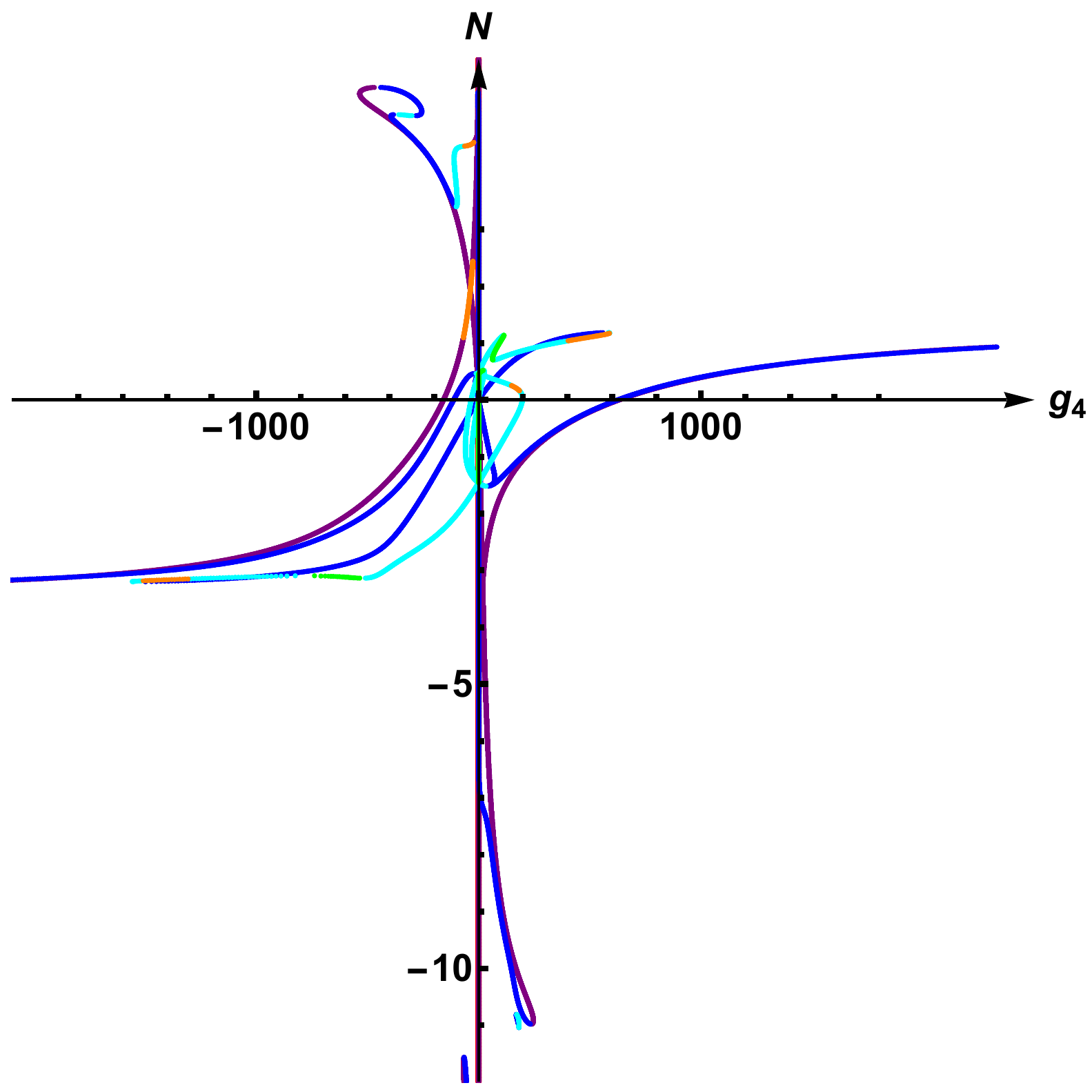}
\caption{The locations of the real perturbative fixed points of the symmetric matrix model in the space of coupling constants as a function of $N$ for small $\epsilon$. The colors indicate the number of stable directions associated with a given fixed point as in figure \ref{MatFlowSym}, with \textcolor{orange}{orange} signifying that $\frac{\partial \beta_i}{\partial g_j}$ has complex eigenvalues.}
\label{actualSymMatFlow}
\end{figure}

We can also study the behaviour of this model for finite $N$ and $\epsilon\ll 1$. For $N> 13.1802 - 57.5808\, \epsilon$ there are three (real, perturbatively accessible) fixed points, which in the large $N$ limit (keeping $\epsilon \ll  \frac{1}{N^2}$) to leading order in $\epsilon$ scale with $N$ as 
\begin{align}
&0=g_1=g_2=g_4
\hspace{10mm}
g_3=\frac{6!(8\pi)^2}{144 N^2}\epsilon
\nonumber
\\
&g_1=144\frac{6!(8\pi)^2}{N^6}\epsilon
\hspace{5mm}
g_2=66\frac{6!(8\pi)^2}{ N^5}\epsilon
\hspace{5mm}
g_3=\frac{6!(8\pi)^2}{144 N^2}\epsilon
\hspace{5mm}
g_4=\frac{6!(8\pi)^2}{3N^3}\epsilon
 \label{eq:sthreeFix}
 \\
& g_1=-144\frac{6!(8\pi)^2}{N^6}\epsilon
\hspace{5mm}
g_2=18\frac{6!(8\pi)^2}{N^5}\epsilon
\hspace{5mm}
g_3=-18\frac{6!(8\pi)^2}{N^6}\epsilon
\hspace{5mm}
g_4=\frac{6!(8\pi)^2}{3N^3}\epsilon
 \nonumber
\end{align}
The first of these three fixed points is the vector model fixed point, which is present generally $N$ in the small $\epsilon$ regime:
\begin{align}
0=g_1=g_2=g_4
\hspace{10mm}
g_3=\frac{6!(8\pi)^2}{48(38+3N+3N^2)}\epsilon
\end{align}
The third fixed point in \eqref{eq:sthreeFix} connects to the large $N$ solution discussed above. 
This fixed point merges with the second fixed point in \eqref{eq:sthreeFix} at a critical point situated at $N(\epsilon) =  13.1802 - 57.5808\, \epsilon$ And so at intermediate values of $N$, only the vector model fixed point exists. But as we keep decreasing $N$ we encounter another critical point at $N(\epsilon)=5.41410+13.7204\,\epsilon$ whence two new fixed points emerge. As we continue to lower $N$, new fixed points appear and disappear as summarized in detail in figures \ref{MatFlowSym} and \ref{actualSymMatFlow}.

\section{Spooky Fixed Points and Limit Cycles}
\label{RGcycle}

As indicated in figure \ref{MatFlowSym}, in the $O(N)$  symmetric traceless model there exist four segments of real, but spooky fixed points as a function of $N$.\footnote{If we allow negative $N$, there is a fifth segment of spooky fixed points at $N\in(-3.148,-3.183)$.} For these fixed points the Jacobian matrix $\left(\frac{\partial \beta_i}{\partial g_j}\right)$
has, in addition to one negative and one positive eigenvalue, a pair of complex conjugate eigenvalues. 
Therefore, there are two complex scaling dimensions (\ref{scaldim}) at these spooky fixed points, so that they correspond to non-unitary CFTs.
The eigenvectors corresponding to the complex eigenvalues have zero norm (a derivation of this fact is given later in this section).
Let us note that, in the $O(N)^2$ model and $O(N)$ model with antisymmetric matrices there are no real fixed points with complex eigenvalues.
The symmetric traceless model provides a simple setting where they occur. In this section we take a close look at the spooky fixed points 
and show that they lead to a Hopf bifurcation and RG limit cycles.  
 
Of the four segments of spooky fixed points with positive $N$, 
three, namely those that fall within the ranges given by $N\in (1.094,2.441)$, $N\in (1.041,1.175)$, and $N\in (0.160,0.253)$, share the property that the complex eigenvalues never become purely imaginary. The number of stable and unstable directions therefore remain the same within these intervals. Something special happens, however, at the integer value $N=2$ that lies within the first interval. Here the two operators with complex dimensions are given by linear combinations of operators $O_i$ 
that vanish by virtue of the linear relations between these operators at $N = 2$.\footnote{This is similar to what happens to evanescent operators when they are continued to an integer dimension.}
As a result, for $N=2$ there are no nearly marginal operators with complex dimensions, as expected.  

\begin{figure}
\begin{center}
\includegraphics[height=30ex]{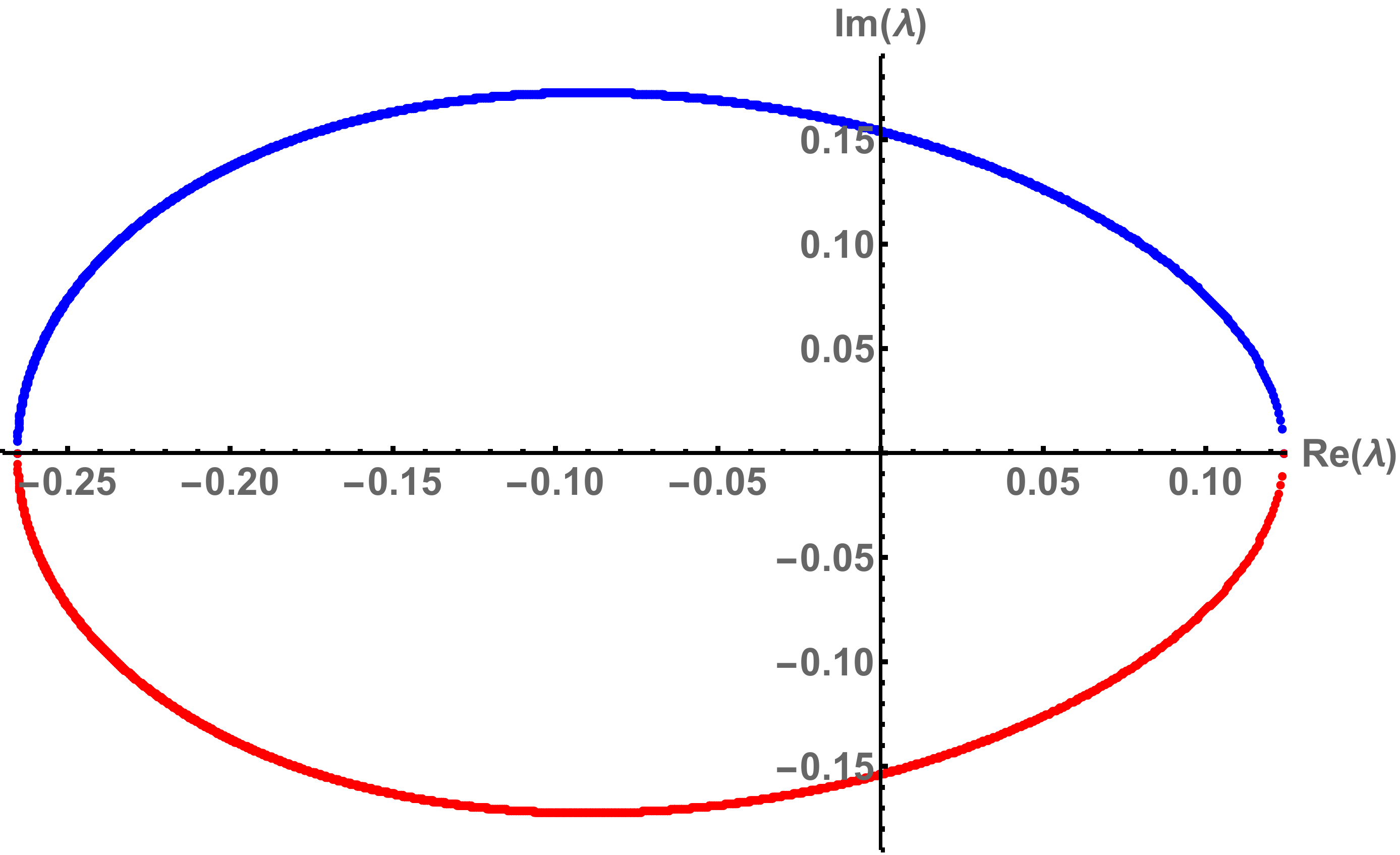}
\end{center}
\caption{The trajectories of the complex eigenvalues of the Jacobian matrix $\left(\frac{\partial \beta_i}{\partial g_j}\right)$ as $N$ is varied from $N_{\text{lower}}$ to $N_{\text{upper}}$.}
\label{trajectories}
\end{figure}

The fourth segment of spooky fixed points  
stands out in that it includes a fixed point with imaginary eigenvalues. This fourth segment lies in the range $N \in (N_{\text{lower}},N_{\text{upper}})$, where, at four-loop level,
\begin{align}
N_{\text{upper}}\approx 4.5339959143 + 1.54247 \epsilon \ , \qquad  N_{\text{lower}} \approx 4.4654144982 + 0.693698 \epsilon\ .
\end{align} 
As $N$ approaches $N_{\text{upper}}$ from above, $\left(\frac{\partial \beta_i}{\partial g_j}\right)$ has one positive and three negative eigenvalues, and two of the negative eigenvalues converge on the same value. 
As $N$ dips below $N_{\text{upper}}$, the two erstwhile identical eigenvalues become complex and form a pair of complex conjugate values.
 As we continue to decrease $N$, the complex conjugate eigenvalues traverse mirrored trajectories in the complex plane until they meet at the same positive value for $N$ equal to $N_{\text{lower}}$. These trajectories are depicted in figure \ref{trajectories}. For a critical value $N=N_{\text{crit}}$ with $N_{\text{lower}} < N < N_{\text{upper}}$, the trajectories intersect the imaginary axis such that the two eigenvalues are purely imaginary. At the two-loop order we find that 
\begin{gather}
N_{\rm crit}\approx 4.47507431683  \ ,
\end{gather}
 and the fixed point is located at 
\begin{gather}
 g_1^* = 158.684\epsilon,\quad g_2^* = -211.383\epsilon, \notag\\ 
 g_3^* =  138.686\epsilon,\quad  g_4^* =-49.4564 \epsilon\ .
\end{gather}
The Jacobian matrix evaluated at this fixed point is 
\begin{gather}
\left( \frac{\partial \beta_i}{\partial g_j}\right) = 
\begin{pmatrix}
	-1.65273 & -1.58311 & 1.33984 & -1.19641 \\
	1.0242 & 0.358518 & -3.24194 & 1.21102 \\
	0.128059 & 0.749009 & 2.9199 & -0.210872 \\
	-0.0618889 & 0.428409 & -0.417582 & -1.20064 \\
\end{pmatrix}
\epsilon
\end{gather}
with eigenvalues $\left\{2 ,-1.57495 , -0.153965 i , 0.153965 i \right\}\epsilon$.
These quantities are subject to further perturbative corrections in powers of $\epsilon$; for example, after including the four-loop corrections
$N_{\rm crit}\approx 4.47507431683  + 3.12476 \epsilon$. The existence of a special spooky fixed point with imaginary eigenvalues is robust under loop corrections that are suppressed by a small expansion parameter, since small perturbations of the trajectories still result in curves that intersect the imaginary axis. In light of the negative value of $g_4^\ast$, one may worry that the potential is unbounded from below at the spooky fixed points. It is not clear how to resolve this question for non-integer $N$, but at the fixed points at $N=4$ and $N=5$ that this spooky fixed point interpolates between, one can explicitly check that the potential is bounded from below.

The appearance of complex eigenvalues changes the behavior of the RG flow around the spooky fixed point. Since the fixed point has one negative eigenvalue for all $N\in (N_{\text{lower}},N_{\text{upper}})$, there is an unstable direction in the space of coupling constants that renders the fixed point IR-unstable. But we can ask the following question: How do the coupling constants flow in the two-dimensional manifold that is invariant under the RG flow and that is tangent to the plane spanned by the eigenvectors of the Jacobian matrix with complex eigenvalues? 

If the real parts of these eigenvalues are non-zero, the spooky fixed point is a {\it focus} and the flow around it is described by spirals steadily moving inwards or outwards from the fixed point. For $N > N_{\text{crit}}$, the real parts are negative and the fixed point is IR-unstable, while for $N < N_{\text{crit}}$ the real parts are positive and the fixed point is stable. By the Hartman-Grobman theorem \cite{hartman1960lemma,grobman1959homeomorphism}, one can locally change coordinates (redefine the coupling constants) such that the beta-functions near the fixed
points are linear. Furthermore, one can get rid of the imaginary part of the eigenvalues in this subspace by a suitable field redefinition\footnote{For instance, in two dimensions with $z =x + i y$, the equation $\dot{z} = (-\alpha + i \omega) z$ can via a change of variable $z \rightarrow z e^{i \frac{\omega}{\alpha} \log|z|}$ be reduced to $\dot{z} = - \alpha z$.}. An analogous statement was given in \cite{Fortin:2012hn}. 
 \begin{figure}
	\centering
	\includegraphics[scale=0.75]{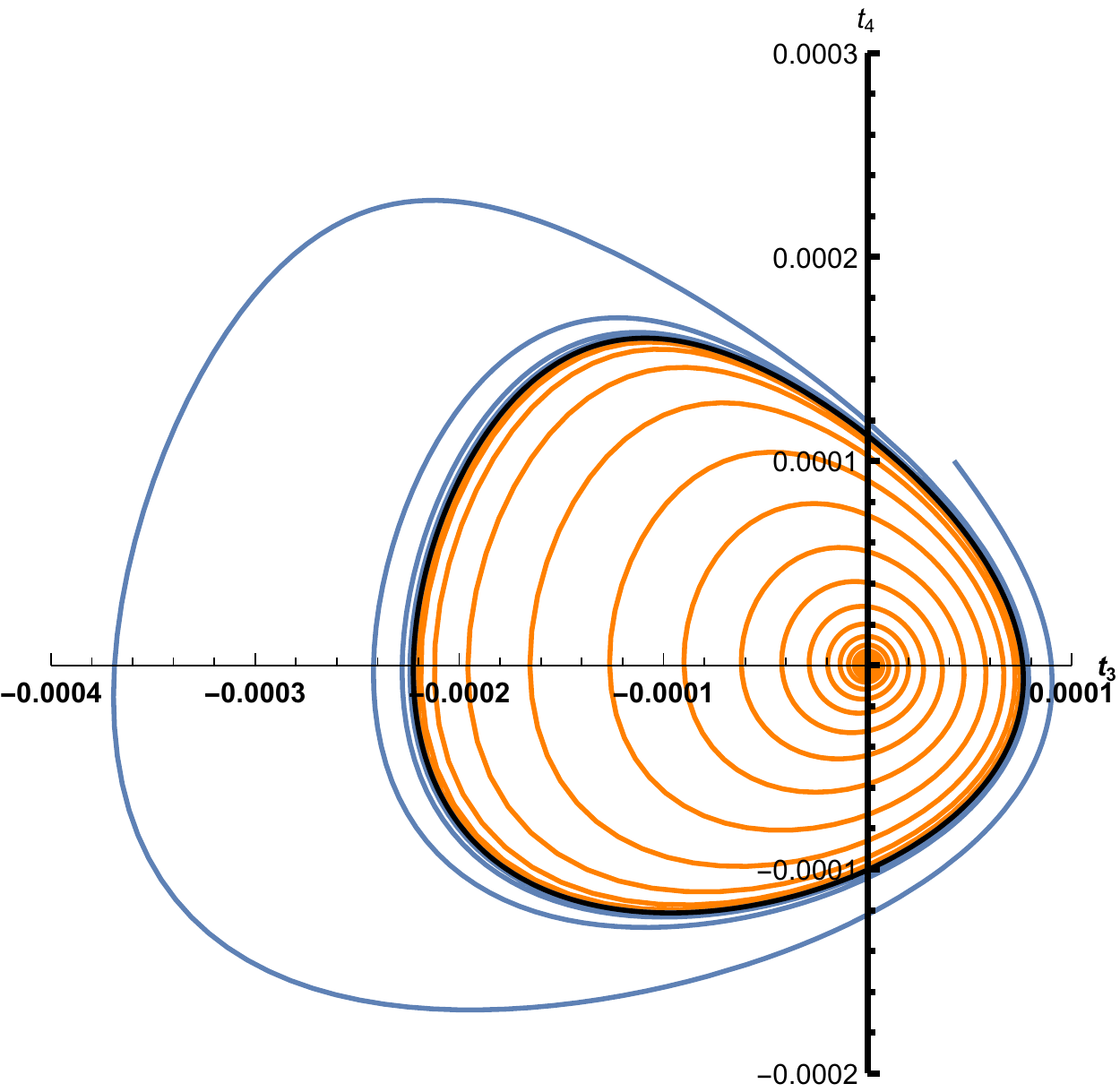}
	\caption{
The RG flow in the invariant manifold tangent to the plane spanned by the eigenvectors with complex eigenvalues in the space of coupling constants for $N=4.476$. In the IR, the \textcolor{myblue}{blue} curve whirls inwards towards a limit cycle marked in \textbf{black}, while the \textcolor{myorange}{orange} curve whirls outwards towards the limit cycle. The coordinates $t_3$ and $t_4$ are given by linear combinations of the couplings $g_1$, $g_2$, $g_3$, and $g_4$ and are defined in appendix \ref{HopfAppendix}. The RG flow on the invariant manifold admits of a description in an infinite expansion in powers of $t_3$ and $t_4$. This plot is drawn retaining terms up to cubic order.	
}

	\label{fig:limitcycle}
\end{figure}

When $N=N_{\text{crit}}$, the real parts of the complex eigenvalues are equal to zero. In this case the equilibrium point is a {\it center}, the Hartman-Grobman theorem is not applicable, and the behavior near the fixed point is controlled by the higher non-linear terms in the autonomous equations. If we consider $N$ as a parameter of the RG flow, $N=N_{\text{crit}}$ corresponds to a bifurcation point, as first introduced by Poincar{\` e}. A standard method of analyzing bifurcations is to reduce the full system to a set of lower dimensional systems by use of the center manifold theorem \cite{guckenheimer2013nonlinear}. Denoting by $\lambda$ the eigenvalues of the Jacobian matrix at a given fixed point, this theorem guarantees the existence of invariant manifolds tangent to the eigenspaces with $\operatorname{Re}\lambda>0$, $\operatorname{Re}\lambda<0$, and $\operatorname{Re}\lambda=0$ respectively. The latter manifold is known as the center manifold, and in general it need neither be unique nor smooth. But when, as in our case, the center at $g^\ast$ is part of a line of fixed points in the space $(g,N)$ that vary smoothly with a parameter $N$, and the complex eigenvalues satisfy
\begin{align}
\kappa = \frac{d}{dN}\text{Re}[\lambda(N)]\,\Big|_{N_{\text{crit}}} \neq 0 \,,
\end{align}
then there exists a unique 3-dimensional center manifold in $(\vec{g},N)$ passing through $(g^\ast,N_{\text{crit}})$. On planes of constant $N$ in this manifold, there exist coordinates $(x,y)$ such that the third order Taylor expansion can be written in the form
\begin{align}
\frac{dx}{dt} =\,& \Big(\kappa\, \delta N+a(x^2+y^2)\Big)x-\Big(\omega+c\, \delta N+b(x^2+y^2)\Big)y\,, \notag
\\
\frac{dy}{dt}=\,& \Big(\omega+c\, \delta N +b(x^2+y^2)\Big)x+\Big(\kappa\, \delta N+a(x^2+y^2) \Big)y\, , \label{eq:41}
\end{align}
where $t=\ln \mu$ and $\delta N = N-N_{\text{crit}}$.
The constant $a$ in these equations is known as the Hopf constant. By a theorem due to Hopf \cite{hopf1942bifurcation}, there exists an IR-attractive limit cycle in the center manifold if $a > 0$, while if $a < 0$ there exists an IR-repulsive limit cycle. To see why this makes sense, let 
$x+i y= r e^{i\phi}$. Then \eqref{eq:41} implies that
\begin{align*}
\frac{dr}{dt}=r\big(\kappa\, \delta N+ar^2\big) , \qquad \frac{d\phi}{dt} = \omega + c\, \delta N +  b r^2\, .
\end{align*}
For the critical point in the symmetric matrix model, $\kappa$ is negative, and in appendix \ref{HopfAppendix} we present an explicit calculation of $a$ and find that $a$ is positive. In the IR ($t\rightarrow -\infty$) the trajectory exponentially approaches a small circle of radius $\sqrt{-\kappa\, \delta N/a}$. We conclude that on analytically continuing in $N$, the RG flow of this QFT contains a periodic orbit in the space of coupling constants, an orbit that is unstable but which in the center manifold constitutes an attractive limit cycle. The periodic orbit exists for small positive values of $\delta N$. This conclusion holds true at all orders in perturbation theory, since the criteria of Hopf's theorem, being topological in nature, are not invalidated by small perturbative corrections. Figure \ref{fig:limitcycle} depicts a numerical plot of RG trajectories approaching the limit cycle. This limit cycle does not look circular because the coordinates used are different from those in \eqref{eq:41}.

Now that we have demonstrated the existence of limit cycles, we should ask about their consistency with the known RG monotonicity theorems. In particular, in 3 dimensions the $F$-theorem has been conjectured and established 
\cite{Myers:2010xs, Jafferis:2011zi, Casini:2012ei}. Furthermore, in perturbative 3-dimensional QFT, one can make a stronger statement that the RG flow is a gradient flow, i.e. 
\begin{equation}
G_{ij} \beta^j  = \frac{\partial F}{\partial g^i}\ , 
\label{gradflow}  
\end{equation}
where $F$ and the metric $G_{ij}$ are functions of the coupling constants which can be calculated perturbatively \cite{Klebanov:2011gs,Giombi:2014xxa,Fei:2015oha,Jack:2015tka,Jack:2016utw}.\footnote{
In  \cite{Jack:2015tka,Jack:2016utw} the terminology $a$-function was used, but we prefer to call it $F$-function instead, since $a$ typically refers to a Weyl anomaly coefficient in $d=4$.} 
At leading order, $G_{ij}$ may be read off from the two-point functions of the nearly marginal operators \cite{Giombi:2014xxa,Fei:2015oha}:
\begin{equation} 
\langle O_i (x) O_j (y)\rangle = \frac{G_{ij}}{|x-y|^6} \ .
\label{Zametric}
\end{equation} 
The $F$-function satisfies the RG equation
\begin{equation}
\mu \frac{\partial}{\partial \mu}F =  \frac{\partial}{\partial t}F=  \beta^i \beta^j  G_{ij}\ .   
\label{Fderiv}
\end{equation}
This shows that, if the metric is positive definite, then $F$ decreases monotonically as the theory flows towards the IR.
These perturbative statements continue to be applicable in $3-\epsilon$ dimensions.

At leading order, the metric $G_{ij}$ is exhibited in appendix \ref{ffun}.
Its determinant is given by 
\begin{equation}
\frac{(N-5) (N-4) (N-3)^2 (N-2)^3 N^{2} (N+1)^3 (N+3) (N+4)^3 (N+6)^2 (N+8) (N+10)}{2654208} \ .
  \end{equation}
This shows that the metric has three zero eigenvalues for $N=2$, two zero eigenvalues for $N=3$, and one zero eigenvalue for $N=4$ and $5$. This is due to the linear relations between operators $O_i$ at these integer values of $N$. For example, for $N=2$ there is only one independent operator.
In the range
$4<N<5$, $\det G_{ij}<0$, the metric has one negative and three positive eigenvalues.
This is what explains the possibility
of RG limit cycles in the range
$ N_{\rm lower} < N < N_{\text{upper}}$. For $N>5$, $G_{ij}$ is positive definite, and for $N<-10$, $G_{ij}$ is negative definite. This is consistent with our observing spooky fixed points only outside of these regimes.\footnote{We have also found the metric
 for the parent $O(N)^2$ theory. In this case it is positive definite for all $N$ except $N\in \{-4,-2,1,2\}$, where there are zero eigenvalues. We further found the metric for the anti-symmetric matrix model. In certain intervals within the range $N\in(-4,5)$ it has both positive and negative eigenvalues, but numerical searches reveal no spooky fixed points in these intervals.} 

In general, the norms of vectors computed with this metric are not positive definite for $N<5$. In particular, we can show that the eigenvectors corresponding to complex eigenvalues of the Jacobian matrix evaluated at real fixed points have zero norm.
Indeed, let us assume that we have a complex eigenvalue $m \in \mathbb{C}$ with eigenvector $u^i$
\begin{gather}
	\frac{\partial \beta^i}{\partial g^j} u^j = m u^i\ .\label{complexeigenvalue}
\end{gather} 
Now let us differentiate the relation \eqref{gradflow} with respect to $g^K$:
\begin{gather}\label{difgradflow}
	\partial_K G_{IJ} \beta^J + G_{IJ} \partial_K \beta^J = \partial_I \partial_K F\ .
\end{gather}
At a spooky fixed point we have $\beta^J(g)=0$ for real couplings $g$. Contracting the relation \eqref{difgradflow} with $u^K$ and $\bar{u}^I$ at a spooky fixed point we get
\begin{gather}
	\bar{u}^I G_{IJ} \partial_K \beta^J u^K = u^K \bar{u}^I \partial_I \partial_K F\ .
\end{gather}
Using \eqref{complexeigenvalue} we arrive at the following relations
\begin{gather}
	m \bar{u}^I u^J G_{IJ} = \bar{u}^I u^J \partial_I \partial_J F \ .
\end{gather}
Since $G_{IJ}$ and $\partial_I \partial_J F$ are real symmetric matrices, the norm $u^2 = G_{IJ} u^I \bar{u}^J$ and $f = \bar{u}^I u^J \partial_I \partial_J F$ are real numbers. If they are not equal to zero, then we must have $m \in \mathbb{R}$, which contradicts our assumption. Therefore, the norm $u^2=0$.
\begin{figure}
\centering
\includegraphics[scale=0.8]{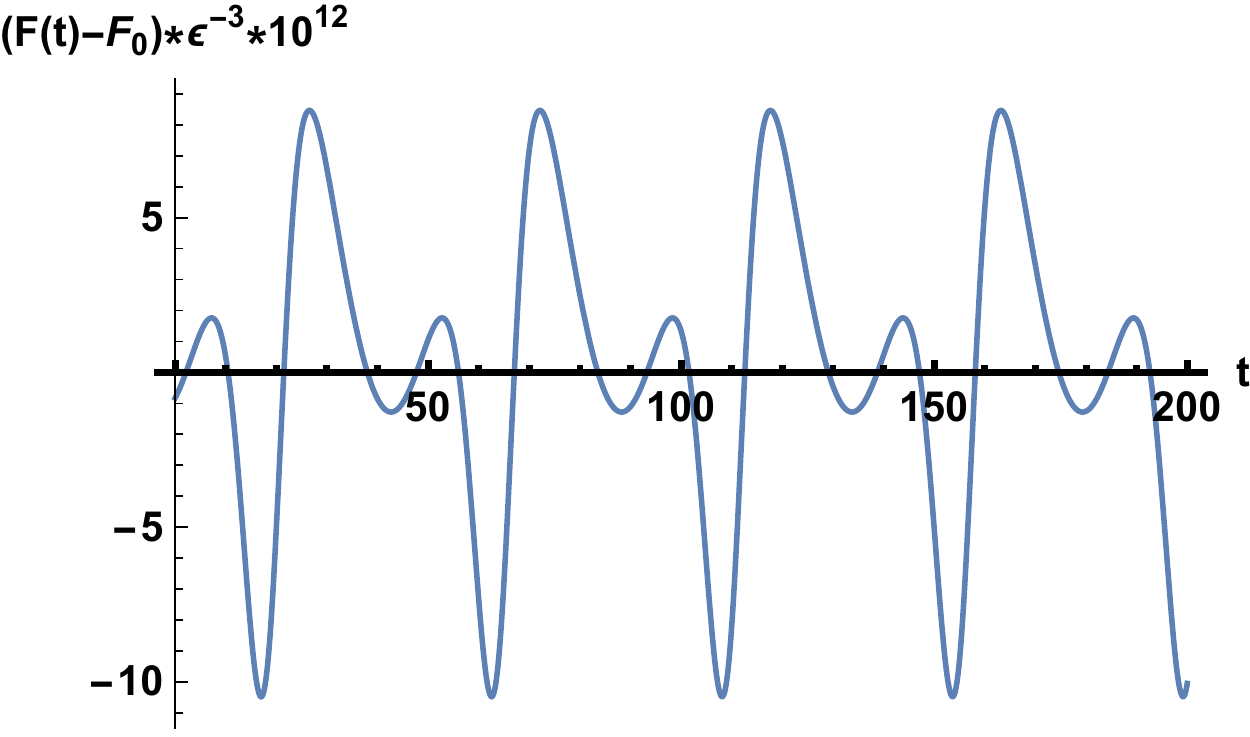}
\caption{The plot of $10^{12} (F(t) - F_{0})/\epsilon^3 $, where $F_0$ is the value at the spooky fixed point, for the cyclic solution found in section \ref{RGcycle} for $N=4.476$. 
}
\label{fig:ft}
\end{figure}

Another consequence of the negative eigenvalues of $G_{ij}$ is that $d F/dt$ can have either sign, as follows from (\ref{Fderiv}).
In fig. \eqref{fig:ft} we plot $F(t)$ for the limit cycle of fig. \ref{fig:limitcycle}, showing that it oscillates. 
This can also be shown analytically 
for a small limit cycle surrounding a fixed point. We may expand around it to find
\begin{equation}
\beta^i (t) = a(t) v^i + \bar a(t) \bar v^i\ ,
\end{equation}
where $v^i$ and $\bar v^i$ are the eigenvectors corresponding to the complex eigenvalues of the Jacobian matrix at the spooky fixed point. 
While $G_{ij} v^i \bar v^j $ vanishes, $G_{ij} v^i v^j \neq 0$. Therefore,
(\ref{Fderiv}) implies that $dF/dt\neq 0$ for a small limit cycle.

\section*{Acknowledgments}

We are grateful to Alexander Gorsky and Wenli Zhao for useful discussions. We also thank the referee for useful comments about evanescent operators. 
This research was supported in part by the US NSF under Grant No. PHY-1914860.

\appendix

\section{The beta functions up to four loops}

In the main text we presented the large $N$ beta functions for the matrix models we have studied. In this appendix we list the full beta functions for any $N$ up to four-loops. Letting $\mu$ denote the renormalization scale, we take the beta function associated with a coupling $g_i$ to be given by
\begin{align}
\beta_{g_i} = \mu \frac{d g_i}{d\mu}
=-2\epsilon g_i + \frac{1}{6!(8\pi)^2}\widetilde\beta_{g_i}^{(2)} + \frac{1}{(6!)^2(8\pi)^4}\widetilde\beta_{g_i}^{(4)}+\mathcal{O}(g^6)\,,
\end{align}
where we have separated out the two-loop contribution $\beta_{g_i}^{(2)}$ and the four-loop contribution $\beta_{g_i}^{(4)}$. The beta functions have been computed by use of the formulas for sextic theories in $d=3-\epsilon$ dimension listed in section \ref{masterSec}.

\subsection{Beta functions for the $O(N)^2$ matrix model}
\newgeometry{left=1cm,right=1cm,top=1cm,bottom=2cm}
\begin{gather}
\widetilde\beta_{g_1}^{(2)}=\,
24(100 + 24 N + 3 N^2) g_1^2 
+ 384  (9 + 4 N) g_1 g_2
+ 3840 g_1 g_3 
+ 64 (32 + N^2) g_2^2,
 \\ \nonumber
\widetilde\beta_{g_2}^{(2)}=\,
144 (8 + 3 N) g_1^2 
+ 96 (38 + 4 N + N^2) g_1 g_2
+ 2304  (1 + N) g_1 g_3
\notag\\ +128(8 + 7 N)  g_2^2 
+ 384 (18 + N^2)  g_2 g_3,
 \\ \nonumber
\widetilde\beta_{g_3}^{(2)}=\,
168 g_1^2 
+ 96  (3 + 2 N) g_1 g_2
+ 1152 g_1 g_3 
+ 32 (21 + 2 N + N^2)  g_2^2
\notag\\+ 768(1 + 2 N)  g_2 g_3 
+ 192(22 + 3 N^2) g_3^2
\end{gather}

\begin{gather}
\widetilde\beta_{g_1}^{(4)}=
-  288\Big(47952 + 4780 \pi^2 + N^4 (17 + \pi^2) + 
    N^3 (372 + 25 \pi^2) + 8 N (3102 + 277 \pi^2) 
    + 
    N^2 (5248 + 412 \pi^2)\Big)  g_1^3  \notag\\    
- 576  \Big(64992 + 6860 \pi^2 + 6 N^3 (104 + 7 \pi^2) + 
    8 N (4728 + 415 \pi^2) + N^2 (5928 + 465 \pi^2)\Big)g_1^2 g_2     
 \notag\\- 1152\Big(48 N (274 + 27 \pi^2) + N^2 (2824 + 225 \pi^2) + 
    4 (7640 + 891 \pi^2)\Big) g_1^2 g_3
\notag\\- 384  \Big(3 N^4 (10 + \pi^2) + 18 N^3 (12 + \pi^2) + 
    48 N (884 + 83 \pi^2) + 112 (867 + 94 \pi^2)      
    + 
    N^2 (10836 + 773 \pi^2)\Big) g_1 g_2^2
 \notag\\ -13824  \Big(3984 + 448 \pi^2 + 2 N^3 (20 + \pi^2) 
+     N^2 (92 + 7 \pi^2) + 8 N (292 + 31 \pi^2)\Big) g_1 g_2 g_3
 \notag\\- 4608\Big(5936 - 8 N^4 + 720 \pi^2 + N^2 (372 + 45 \pi^2)\Big)  g_1 g_3^2                  
-  \frac{512}{3}  \Big(N^3 (960 + 46 \pi^2) + 64 (900 + 97 \pi^2) 
  \notag\\ + N^2 (1704 + 137 \pi^2) + 16 N (2124 + 203 \pi^2)\Big)g_2^3 
-9216 \Big(4 N^4 + 384 (9 + \pi^2) + N^2 (248 + 21 \pi^2)\Big) g_2^2 g_3,
\end{gather}

\begin{gather}
\widetilde\beta_{g_2}^{(4)}=
-  432 \Big(20400 + 2260 \pi^2 + 2 N^3 (90 + 7 \pi^2) + 
    12 N (940 + 91 \pi^2) + N^2 (1740 + 151 \pi^2)\Big) g_1^3 
  \notag\\- 288  \Big(3 N^4 (10 + \pi^2) + 6 N^3 (56 + 5 \pi^2) + 
    16 (6408 + 683 \pi^2) + N^2 (11184 + 995 \pi^2) 
       + 
    N (46896 + 4516 \pi^2)\Big)g_1^2 g_2
 \notag\\- 1728  \Big(N^3 (248 + 22 \pi^2) + N^2 (1380 + 109 \pi^2) + 
    8 N (1510 + 127 \pi^2) + 4 (4132 + 401 \pi^2)\Big) g_1^2 g_3
 \notag\\ - 384 \Big(2 N^3 (534 + 49 \pi^2) + N^2 (5148 + 443 \pi^2) + 
    8 (8922 + 923 \pi^2) + N (48384 + 4444 \pi^2)\Big) g_1 g_2^2     
 \notag\\- 4608  \Big(2 N^4 (6 + \pi^2) + 6 N^3 (8 + \pi^2) + 
    6 N (948 + 77 \pi^2) + N^2 (2748 + 197 \pi^2) 
   + 
    2 (8112 + 841 \pi^2)\Big) g_1 g_2 g_3
 \notag\\ -27648  (1 + N) \Big(N^2 (62 + 3 \pi^2) + 
    2 (532 + 51 \pi^2)\Big) g_1 g_3^2
 \notag\\ -128 \Big(95152 + 10024 \pi^2 + 36 N^3 (6 + \pi^2) + 
    2 N^4 (36 + 7 \pi^2) + 24 N (1264 + 113 \pi^2) 
    +  N^2 (14804 + 1179 \pi^2)\Big) g_2^3 
 \notag\\- 768  \Big(2 N^4 \pi^2 + N^5 \pi^2 + 134 N^2 (12 + \pi^2) + 
    16 N^3 (102 + 7 \pi^2) + 8 (4308 + 433 \pi^2) 
    + 
    8 N (4584 + 437 \pi^2)\Big) g_2^2 g_3
 \notag\\- 13824 \Big(4816 + 512 \pi^2 + N^4 (18 + \pi^2) + 
    N^2 (644 + 57 \pi^2)\Big) g_2 g_3^2 
\end{gather}

\begin{gather}
\widetilde\beta_{g_3}^{(4)}=
- 432 \Big (2760 + 380 \pi^2 + N^2 (210 + 23 \pi^2) + 
    4 N (270 + 31 \pi^2)\Big) g_1^3
 \notag\\-
576  \Big(7308 + 776 \pi^2 + N^3 (78 + 8 \pi^2) + 
    N^2 (483 + 45 \pi^2) + 6 N (766 + 83 \pi^2)\Big) g_1^2 g_2
\notag\\-576  \Big(-48 N^3 - 8 N^4 + 6 N (836 + 81 \pi^2) + 
    6 (1984 + 189 \pi^2)  + N^2 (1676 + 207 \pi^2)\Big) g_1^2 g_3
\notag\\- 768 \Big(8772 + 894 \pi^2 + N^4 (6 + \pi^2) + 
    N^3 (36 + 5 \pi^2) + 10 N (336 + 31 \pi^2) 
    + 
    N^2 (1269 + 140 \pi^2)\Big) g_1 g_2^2 
\notag\\- 2304 \Big(6096 + 550 \pi^2 + 2 N^3 (84 + 17 \pi^2) + 
    N^2 (432 + 59 \pi^2) + N (6312 + 554 \pi^2)\Big) g_1 g_2 g_3 
\notag\\- 1152 \Big(18 N^3 \pi^2 + 15 N^4 \pi^2 + 
    96 N (35 + 3 \pi^2) + 8 N^2 (443 + 36 \pi^2) + 
    8 (1876 + 177 \pi^2)\Big)  g_1 g_3^2
 \notag\\- 384  \Big(41328 + 4192 \pi^2 + 2 N^4 (54 + 19 \pi^2) + 
    N^3 (216 + 38 \pi^2) + 8 N (1536 + 125 \pi^2)
     + 
    3 N^2 (3932 + 323 \pi^2)\Big) g_2^2 g_3
\notag\\- \frac{128}{3}\Big(49104 + 4784 \pi^2 + 4 N^4 \pi^2 + N^5 \pi^2 + 
    12 N^2 (487 + 42 \pi^2) + N^3 (2136 + 281 \pi^2) 
    + 
    12 N (4552 + 425 \pi^2)\Big)  g_2^3 
\notag\\ - 3456  (1 + 2 N) \Big(N^4 \pi^2 + 112 (32 + 3 \pi^2) + 
    4 N^2 (88 + 7 \pi^2)\Big) g_2 g_3^2
\notag\\ - 1152 \Big(N^6 \pi^2 + N^4 (424 + 34 \pi^2) + 
    32 (826 + 85 \pi^2) + N^2 (6864 + 620 \pi^2)\Big)g_3^3 
\end{gather}

\subsection{Beta functions for the anti-symmetric matrix model}

\begin{gather}
\widetilde\beta_{g_1}^{(2)}
= 6 (112 - 3 N + 3 N^2)g_1^2 + 384 (-1 + 2 N) g_1 g_2
+3840 g_1 g_3  + 32  (64 - N + N^2)g_2^2 
\\ \nonumber
\widetilde\beta_{g_2}^{(2)}
=
54 (-1 + 2 N)g_1^2 + 24  (68 - N + N^2)g_1 g_2+ 576  (-1 + 2 N)g_1 g_3 + 224  (-1 + 2 N)g_2^2
+ 
192  (36 - N + N^2) g_2 g_3
\\  \nonumber
\widetilde\beta_{g_3}^{(2)}
=
42 g_1^2+ (-24 + 48 N)g_1 g_2  + 576 g_1 g_3  + 
 8  (40 - N + N^2)g_2^2 + 384  (-1 + 2 N) g_2 g_3
 + 96 (44 - 3 N + 3 N^2)g_3^2 
\end{gather}

\begin{gather}
\widetilde\beta_{g_1}^{(4)}
=
-  9 \Big(-4 N^3 (17 + \pi^2) + 2 N^4 (17 + \pi^2) + 
    32 (3209 + 293 \pi^2) - N (10928 + 861 \pi^2) +   
    N^2 (10962 + 863 \pi^2)\Big)  g_1^3
\nonumber \\ 
-  72  (-1 + 2 N) \Big(-3 N (104 + 7 \pi^2) + 
     3 N^2 (104 + 7 \pi^2) + 4 (4896 + 413 \pi^2)\Big) g_1^2 g_2
- 288  \Big(-N (2824 + 225 \pi^2) 
\notag\\+ N^2 (2824 + 225 \pi^2) + 
    4 (7804 + 945 \pi^2)\Big) g_1^2 g_3
- 48 \Big(198048 + 21616 \pi^2 - 6 N^3 (10 + \pi^2) + 
    3 N^4 (10 + \pi^2) 
   \notag \\ 
  +2 N^2 (10479 + 746 \pi^2)  - 
    N (20928 + 1489 \pi^2)\Big)g_1 g_2^2 
-  3456  (-1 + 2 N) \Big(-N (20 + \pi^2) + N^2 (20 + \pi^2) \notag
\\+
    8 (292 + 31 \pi^2)\Big) g_1 g_2 g_3 
-  2304 \Big(8 N^3 - 4 N^4 - 3 N (124 + 15 \pi^2) + 
    N^2 (368 + 45 \pi^2) + 32 (371 + 45 \pi^2)\Big) g_1 g_3^2 
\nonumber \\ 
-  \frac{128}{3} (-1 + 2 N) \Big(33984 + 3248 \pi^2 - N (480 + 23 \pi^2) + 
    N^2 (480 + 23 \pi^2)\Big)g_2^3  
\nonumber \\ 
-4608  \Big(-4 N^3 + 2 N^4 + 768 (9 + \pi^2) -  N (248 + 21 \pi^2) 
+ N^2 (250 + 21 \pi^2)\Big)g_2^2 g_3
\end{gather}

\begin{gather}
\widetilde\beta_{g_2}^{(4)}
=
-27  (-1 + 2 N) \Big(5760 + 557 \pi^2 - N (90 + 7 \pi^2) + 
    N^2 (90 + 7 \pi^2)\Big) g_1^3
- 18  \Big(-6 N^3 (10 + \pi^2) 
\nonumber\\+ 3 N^4 (10 + \pi^2) + 
    34 N^2 (579 + 52 \pi^2) - N (19656 + 1765 \pi^2)    
    + 
    4 (49956 + 5437 \pi^2)\Big)g_1^2 g_2  
\nonumber\\- 216 (-1 + 2 N) \Big(9536 + 830 \pi^2 - N (124 + 11 \pi^2) + 
    N^2 (124 + 11 \pi^2)\Big) g_1^2 g_3 
- 48 (-1 + 2 N) \Big(39744 + 3739 \pi^2 
\nonumber\\-N (534 + 49 \pi^2) + 
    N^2 (534 + 49 \pi^2)\Big)  g_1 g_2^2
- 288 \Big(-8 N^3 (6 + \pi^2) + 4 N^4 (6 + \pi^2) + 
    3 N^2 (3608 + 259 \pi^2) 
\nonumber\\- N (10800 + 773 \pi^2) 
     + 
    4 (25284 + 2719 \pi^2)\Big) g_1 g_2 g_3     
- 3456 (-1 + 2 N) \Big(-N (62 + 3 \pi^2) + 
    N^2 (62 + 3 \pi^2) 
\nonumber\\+ 4 (532 + 51 \pi^2)\Big)  g1 g3^2 
- 32  \Big(-2 N^3 (36 + 7 \pi^2) + N^4 (36 + 7 \pi^2) - 
    4 N (3656 + 291 \pi^2) + N^2 (14660 + 1171 \pi^2)     
\nonumber\\+ 4 (38180 + 4109 \pi^2)\Big) g_2^3       
- 48 (-1 + 2 N) \Big(-2 N^3 \pi^2 + N^4 \pi^2 - 
    32 N (102 + 7 \pi^2) + 3 N^2 (1088 + 75 \pi^2)   
\nonumber\\+ 
    32 (4584 + 437 \pi^2)\Big) g_2^2 g_3       
- 3456 \Big(-2 N^3 (18 + \pi^2) + N^4 (18 + \pi^2) + 
    64 (301 + 32 \pi^2) - 2 N (644 + 57 \pi^2)   
\nonumber\\+
    N^2 (1306 + 115 \pi^2)\Big)  g_2 g_3^2
\end{gather}

\begin{gather}
\widetilde\beta_{g_3}^{(4)}
=
-  27  \Big(2760 + 422 \pi^2 - N (210 + 23 \pi^2) + 
    N^2 (210 + 23 \pi^2)\Big) g_1^3   
-18  (-1 + 2 N) \Big(-2 N (39 + 4 \pi^2) 
\nonumber\\+
    N^2 (78 + 8 \pi^2) + 75 (96 + 11 \pi^2)\Big) g_1^2 g_2
- 72  \Big(16724 + 8 N^3 - 4 N^4 + 1647 \pi^2 + 
   3 N^2 (592 + 69 \pi^2) - \\-N (1780 + 207 \pi^2)\Big) g_1^2 g_3
- 12  \Big(-8 N^3 (6 + \pi^2) + 4 N^4 (6 + \pi^2) + 
    256 (402 + 43 \pi^2) + 3 N^2 (3184 + 351 \pi^2)  
\nonumber\\-
    N (9528 + 1049 \pi^2)\Big) g_1 g_2^2      
- 288 (-1 + 2 N) \Big(5952 + 518 \pi^2 - N (84 + 17 \pi^2) + 
    N^2 (84 + 17 \pi^2)\Big) g_1 g_2 g_3 
\nonumber\\- 144 \Big(-30 N^3 \pi^2 + 15 N^4 \pi^2 + 
    224 (253 + 24 \pi^2) - N (7088 + 567 \pi^2) + 
    N^2 (7088 + 582 \pi^2)\Big) g_1 g_3^2 
\nonumber\\- \frac{4}{3} (-1 + 2 N) \Big(-2 N^3 \pi^2 + N^4 \pi^2 - 
    24 N (178 + 23 \pi^2) + N^2 (4272 + 553 \pi^2) 
    + 
    8 (24672 + 2359 \pi^2)\Big)g_2^3     
\nonumber\\- 96  \Big(75984 + 7828 \pi^2 - 2 N^3 (54 + 19 \pi^2) + 
    N^4 (54 + 19 \pi^2) + 3 N^2 (3914 + 327 \pi^2) 
    - 
    2 N (5844 + 481 \pi^2)\Big) g_2^2 g_3    
\nonumber\\-432 (-1 + 2 N) \Big(-2 N^3 \pi^2 + N^4 \pi^2 + 
    448 (32 + 3 \pi^2) - 8 N (88 + 7 \pi^2) + 
    N^2 (704 + 57 \pi^2)\Big) g_2 g_3^2   
\nonumber\\-144\Big(-3 N^5 \pi^2 + N^6 \pi^2 + N^4 (848 + 71 \pi^2) + 
    256 (826 + 85 \pi^2) - N^3 (1696 + 137 \pi^2)    
    - 
    16 N (1716 + 155 \pi^2) 
\nonumber\\+ 4 N^2 (7076 + 637 \pi^2)\Big)  g_3^3 
\end{gather}

\subsection{Beta functions for the symmetric traceless matrix model}

\begin{gather}
\widetilde\beta_{g_1}^{(2)}
=\,
6  \frac{2400 - 1200 N + 250 N^2 + 51 N^3 + 3 N^4}{N^2} g_1^2
+ 384  \frac{2 N^2+ 10 N -35}{N} g_1 g_2
+3840 g_1 g_3 
\nonumber\\+864 \frac{-20 + 5 N + N^2}{N}g_1 g_4
+32 (62 + N + N^2) g_2^2 
+4608 g_2 g_4 
+ 2592 g_4^2\\
\widetilde\beta_{g_2}^{(2)}
=\,
18 \frac{- 150+35N + 6 N^2}{N} g_1^2 
+ 24 \frac{480 - 120 N + 66 N^2 + 9 N^3 + N^4}{N^2} g_1 g_2 
+ 576 \frac{-10 + 5 N + 2 N^2}{N} g_1 g_3
\nonumber \\+216  \frac{80 - 20 N + N^2}{N^2}g_1 g_4
+ 32 \frac{- 132+39N + 14 N^2}{N} g_2^2 
+  192 (34 + N + N^2)  g_2 g_3
+ 288\frac{-40 + 3 N + N^2}{N} g_2 g_4 
\nonumber\\+3456 g_3 g_4
+324 \frac{-24 + 2 N + N^2}{N} g_4^2 \\
\widetilde\beta_{g_3}^{(2)}
=\,
42 g_1^2 
+ 576 g_1 g_3 
+ 24\frac{- 30 + 7N + 2 N^2}{N}  g_1 g_2 
-\frac{1080}{N} g_1 g_4 
+ 384\frac{-6 + 3 N + 2 N^2}{N}  g_2 g_3
\nonumber\\
- 288  \frac{-24 + 3 N + N^2}{N^2} g_2 g_4
+ 96 (38 + 3 N + 3 N^2)  g_3^2
+ 8 \frac{288 - 36 N + 30 N^2 + 5 N^3 + N^4}{N^2} g_2^2
 - \frac{3456}{N} g_3 g_4
\nonumber\\-324  \frac{-16 + 2 N + N^2}{N^2} g_4^2\\
\widetilde\beta_{g_4}^{(2)}
=\,
24  \frac{- 200 - 75N^2+15N^3 + 3 N^4}{N^3} g_1^2
+ 192 \frac{10 - 5 N + N^2}{N^2} g_1 g_2
+ 12  \frac{160 - 120 N + 34 N^2 + 15 N^3 + 3 N^4}{N^2}g_1 g_4
\nonumber\\-32  \frac{62 + N + N^2}{N} g_2^2
+384 \frac{-15 + 3 N + N^2}{N}  g_2 g_4
+3840 g_3 g_4
+ 6  \frac{-704 + 60 N + 28 N^2 + 3 N^3 + N^4}{N} g_4^2
\end{gather}

\begin{gather}
\widetilde\beta_{g_1}^{(4)}=\,
- \frac{9}{N^4} \Big(2 N^8 (17 + \pi^2) + 4 N^7 (389 + 26 \pi^2) + 
    38400 (1252 + 135 \pi^2)
    - 19200 N (2159 + 225 \pi^2)  
\nonumber\\-
    60 N^4 (7338 + 455 \pi^2) - 1200 N^3 (4896 + 587 \pi^2)
    +  N^6 (38822 + 3167 \pi^2) 
    + 800 N^2 (30564 + 3215 \pi^2) 
\nonumber\\+
    N^5 (279004 + 28019 \pi^2)\Big) g_1^3
- \frac{216 }{N^3} \Big(2 N^6 (104 + 7 \pi^2) + 4320 N (522 + 55 \pi^2) + 
    10 N^3 (-1416 + 131 \pi^2) 
\nonumber\\+ N^5 (4264 + 331 \pi^2) - 
    960 (3344 + 375 \pi^2) + 5 N^4 (7096 + 681 \pi^2) - 
    40 N^2 (16616 + 1977 \pi^2)\Big)  g_1^2 g_2
\nonumber\\-\frac{288}{N^2}\Big(1920 (388 + 45 \pi^2) + 30 N^2 (1184 + 171 \pi^2)+ 
    N^4 (2824 + 225 \pi^2) - 120 N (3628 + 405 \pi^2)      
\nonumber \\+ N^3 (29128 + 2817 \pi^2)\Big)  g_1^2 g_3 
- \frac{54}{N^3} \Big(12 N^6 (95 + 7 \pi^2) - 20 N^3 (9940 + 471 \pi^2) + 
    3 N^5 (6608 + 561 \pi^2) 
\nonumber \\- 320 N^2 (9175 + 1104 \pi^2)   
    - 
    1280 (13166 + 1485 \pi^2) + 640 N (17947 + 1890 \pi^2) + 
    N^4 (137468 + 13785 \pi^2)\Big)  g_1^2 g_4    
\nonumber\\- \frac{48}{N^2} \Big(3 N^6 (10 + \pi^2) + 6 N^5 (82 + 7 \pi^2) + 
    9504 (756 + 85 \pi^2) - 2160 N (1581 + 172 \pi^2)     
    + 
    2 N^4 (10971 + 763 \pi^2) 
\nonumber\\+2 N^2 (37896 + 6847 \pi^2) + 
    N^3 (177600 + 16271 \pi^2)\Big)  g_1 g_2^2    
- \frac{3456}{N} \Big(2 N^4 (20 + \pi^2) + N^3 (244 + 17 \pi^2) - 
    24 (2372 
    \nonumber\\+ 275 \pi^2) + N^2 (4108 + 447 \pi^2) 
    + 
    2 N (8588 + 975 \pi^2)\Big)  g_1 g_2 g_3    
- \frac{1728}{N^2} \Big(518304 + 58320 \pi^2 + N^2 (-3072 + 41 \pi^2) 
\nonumber\\+
    2 N^4 (555 + 49 \pi^2) - 48 N (4834 + 525 \pi^2)   
     + 
    N^3 (10782 + 1037 \pi^2)\Big)  g_1 g_2 g_4    
- 2304 \Big(-8 N^3 - 4 N^4 
\nonumber\\+ N^2 (384 + 45 \pi^2) + 
    N (388 + 45 \pi^2) + 6 (1852 + 225 \pi^2)\Big) g_1 g_3^2    
- \frac{13824}{N} \Big(-47 (388 + 45 \pi^2) + N^2 (1120 + 117 \pi^2) 
\nonumber\\+
    N (4844 + 540 \pi^2)\Big)  g_1 g_3 g_4
- \frac{108}{N^2} \Big(301 N^5 + 41 N^6 + N^4 (10882 + 1053 \pi^2) - 
    4 N^2 (26366 + 1755 \pi^2) 
    \nonumber\\+ 128 (41992 + 4725 \pi^2) 
    +     N^3 (98224 + 9801 \pi^2) - 
    16 N (140912 + 15255 \pi^2)\Big)  g_1 g_4^2        
\nonumber\\- \frac{128}{3N}\Big(N^4 (960 + 46 \pi^2) + 108 N (2220 + 241 \pi^2) + 
    N^3 (4848 + 343 \pi^2) - 324 (3352 + 375 \pi^2)     
  \nonumber  \\+  7 N^2 (7584 + 749 \pi^2)\Big)  g_2^3     
- 4608 \Big(6424 + 4 N^3 + 2 N^4 + 726 \pi^2 + 
    3 N (80 + 7 \pi^2) + N^2 (242 + 21 \pi^2)\Big) g_2^2 g_3     
\nonumber\\-\frac{288}{N} \Big(N^4 (120 + 7 \pi^2) + N^3 (660 + 43 \pi^2) + 
    6 N^2 (4426 + 437 \pi^2) + 24 N (4957 + 534 \pi^2)   
    - 
    48 (12638 + 1413 \pi^2)\Big)  g_2^2 g_4    
    \nonumber\\
- 20736\Big(3368 + 378 \pi^2 + N (44 + 3 \pi^2) + 
    N^2 (44 + 3 \pi^2)\Big)  g_2 g_3 g_4     
- \frac{1296}{N} \Big(N^4 (32 + 3 \pi^2) + N^3 (96 + 9 \pi^2) 
\nonumber \\-
    896 (188 + 21 \pi^2) + 12 N (2440 + 261 \pi^2) 
    + 
    N^2 (7184 + 716 \pi^2)\Big)  g_2 g_4^2
-4478976 (9 + \pi^2)  g_3 g_4^2   
\nonumber\\- \frac{1944}{N}\Big(  36 N^3 + 12 N^4 + 96 N (75 + 8 \pi^2) - 
    192 (242 + 27 \pi^2) + N^2 (2028 + 203 \pi^2)\Big) g_4^3 
\end{gather}

\begin{gather}
\widetilde\beta_{g_2}^{(4)}=\,
 - \frac{27}{N^5} \Big(2 N^8 (90 + 7 \pi^2) + 24000 N (20 + 9 \pi^2) - 
    96000 (28 + 9 \pi^2) - 3200 N^2 (1443 + 170 \pi^2) 
\nonumber\\-
    5 N^5 (8372 + 209 \pi^2) + N^7 (3750 + 323 \pi^2) - 
    100 N^4 (6542 + 785 \pi^2) 
    + 400 N^3 (7120 + 797 \pi^2) 
\nonumber\\+
    N^6 (28350 + 3133 \pi^2)\Big) g_1^3 
- \frac{18}{N^4}  \Big(3 N^8 (10 + \pi^2) - 309600 N (32 + 3 \pi^2) + 
    86400 (212 + 23 \pi^2) + N^7 (732 + 66 \pi^2)    
\nonumber\\-
    8 N^4 (13443 + 482 \pi^2) + N^6 (25770 + 2294 \pi^2) - 
    120 N^3 (29800 + 3209 \pi^2)   
    + 240 N^2 (46680 + 4993 \pi^2) 
\nonumber\\+
    N^5 (200436 + 19313 \pi^2)\Big)g_1^2 g_2    
- \frac{216}{N^3} \Big(-7200 (88 + 7 \pi^2) + N^6 (248 + 22 \pi^2) + 
    600 N (832 + 63 \pi^2) 
\nonumber\\+N^5 (3132 + 251 \pi^2)    
    + 
    10 N^3 (2672 + 359 \pi^2) - 40 N^2 (6504 + 677 \pi^2) + 
    3 N^4 (8852 + 717 \pi^2)\Big) g_1^2 g_3     
\nonumber\\- \frac{162}{N^4} \Big(32000 (92 + 9 \pi^2) - 16000 N (104 + 9 \pi^2) + 
    N^6 (1418 + 155 \pi^2) + 320 N^2 (4644 + 521 \pi^2) 
    + 
    N^5 (12998 
\nonumber\\+ 1497 \pi^2) - 20 N^3 (19900 + 2381 \pi^2) - 
    N^4 (42584 + 3431 \pi^2)\Big)g1^2 g_4     
- \frac{48}{N^3}\Big(2 N^6 (534 + 49 \pi^2) + 1620 N (2044 
\nonumber\\+ 193 \pi^2) 			-  2160 (3060 + 319 \pi^2) + N^5 (11898 + 1033 \pi^2) 
-     72 N^2 (14522 + 1613 \pi^2) + 3 N^3 (32656 + 4997 \pi^2) 
\nonumber\\+
    N^4 (91914 + 8441 \pi^2)\Big)  g_1 g_2^2 
- \frac{288}{N^2}  \Big(4 N^6 (6 + \pi^2) + 16 N^5 (15 + 2 \pi^2) + 
    720 (1400 + 139 \pi^2) + 5 N^4 (2184 + 149 \pi^2)   
\nonumber\\-
    180 N (2568 + 229 \pi^2) + N^3 (52944 + 4095 \pi^2) + 
    N^2 (47952 + 6716 \pi^2)\Big) g_1 g_2 g_3
- \frac{216}{N^3} \Big(-36 N^3 (626 + 7 \pi^2) 
\nonumber\\+ 2 N^6 (69 + 8 \pi^2) + 
    9 N^5 (194 + 19 \pi^2) - 11520 (358 + 37 \pi^2)   
    + 
    480 N (4018 + 381 \pi^2) + N^4 (15348 + 1549 \pi^2) 
\nonumber\\-
    8 N^2 (53892 + 6641 \pi^2)\Big) g_1 g_2 g_4 
- \frac{3456}{N}  \Big(-10 + 5 N + 2 N^2) (N (62 + 3 \pi^2) + 
    N^2 (62 + 3 \pi^2) + 6 (334 + 33 \pi^2)\Big) g_1 g_3^2
\nonumber\\
- \frac{2592}{N^2}  \Big(N^4 (492 + 35 \pi^2) + 160 (988 + 99 \pi^2) - 
    2 N^2 (2248 + 179 \pi^2) + N^3 (3084 + 227 \pi^2)
\nonumber\\-
    20 N (3176 + 291 \pi^2)\Big) g_1 g_3 g_4
- \frac{162}{N^3} \Big( 36 N^4 (163 + 17 \pi^2) + N^6 (172 + 21 \pi^2) + 
    7 N^5 (182 + 27 \pi^2) + 1600 N (1028 
\nonumber\\+ 99 \pi^2) 
    - 
    640 (6016 + 621 \pi^2) - 24 N^2 (9098 + 1363 \pi^2) - 
    2 N^3 (17432 + 1635 \pi^2)\Big) g_1 g_4^2 
- \frac{32}{N^2}  \Big(N^6 (36 + 7 \pi^2) 
\nonumber\\+ N^5 (288 + 50 \pi^2) + 
    2 N^4 (7186 + 545 \pi^2) + 8 N^3 (8863 + 756 \pi^2) 
    + 
    216 (11996 + 1265 \pi^2) - 54 N (13992 + 1337 \pi^2) 
\nonumber\\+
    N^2 (30720 + 5671 \pi^2)\Big)g_2^3    
- \frac{48}{N} \Big(13 N^5 \pi^2 + 2 N^6 \pi^2 + 
    8 N^4 (816 + 53 \pi^2) - 720 (2872 + 295 \pi^2) + 
    16 N^2 (15324 
\nonumber\\+ 1499 \pi^2) 
    + N^3 (22656 + 1637 \pi^2) + 
    36 N (17072 + 1737 \pi^2)\Big)  g_2^2 g_3    
- \frac{288}{N^2} \Big(2 N^4 (771 + 59 \pi^2) - 144 N (2041 + 200 \pi^2)   
\nonumber\\   - 4 N^2 (9261 + 824 \pi^2) + N^3 (9918 + 829 \pi^2) 
+ 144 (7988 + 839 \pi^2)\Big) g_2^2 g_4 -  3456 \Big(2 N^3 (18 + \pi^2) + N^4 (18 + \pi^2) 
\nonumber \\ +2 N (608 + 55 \pi^2) + N^2 (1234 + 111 \pi^2) 
    + 8 (2095 + 228 \pi^2)\Big) g_2 g_3^2  
- \frac{1728}{N} \Big(2 N^4 (24 + \pi^2) + N^3 (192 + 11 \pi^2) 
\nonumber\\+
    6 N (3168 + 281 \pi^2) - 48 (3112 + 319 \pi^2)   
    + 
    N^2 (5952 + 481 \pi^2)\Big) g_2 g_3 g_4
- \frac{324}{N^2}  \Big(N^6 (2 + \pi^2) + 2 N^5 (5 + 3 \pi^2) 
\nonumber\\+
    N^4 (1628 + 171 \pi^2) - 16 N^2 (3517 + 339 \pi^2)  
    + 
    192 (7024 + 735 \pi^2) - 32 N (9160 + 933 \pi^2) 
\nonumber \\+ N^3 (9376 + 974 \pi^2)\Big) g_2 g_4^2            
-20736  \Big(N (62 + 3 \pi^2) + N^2 (62 + 3 \pi^2) + 
    6 (334 + 33 \pi^2)\Big) g_3^2 g_4
\nonumber\\- \frac{2592}{N} \Big(N^4 (28 + 3 \pi^2) + 198 N (32 + 3 \pi^2) +
    N^3 (84 + 9 \pi^2) + N^2 (2572 + 249 \pi^2)
     - 
    8 (7960 + 813 \pi^2)\Big) g_3 g_4^2 
\nonumber\\-\frac{972}{N^2} \Big(20 N^5 + 4 N^6 + N^4 (146 + 21 \pi^2) + 
    4 N^3 (170 + 27 \pi^2) - 32 N (1082 + 117 \pi^2)   
    + 
    128 (1526 + 159 \pi^2) 
\nonumber\\ - 2 N^2 (4736 + 513 \pi^2)\Big)g_4^3   
\end{gather}

\begin{gather}
\widetilde\beta_{g_3}^{(4)}=\,
- \frac{9}{N^6}\Big(432000 \pi^2 + 72000 N^2 (8 + 3 \pi^2) + 
    2400 N^4 (214 + 35 \pi^2) - 600 N^5 (312 + 41 \pi^2) 
    + N^8 (630 + 69 \pi^2) 
\nonumber\\-2 N^6 (1200 + 137 \pi^2) + 
    N^7 (7110 + 813 \pi^2)\Big)  g_1^3 
- \frac{18}{N^5}\Big(28800 N (2 + 3 \pi^2) + 4 N^8 (39 + 4 \pi^2) - 
    14400 (112 + 27 \pi^2) 
\nonumber\\- 7200 N^2 (292 + 27 \pi^2) 
    + 
    6 N^7 (361 + 34 \pi^2) + 240 N^3 (4272 + 427 \pi^2) - 
    20 N^4 (11754 + 869 \pi^2) 
    + 2 N^6 (8379 + 883 \pi^2) 
\nonumber\\+ 
    N^5 (516 + 1189 \pi^2)\Big) g_1^2 g_2 
+ \frac{72}{N^4}\Big(56 N^7 + 4 N^8 + N^4 (2220 - 519 \pi^2) + 
    1200 N (-32 + 9 \pi^2) - 1200 (136 + 27 \pi^2) 
\nonumber\\+
    240 N^3 (614 + 39 \pi^2) - 300 N^2 (896 + 75 \pi^2) - 
    N^6 (1576 + 207 \pi^2)
     - N^5 (12524 + 1179 \pi^2)\Big)  g_1^2 g_3 
\nonumber\\+\frac{162}{N^5} \Big(N^6 (302 + \pi^2) - 4800 N (2 + 3 \pi^2) + 
    19200 (14 + 3 \pi^2) + 1600 N^2 (215 + 18 \pi^2)
     - 
    480 N^3 (321 + 31 \pi^2) 
\nonumber\\+ 20 N^4 (587 + 74 \pi^2) + 
    N^5 (5174 + 314 \pi^2)\Big) g_1^2 g_4 
- \frac{12}{N^4} \Big(4 N^8 (6 + \pi^2) + 48 N^7 (7 + \pi^2) - 
    8640 N (212 + 19 \pi^2) 
\nonumber\\+12960 (352 + 35 \pi^2) 
    + 
    16 N^4 (-1119 + 188 \pi^2) - 192 N^3 (3093 + 241 \pi^2) + 
    3 N^6 (3440 + 367 \pi^2) 
\nonumber\\+144 N^2 (16952 + 1693 \pi^2) + 
    N^5 (55992 + 4967 \pi^2)\Big)  g_1 g_2^2
- \frac{288}{N^3} \Big(2880 N (35 + 3 \pi^2) + 2 N^6 (84 + 17 \pi^2) 
\nonumber\\-  360 (440 + 39 \pi^2) + 24 N^3 (770 + 69 \pi^2) 
    + 
    N^5 (1116 + 169 \pi^2) - 6 N^2 (10352 + 953 \pi^2) + 
    N^4 (10956 + 985 \pi^2)\Big) g_1 g_2 g_3 
\nonumber\\-\frac{216}{N^4}  \Big(3 N^6 (40 + \pi^2) + 9 N^5 (78 + \pi^2) + 
    3840 (191 + 18 \pi^2) - 960 N (317 + 27 \pi^2) 
    - 
    24 N^3 (2503 + 192 \pi^2) 
\nonumber\\+96 N^2 (3303 + 352 \pi^2) - 
    N^4 (15564 + 995 \pi^2)\Big) g_1 g_2 g_4
- \frac{144}{N^2}\Big(66 N^5 \pi^2 + 15 N^6 \pi^2 + 
    600 (112 + 9 \pi^2) 
\nonumber\\+16 N^4 (443 + 33 \pi^2) - 
    60 N (1120 + 87 \pi^2) 
    + 6 N^2 (5216 + 567 \pi^2) + 
    N^3 (20528 + 1377 \pi^2)\Big)  g_1 g_3^2 
\nonumber\\
+ \frac{864}{N^3} \Big(84 N^5 + 12 N^6 + N^4 (148 - 27 \pi^2) + 
    N^3 (1628 + 57 \pi^2) + 80 (956 + 81 \pi^2)
     - 
    20 N (2552 + 207 \pi^2) 
\nonumber\\+
    2 N^2 (9808 + 1017 \pi^2)\Big)  g_1 g_3 g_4
- \frac{162}{N^4} \Big(3 N^5 (-208 + 7 \pi^2) + N^6 (-94 + 9 \pi^2) + 
    1920 (368 + 33 \pi^2) 
\nonumber\\-480 N (632 + 51 \pi^2) 
    - 
    24 N^3 (1022 + 81 \pi^2) - 2 N^4 (5156 + 357 \pi^2) + 
    16 N^2 (14228 + 1707 \pi^2)\Big)  g_1 g_4^2    
\nonumber\\- \frac{4}{3N^3} \Big(21 N^7 \pi^2 + 2 N^8 \pi^2 + 
    17280 N (465 + 41 \pi^2) + 576 N^3 (433 + 80 \pi^2)
     - 
    432 N^2 (1464 + 205 \pi^2) 
\nonumber\\+ 4 N^6 (2136 + 277 \pi^2) - 
    2592 (9552 + 971 \pi^2) 
    + 12 N^4 (28836 + 2929 \pi^2) + 
    N^5 (59568 + 5281 \pi^2)\Big) g_2^3
\nonumber\\ -\frac{96}{N^2} \Big(N^6 (54 + 19 \pi^2) + 4 N^5 (81 + 19 \pi^2) - 
    108 N (1264 + 103 \pi^2) + 540 (1112 + 109 \pi^2) 
    + 
    15 N^2 (2064 + 281 \pi^2) 
\nonumber\\+ 6 N^3 (5540 + 431 \pi^2) + 
    2 N^4 (5655 + 454 \pi^2)\Big)  g_2^2 g_3
- \frac{72}{N^3} \Big(2 N^6 (24 + \pi^2) + 24 N^2 (492 + \pi^2) + 
    N^5 (384 + 19 \pi^2) 
\nonumber\\- N^4 (2952 + 35 \pi^2) 
    - 
    1728 (1150 + 117 \pi^2) - 6 N^3 (4492 + 271 \pi^2) + 
    144 N (3942 + 367 \pi^2)\Big)  g_2^2 g_4
\nonumber\\- \frac{432}{N} (-6 + 3 N + 2 N^2) \Big(2 N^3 \pi^2 + N^4 \pi^2 + 
    N (704 + 52 \pi^2) + N^2 (704 + 53 \pi^2)
     + 
    4 (3232 + 309 \pi^2)\Big)  g_2 g_3^2
\nonumber\\-\frac{1728}{N^2} \Big(27 N^3 (4 + \pi^2) + 4 N^4 (3 + 2 \pi^2) - 
    174 N (104 + 9 \pi^2) + 168 (580 + 57 \pi^2) 
    - 
    3 N^2 (1720 + 137 \pi^2)\Big) g_2 g_3 g_4 
\nonumber\\-\frac{324}{N^3}  \Big(4 N^6 + N^5 (50 + \pi^2) - N^4 (1060 + 91 \pi^2) + 
    16 N^2 (1289 + 133 \pi^2) - 2 N^3 (3704 + 339 \pi^2) 
\nonumber\\+
    16 N (9704 + 975 \pi^2) - 64 (9968 + 1017 \pi^2)\Big) g_2 g_4^2
-  144\Big(3 N^5 \pi^2 + N^6 \pi^2 + N^4 (848 + 65 \pi^2) + 
    N^3 (1696 + 125 \pi^2) 
\nonumber\\+ 4 N (6016 + 555 \pi^2) 
    + 
    24 (6664 + 711 \pi^2) + N^2 (24912 + 2282 \pi^2)\Big)  g_3^3 
-\frac{864}{N} \Big(-57056 + 752 N - 368 N^2 
\nonumber\\ - 5472 \pi^2 + 
     84 N \pi^2 + 9 N^3 \pi^2 +3 N^4 \pi^2\Big) g_3^2 g_4 
+ \frac{432}{N^2}\Big(20 N^5 + 4 N^6 - N^4 (52 + 9 \pi^2) - 
    2 N^3 (212 + 27 \pi^2)
\nonumber\\+ 10 N^2 (1580 + 153 \pi^2) 
    + 
    4 N (10064 +  963 \pi^2) - 16 (17792 + 1755 \pi^2)\Big)  g_3 g_4^2 
- \frac{486}{N^3}  \Big(N^6 (-8 + \pi^2) - 10 N^4 (34 + 3 \pi^2) 
\nonumber\\
+ 
    N^5 (-40 + 6 \pi^2) 
-16 N^3 (94 + 13 \pi^2)
     - 
    512 (400 + 41 \pi^2) + 64 N (632 + 71 \pi^2) + 
    8 N^2 (1354 + 163 \pi^2)\Big) g_4^3   
\end{gather}

\begin{gather}
\widetilde\beta_{g_4}^{(4)}=\,
- \frac{9}{N^5} \Big(-13824000 + 6912000 N + N^8 (360 + 22 \pi^2) - 
    2400 N^4 (448 + 43 \pi^2) 
    + 9600 N^3 (395 + 49 \pi^2) 
\nonumber\\- 
    60 N^5 (1672 + 65 \pi^2) - 3200 N^2 (2352 + 275 \pi^2) 
    + 
    N^7 (6660 + 489 \pi^2) + N^6 (41220 + 4241 \pi^2)\Big)  g_1^3
\nonumber\\- \frac{72}{N^4} \Big(172800 (14 + \pi^2) - 14400 N (68 + 3 \pi^2) - 
    2 N^4 (14844 + 131 \pi^2) + N^6 (1848 + 191 \pi^2) 
    + 
    480 N^2 (2120 + 243 \pi^2) 
\nonumber\\+ 5 N^5 (3504 + 415 \pi^2) - 
    120 N^3 (3640 + 423 \pi^2)\Big)  g_1^2 g_2
- \frac{10368}{N^3} \Big(800 N + 20 N^3 (17 + 3 \pi^2) - 
    200 (28 + 3 \pi^2) 
\nonumber\\+3 N^4 (32 + 5 \pi^2) - 
    5 N^2 (392 + 51 \pi^2)\Big)  g_1^2 g_3
- \frac{18}{N^4}\Big(670 N^7 + 41 N^8 + 38400 (320 + 27 \pi^2) - 
    9600 N (467 + 27 \pi^2) 
\nonumber\\- 6 N^4 (32878 + 417 \pi^2) 
    + 
    N^6 (14413 + 801 \pi^2) - 120 N^3 (17800 + 1731 \pi^2) + 
    240 N^2 (22264 + 2481 \pi^2)
\nonumber\\   + N^5 (95816 + 8415 \pi^2)\Big) g_1^2 g_4 
-\frac{3456}{N^2} \Big(3 N^3 (-28 + \pi^2) + N^4 (-4 + 3 \pi^2) + 
     240 (52 + 5 \pi^2) + 2 N^2 (-148 + 9 \pi^2)
\nonumber\\-
     40 N (152 + 15 \pi^2)\Big)  g_1 g_2 g_3
- \frac{576}{N^2} \Big(960 (94 + 9 \pi^2) - 240 N (262 + 27 \pi^2) + 
    6 N^2 (728 + 153 \pi^2)
\nonumber\\ + N^4 (1160 + 153 \pi^2) 
    + 
    N^3 (6392 + 801 \pi^2)\Big)  g_1 g_3 g_4
- \frac{48}{N^3} \Big(51840 N (15 + \pi^2) - 34560 (57 + 5 \pi^2) + 
    N^6 (48 + 7 \pi^2) 
\nonumber\\+ N^5 (228 + 43 \pi^2) 
    + 
    18 N^4 (346 + 101 \pi^2) + 4 N^3 (2526 + 1579 \pi^2) - 
    8 N^2 (54528 + 6997 \pi^2)\Big)  g_1 g_2^2
\nonumber\\-\frac{432}{N^3} \Big(7 N^6 (16 + \pi^2) + 10 N^5 (102 + 7 \pi^2) - 
    1920 (296 + 27 \pi^2) + 480 N (521 + 39 \pi^2) 
    + 
    8 N^3 (437 + 221 \pi^2) 
\nonumber\\+ N^4 (5556 + 727 \pi^2) - 
    4 N^2 (28812 + 3317 \pi^2)\Big)  g_1 g_2 g_4
- \frac{108}{N^3}\Big(2514 N^5 + 318 N^6 + 88 N^3 (-241 + 18 \pi^2)  
\nonumber\\-5120 (289 + 27 \pi^2) + 7 N^4 (1084 + 99 \pi^2) 
    + 
    320 N (2350 + 189 \pi^2) - 
    16 N^2 (16366 + 1557 \pi^2)\Big)  g_1 g_4^2 
+\\+ \frac{128}{3N^2} \Big(-63504 (10 + \pi^2) + N^4 (528 + \pi^2) + 
    324 N (532 + 53 \pi^2) + N^3 (3120 + 163 \pi^2) 
    + 
    8 N^2 (4314 + 337 \pi^2)\Big) g_2^3 
\nonumber\\+ \frac{4608}{N} \Big(6424 + 4 N^3 + 2 N^4 + 726 \pi^2 + 
    3 N (80 + 7 \pi^2) + N^2 (242 + 21 \pi^2)\Big)  g_2^2 g_3
- \frac{144}{N^2} \Big(N^6 (2 + \pi^2) + N^5 (12 + 7 \pi^2) 
\nonumber\\+
    9408 (88 + 9 \pi^2) - 6 N^2 (3476 + 93 \pi^2)
 +    N^4 (2134 + 239 \pi^2)
     - 48 N (4495 + 429 \pi^2) + 
    N^3 (12708 + 1351 \pi^2)\Big)  g_2^2 g_4
\nonumber\\-\frac{6912}{N} \Big(N^4 (4 + \pi^2) + 4 N^3 (6 + \pi^2) - 
    72 (184 + 21 \pi^2) + 3 N^2 (308 + 39 \pi^2) 
    + 
    N (2488 + 306 \pi^2)\Big)  g_2 g_3 g_4
\nonumber\\-\frac{2592}{N^2}  \Big(68224 + 12 N^5 + 2 N^6 + 7104 \pi^2 - 
    224 N^2 (15 + \pi^2) + N^4 (126 + 11 \pi^2) 
+ 
    N^3 (644 + 57 \pi^2) 
\nonumber\\-16 N (1043 + 93 \pi^2)\Big)  g_2 g_4^2
-  2304 \Big(-8 N^3 - 4 N^4 + N^2 (384 + 45 \pi^2) + 
    N (388 + 45 \pi^2) + 6 (1852 + 225 \pi^2)\Big) g_3^2 g_4 
\nonumber\\- \frac{864}{N}\Big(27 N^3 (8 + \pi^2) + 9 N^4 (8 + \pi^2) + 
    4 N^2 (808 + 99 \pi^2) - 32 (2584 + 297 \pi^2)
     + 
    N (8096 + 972 \pi^2)\Big) g_3 g_4^2 
\nonumber\\-\frac{108}{N^2} \Big(13 N^5 + 5 N^6 + N^4 (1198 + 45 \pi^2) + 
    4 N^3 (1723 + 63 \pi^2) - 8 N^2 (6253 + 414 \pi^2) 
    - 
    32 N (5692 + 459 \pi^2) 
\nonumber\\+128 (6388 + 675 \pi^2)\Big) g_4^3
\end{gather}

\section{The $F$-function and metric for the symmetric traceless model}
\label{ffun}

Working up to the two-loop order, we find that the $F$-function which enters the gradient flow expression (\ref{gradflow}) is given by $F= F^{(1)} + F^{(2)}$, where
\begin{gather}
	F^{(1)} =  -\frac{\epsilon }{576 N^3} \notag\\ 
	\times \Big[  \left(2 N^2 \left(48 g_2 \left(4 g_3 N^5+\left(10 g_3+3 g_4\right) N^4+3 \left(6 g_3+5 g_4\right) N^3+6 \left(4 g_3-7
		g_4\right) N^2 \right.\right.\right. 
		\notag\\ \left.\left.\left. - 72 \left(g_3+2 g_4\right) N+288 g_4\right)+4 g_2^2 \left(N^6+6 N^5+45 N^4+124 N^3-168 N^2-720 N+1296\right) \right.\right. 
		\notag\\ \left.\left. + 3
		\left(\left(16 g_3^2+3 g_4^2\right) N^6+\left(32 g_3^2+15 g_4^2\right) N^5+24 \left(6 g_3^2+g_4^2\right) N^4+4 \left(32 g_3^2+48 g_4
		g_3+15 g_4^2\right) N^3  \right.\right.\right. 
\notag\\ \left.\left.\left. + 96 \left(2 g_3^2+4 g_4 g_3-5 g_4^2\right) N^2-192 g_4 \left(8 g_3+7 g_4\right) N+3072 g_4^2\right)\right) \right. 
\notag\\ \left. +12
		g_1 N \left(9 g_4 N^6+\left(80 g_3+63 g_4\right) N^5+\left(272 g_3-42 g_4\right) N^4-120 \left(2 g_3+7 g_4\right) N^3  \right.\right. 
\notag\\ \left.\left. -240 \left(4
		g_3-g_4\right) N^2+4 g_2 \left(2 N^6+15 N^5+11 N^4-140 N^3+720 N-720\right)+960 \left(g_3+4 g_4\right) N-3840 g_4\right) \right.  
\notag \\ \left. + 3 g_1^2
		\left(N^8+14 N^7+83 N^6+46 N^5-960 N^4+4800 N^2-9600 N+9600\right)\right)\Big]
\end{gather}
and $ F^{(2)} $ may be written in terms of the 3-point functions in the free theory in $d=3$ \cite{Giombi:2014xxa,Fei:2015oha}:
\begin{equation}
F^{(2)}\sim C_{ijk} g^i g^j g^k \ , \qquad  \langle O_i (x) O_j (y) O_k (z) \rangle = {C_{ijk} \over |x-y|^3 |x-z|^3 |y-z|^3 }\ .
\end{equation}

Explicitly, we find
\begin{gather}
	 F^{(2)}= \frac{3}{13271040 N^5 \pi ^2} 
	 \notag\\
	 \times\Big[\big(3 N^{12}+93 N^{11}+1717 N^{10}+13103 N^9+15072 N^8-227572 N^7-326400 N^6
	 \notag\\ +2596800 N^5-758400 N^4-12288000 N^3+29952000 N^2-40704000 N+29184000\big) g_1^3
	 \notag\\+\frac{16}{3} N g_2 \big(27 \big(6 N^{10}+109 N^9+878 N^8+1885 N^7-10882 N^6-28000 N^5+122880 N^4+28800 N^3
	 \notag\\- 672000
	N^2+1411200 N-1094400\big) g_1^2+9 N \big(N^{10}+15 N^9+405 N^8+3493 N^7+8634 N^6
	\notag\\-30684 N^5-102504 N^4+351168 N^3+408960 N^2-2194560 N+1969920\big) g_2 g_1
	\notag\\+8 N^2 \big(26 N^8+219 N^7+1446 N^6+5399 N^5-714 N^4-57456 N^3+30240 N^2+343440 N-443232\big) g_2^2\big)
	\notag\\+192 N^2 g_3
	\big(2 \big(\big(N^8+7 N^7+181 N^6+757 N^5+1990 N^4+3832 N^3-7296 N^2-27504 N+49248\big) g_2^2
	\notag\\+12 N \big(6 N^6+21 N^5+118 N^4+253 N^3+270 N^2+348 N-1368\big) g_3 g_2+4 N^2 \big(3 N^6+9 N^5
	\notag\\+71 N^4+127 N^3+402 N^2+340 N+456\big) g_3^2\big) N^2+12 g_1 \big(\big(2 N^8+17
	N^7+174 N^6+773 N^5
	\notag\\+162 N^4-6176 N^3+240 N^2+28080 N-27360\big) g_2+2 N \big(15 N^6+66 N^5+196 N^4+421 N^3
	\notag\\ -570 N^2-2100 N+2280\big) g_3\big) N+3 \big(29 N^8+310 N^7+997 N^6-1612 N^5-10020 N^4
	\notag\\+15600 N^3+38400 N^2-112800 N+91200\big) g_1^2\big)-18 N \big(-N^2-2 N+8\big) g_4
	\big(\big(N^8+6 N^7+47 N^6
	\notag\\+198 N^5+1428 N^4+7416 N^3-32512 N^2-121344 N+311296\big) g_4^2 N^2+32 \big(\big(N^6+7 N^5+113 N^4
	\notag\\+629 N^3-1470 N^2-7920 N+16416\big) g_2^2+24 N \big(N^4+4 N^3+41 N^2+114 N-456\big) g_3 g_2
	\notag\\+48 N^2 \big(3 N^2+3 N+38\big) g_3^2\big) N^2+96
	\big(\big(N^6+6 N^5+46 N^4+225 N^3-728 N^2-3192 N+7296\big) g_2
\notag\\	
	+2 N \big(5 N^4+15 N^3+86 N^2+228 N-1216\big) g_3\big) g_4 N^2+192 g_1 \big(\big(7 N^6+65 N^5+52 N^4-964 N^3-650 N^2
	\notag\\+7680 N-9120\big) g_2++2 N \big(27 N^4+141 N^3-190 N^2-1140 N+1520\big) g_3\big) N+3
	\big(3 N^8+24 N^7
	\notag\\+325 N^6+2364 N^5-100 N^4-41712 N^3-10240 N^2+318720 N-389120\big) g_1 g_4 N+3 \big(21 N^8+294 N^7
	\notag\\+1599 N^6+30 N^5-27920 N^4+209600 N^2-499200 N+486400\big) g_1^2\Big]
 \end{gather}
The metric $G_{ij}$ is given by  
\begin{gather}
	G_{11} = \frac{1}{192 N^3} \left(N^8+14 N^7+83 N^6+46 N^5-960 N^4+4800 N^2-9600 N+9600\right), \notag\\
	G_{12} = G_{21} =  \frac{1}{24 N^2} \left(2 N^6+15 N^5+11 N^4-140 N^3+720 N-720\right), \notag\\
	G_{13}=G_{23} =  \frac{1}{6 N} \left(5 N^4+17 N^3-15 N^2-60 N+60\right), \notag\\
	G_{14}= G_{41}=\frac{1}{32 N^2} (N-2) (N+4) \left(3 N^4+15 N^3-20 N^2-120 N+160\right), \notag \\
	G_{22}= \frac{1}{72 N} \left(N^6+6 N^5+45 N^4+124 N^3-168 N^2-720 N+1296\right), \notag\\
	G_{23}=G_{32}=\frac{1}{6}\left(2 N^4+5 N^3+9 N^2+12 N-36\right), \notag\\
	G_{24}=G_{42}=\frac{1}{4 N} (N-2) (N+4) \left(N^2+3 N-12\right), \notag \\
	G_{33} = \frac{1}{6} N^3 \left(N^4+2 N^3+9 N^2+8 N+12\right),\quad G_{34}=G_{43}= (N-2) N^3 (N+4), \notag \\
	G_{44}=\frac{1}{32 N} (N-2)^2 (N+4)^2 \left(N^2+N+16\right).
\end{gather}
At this order it is independent of the couplings $g^i$ and is proportional to the matrix of two-point functions (\ref{Zametric}) in the free theory in $d=3$.

\restoregeometry

\section{Calculating the Hopf constant}
\label{HopfAppendix}

In this appendix we compute the Hopf constant $a$ at two loops. Introducing rescaled couplings $g_i = 720 (8\pi)^2\epsilon\, \mathpzc{g}_i$, the beta functions at the critical value $N=N_{\text{crit}}=4.475$ in units of $\epsilon$ become 
\begin{gather*}
\beta_{\mathpzc{g}_1} =-2\mathpzc{g} _1+\left(2339.99 \mathpzc{g}_1+4273.55 \mathpzc{g} _2+3840. \mathpzc{g} _3+4325.08 \mathpzc{g} _4\right) \mathpzc{g} _1+2768.04 \mathpzc{g} _2^2+2592. \mathpzc{g} _4^2+4608. \mathpzc{g} _2 \mathpzc{g} _4
\notag\\
\beta_{\mathpzc{g}_2} = 
-2\mathpzc{g} _2 +
\left(509.966
	\mathpzc{g} _1+2962.93 \mathpzc{g} _2+6748.16 \mathpzc{g} _3+113.519 \mathpzc{g} _4\right)\mathpzc{g} _1+ \left(3456. \mathpzc{g} _3+360.299  \mathpzc{g} _4\right) \mathpzc{g} _4
\notag\\
	+ \left(2308.94 \mathpzc{g} _2+11232.3 \mathpzc{g} _3-421.438 \mathpzc{g}
	_4\right) \mathpzc{g} _2
\notag\\
\beta_{\mathpzc{g}_3} = 
-2 \mathpzc{g} _3+
\left(42 \mathpzc{g} _1+221.912 \mathpzc{g} _2+576. \mathpzc{g} _3-241.337 \mathpzc{g} _4\right) \mathpzc{g} _1+10704.4 \mathpzc{g} _3^2-209.942\mathpzc{g} _4^2 -772.278 \mathpzc{g} _3 \mathpzc{g} _4 
\notag\\ 
	+ \left(629.906\mathpzc{g}
	_2+4074.01 \mathpzc{g} _3-135.923 \mathpzc{g} _4\right)\mathpzc{g} _2
	\notag\\
\beta_{\mathpzc{g}_4} =
-2\mathpzc{g} _4+ 	
	\left(226.417 \mathpzc{g}
	_1+73.3524 \mathpzc{g} _2+1708.55 \mathpzc{g} _4\right) \mathpzc{g} _1-618.547\mathpzc{g} _2^2+\left(1583.3 \mathpzc{g} _2+3840. \mathpzc{g} _3+1066.11 \mathpzc{g}
	_4\right)\mathpzc{g} _4 
\end{gather*}
These beta functions have a fixed point at
\begin{align}
\mathpzc{g}^\ast(N_{\text{crit}}) = 10^{-4}\cdot\left(3.48916, -4.64792, 3.04945, -1.08745\right).
\end{align}
Letting $V=(v_1,v_2,v_3,\overline{v}_3)$ be the matrix of eigenvectors $v_i$ of the stability matrix $\left(\frac{\partial \beta_{\mathpzc{g}_i}}{\partial \mathpzc{g}_j}\right)$ evaluated at this fixed point,
\begin{gather}
	V^{-1}\left(\frac{\partial \beta_{\mathpzc{g}_i}}{\partial \mathpzc{g}_j}\right) V  = \operatorname{diag}\left(2, -1.57495,  -0.153965i, 0.153965 i \right).
\end{gather}
One can check that these eigenvalues change on varying $N$. In particular, the real parts of the complex eigenvalues change linearly with $N$ for $N$ close to $N_{\text{crit}}$.  
Changing to variables $t_1 = v_1\cdot \mathpzc{g}$, $t_2 = v_2\cdot \mathpzc{g}$, $t_3 = \Re[v_3\cdot \mathpzc{g}]$, $t_4 = \Im[v_3\cdot \mathpzc{g}]$, we get the equations
\begin{gather}
\beta_{t_1} = 2 t_1-3006.27 t_1^2-635.361 t_2^2-4.22379 t_3^2+4.22379 t_4^2+7.65924 t_3 t_4 \notag\\
\beta_{t_2} = -1.57495 t_2+\left(-638.903 t_1+1471.36 t_2-96.8862 t_3+72.0709 t_4\right) t_2+ \notag\\ + 1.0131 t_3^2-0.34628
t_4^2-1.37241 t_3 t_4 \notag\\
\beta_{t_3}= -0.153965 t_4+ \left(231.430 t_4-3006.27 t_3\right)  t_1+\left(-31746.2 t_2+1284.37 t_3-347.122 t_4\right) t_2 \notag\\
-49.5972 t_3^2+492.731 t_4^2+178.686 t_3 t_4  
\notag\\
\beta_{t_4} =0.153965 t_3+\left(-231.43 t_3-3006.27 t_4\right) t_1+\left(638.003 t_2  +730.144 t_4-82.7131 t_3\right) t_2\notag\\
+ 8.73689 t_3^2+823.772 t_4^2+153.731 t_3 t_4\,.
\end{gather}
We wish to study the RG flow in the manifold that is tangent to the center eigenspace. We cannot simply set $t_1$ and $t_2$ to zero, since this plane is not invariant under the RG flow: the $t_3^2$, $t_4^2$, and $t_3t_3$ terms in $\beta_{t_1}$ and $\beta_{t_2}$ generate a flow in $t_1$ and $t_2$. But by introducing new variables with $t_1$ and $t_2$ suitably shifted,
\begin{align}
	 &u_1=t_1- 1.77501 t_3^2+4.3762 t_4 t_3+1.77501 t_4^2, \\ 
	 &u_2= t_2-0.709414 t_3^2+0.676770 t_4 t_3+0.286027 t_4^2\,,
\end{align}
the $t_3^2$, $t_4^2$, and $t_3t_3$ terms in $\beta_{u_1}$ and $\beta_{u_2}$ cancel out. While $\beta_{u_1}$ and $\beta_{u_2}$ do couple to $t_3$ and $t_4$ at third order, one can introduce new variables yet again and shift $u_1$ and $u_2$ by cubic terms in $t_3$ and $t_4$ to remove this third order coupling. This procedure may be iterated indefinitely to obtain a coordinate expansion of the center manifold to arbitrary order, in accordance with the center manifold theorem. We will content ourselves with the cubic approximation of the center manifold, which consists of the surface $u_1=u_2=0$, since this approximation suffices to determine the Hopf constant. Eliminating $t_1$ and $t_2$ in favour of $u_1$ and $u_2$ in the equations for $\beta_{t_3}$ and $\beta_{t_4}$, setting $u_1$ and $u_2$ to zero, and discarding unreliable quartic terms gives
\begin{gather}
	\beta_{t_3} =  -49.5972 t_3^2+178.686 t_4 t_3+492.731 t_4^2-0.153965 t_4 \notag\\ -4425.01 \left(1. t_3^3-2.81386 t_4 t_3^2-0.947101 t_4^2 t_3+0.0703961 t_4^3\right) \notag\\
	\beta_{t_4} = 8.73689 t_3^2+153.731 t_4 t_3+0.153965
	t_3+823.772 t_4^2\notag\\ - 469.468 \left(1. t_3^3+7.98654 t_4 t_3^2-27.8962 t_4^2 t_3-10.9216 t_4^3\right).
\end{gather}
From these equations the Hopf constant can be directly obtained by the use of equation (3.4.11) in \cite{guckenheimer2013nonlinear} or by the equivalent formula in \cite{chow1977integral}. We find that 
\begin{gather}
	a \approx 620479 0 
\end{gather}
so that Hopf's theorem guarantees the existence of a periodic orbit that is IR-attractive in the center manifold, implying that if we fine-tune the couplings in the vicinity of $N_{\text{crit}}$, there is a cyclic solution to the beta functions that comes back precisely to itself.

\bibliographystyle{ssg}
\bibliography{biblio}

\begingroup\raggedright\begin{thebibliography}{10}

\bibitem{Wilson:1973jj}
K.~Wilson and J.~B. Kogut, ``{The Renormalization group and the epsilon
  expansion},'' {\em Phys. Rept.} {\bf 12} (1974) 75--199.

\bibitem{Gukov:2016tnp}
S.~Gukov, ``{RG Flows and Bifurcations},'' {\em Nucl. Phys. B} {\bf 919} (2017)
  583--638, \href{https://arxiv.org/abs/1608.06638}{{\tt 1608.06638}}.

\bibitem{Glazek:2002hq}
S.~D. Glazek and K.~G. Wilson, ``{Limit cycles in quantum theories},'' {\em
  Phys. Rev. Lett.} {\bf 89} (2002) 230401,
  \href{https://arxiv.org/abs/hep-th/0203088}{{\tt hep-th/0203088}}. [Erratum:
  Phys.Rev.Lett. 92, 139901 (2004)].

\bibitem{Braaten_2004}
E.~Braaten and D.~Phillips, ``Renormalization-group limit cycle for the
  $\frac{1}{r^2}$-potential,'' {\em Physical Review A} {\bf 70} (Nov, 2004).

\bibitem{Gorsky:2013yba}
A.~Gorsky and F.~Popov, ``{Atomic collapse in graphene and cyclic
  renormalization group flow},'' {\em Phys. Rev. D} {\bf 89} (2014), no.~6
  061702, \href{https://arxiv.org/abs/1312.7399}{{\tt 1312.7399}}.

\bibitem{Dawid:2017ahd}
S.~M. Dawid, R.~Gonsior, J.~Kwapisz, K.~Serafin, M.~Tobolski, and S.~G\l{}azek,
  ``{Renormalization group procedure for potential $-g/r^2$},'' {\em Phys.
  Lett. B} {\bf 777} (2018) 260--264,
  \href{https://arxiv.org/abs/1704.08206}{{\tt 1704.08206}}.

\bibitem{efimov1970energy}
V.~Efimov, ``Energy levels arising from resonant two-body forces in a
  three-body system,'' {\em Physics Letters B} {\bf 33} (1970), no.~8 563--564.

\bibitem{Bulycheva:2014twa}
K.~Bulycheva and A.~Gorsky, ``{Limit cycles in renormalization group
  dynamics},'' {\em Usp. Fiz. Nauk} {\bf 184} (2014), no.~2 182--193,
  \href{https://arxiv.org/abs/1402.2431}{{\tt 1402.2431}}.

\bibitem{Fortin:2012cq}
J.-F. Fortin, B.~Grinstein, and A.~Stergiou, ``{Limit Cycles in Four
  Dimensions},'' {\em JHEP} {\bf 12} (2012) 112,
  \href{https://arxiv.org/abs/1206.2921}{{\tt 1206.2921}}.

\bibitem{Fortin:2012hn}
J.-F. Fortin, B.~Grinstein, and A.~Stergiou, ``{Limit Cycles and Conformal
  Invariance},'' {\em JHEP} {\bf 01} (2013) 184,
  \href{https://arxiv.org/abs/1208.3674}{{\tt 1208.3674}}.

\bibitem{Luty:2012ww}
M.~A. Luty, J.~Polchinski, and R.~Rattazzi, ``{The $a$-theorem and the
  Asymptotics of 4D Quantum Field Theory},'' {\em JHEP} {\bf 01} (2013) 152,
  \href{https://arxiv.org/abs/1204.5221}{{\tt 1204.5221}}.

\bibitem{Cardy:1988cwa}
J.~L. Cardy, ``{Is There a c Theorem in Four-Dimensions?},'' {\em Phys. Lett.
  B} {\bf 215} (1988) 749--752.

\bibitem{Jack:1990eb}
I.~Jack and H.~Osborn, ``{Analogs for the $c$ Theorem for Four-dimensional
  Renormalizable Field Theories},'' {\em Nucl. Phys. B} {\bf 343} (1990)
  647--688.

\bibitem{Komargodski:2011vj}
Z.~Komargodski and A.~Schwimmer, ``{On Renormalization Group Flows in Four
  Dimensions},'' {\em JHEP} {\bf 12} (2011) 099,
  \href{https://arxiv.org/abs/1107.3987}{{\tt 1107.3987}}.

\bibitem{Morozov:2003ik}
A.~Morozov and A.~J. Niemi, ``{Can renormalization group flow end in a big
  mess?},'' {\em Nucl. Phys. B} {\bf 666} (2003) 311--336,
  \href{https://arxiv.org/abs/hep-th/0304178}{{\tt hep-th/0304178}}.

\bibitem{Curtright:2011qg}
T.~L. Curtright, X.~Jin, and C.~K. Zachos, ``{RG flows, cycles, and c-theorem
  folklore},'' {\em Phys. Rev. Lett.} {\bf 108} (2012) 131601,
  \href{https://arxiv.org/abs/1111.2649}{{\tt 1111.2649}}.

\bibitem{Binder:2019zqc}
D.~J. Binder and S.~Rychkov, ``{Deligne Categories in Lattice Models and
  Quantum Field Theory, or Making Sense of $O(N)$ Symmetry with Non-integer
  $N$},'' {\em JHEP} {\bf 04} (2020) 117,
  \href{https://arxiv.org/abs/1911.07895}{{\tt 1911.07895}}.

\bibitem{Hogervorst:2015akt}
M.~Hogervorst, S.~Rychkov, and B.~C. van Rees, ``{Unitarity violation at the
  Wilson-Fisher fixed point in 4-$\epsilon$ dimensions},'' {\em Phys. Rev. D}
  {\bf 93} (2016), no.~12 125025, \href{https://arxiv.org/abs/1512.00013}{{\tt
  1512.00013}}.

\bibitem{hopf1942bifurcation}
E.~Hopf, ``Bifurcation of a periodic solution from a stationary solution of a
  system of differential equations,'' {\em Berlin Mathematische Physics Klasse,
  Sachsischen Akademic der Wissenschaften Leipzig} {\bf 94} (1942) 3--32.

\bibitem{Myers:2010xs}
R.~C. Myers and A.~Sinha, ``{Seeing a c-theorem with holography},'' {\em Phys.
  Rev. D} {\bf 82} (2010) 046006, \href{https://arxiv.org/abs/1006.1263}{{\tt
  1006.1263}}.

\bibitem{Jafferis:2011zi}
D.~L. Jafferis, I.~R. Klebanov, S.~S. Pufu, and B.~R. Safdi, ``{Towards the
  F-Theorem: N=2 Field Theories on the Three-Sphere},'' {\em JHEP} {\bf 06}
  (2011) 102, \href{https://arxiv.org/abs/1103.1181}{{\tt 1103.1181}}.

\bibitem{Klebanov:2011gs}
I.~R. Klebanov, S.~S. Pufu, and B.~R. Safdi, ``{F-Theorem without
  Supersymmetry},'' {\em JHEP} {\bf 10} (2011) 038,
  \href{https://arxiv.org/abs/1105.4598}{{\tt 1105.4598}}.

\bibitem{Casini:2012ei}
H.~Casini and M.~Huerta, ``{On the RG running of the entanglement entropy of a
  circle},'' {\em Phys. Rev. D} {\bf 85} (2012) 125016,
  \href{https://arxiv.org/abs/1202.5650}{{\tt 1202.5650}}.

\bibitem{Giombi:2014xxa}
S.~Giombi and I.~R. Klebanov, ``{Interpolating between $a$ and $F$},'' {\em
  JHEP} {\bf 03} (2015) 117, \href{https://arxiv.org/abs/1409.1937}{{\tt
  1409.1937}}.

\bibitem{Fei:2015oha}
L.~Fei, S.~Giombi, I.~R. Klebanov, and G.~Tarnopolsky, ``{Generalized
  $F$-Theorem and the $\epsilon$ Expansion},'' {\em JHEP} {\bf 12} (2015) 155,
  \href{https://arxiv.org/abs/1507.01960}{{\tt 1507.01960}}.

\bibitem{Jack:2015tka}
I.~Jack, D.~Jones, and C.~Poole, ``{Gradient flows in three dimensions},'' {\em
  JHEP} {\bf 09} (2015) 061, \href{https://arxiv.org/abs/1505.05400}{{\tt
  1505.05400}}.

\bibitem{Jack:2016utw}
I.~Jack and C.~Poole, ``{$a$-function in three dimensions: Beyond the leading
  order},'' {\em Phys. Rev. D} {\bf 95} (2017), no.~2 025010,
  \href{https://arxiv.org/abs/1607.00236}{{\tt 1607.00236}}.

\bibitem{Klebanov:2018fzb}
I.~R. Klebanov, F.~Popov, and G.~Tarnopolsky, ``{TASI Lectures on Large $N$
  Tensor Models},'' {\em PoS} {\bf TASI2017} (2018) 004,
  \href{https://arxiv.org/abs/1808.09434}{{\tt 1808.09434}}.

\bibitem{Eguchi:1982nm}
T.~Eguchi and H.~Kawai, ``{Reduction of Dynamical Degrees of Freedom in the
  Large N Gauge Theory},'' {\em Phys. Rev. Lett.} {\bf 48} (1982) 1063.

\bibitem{Kachru:1998ys}
S.~Kachru and E.~Silverstein, ``{4-D conformal theories and strings on
  orbifolds},'' {\em Phys. Rev. Lett.} {\bf 80} (1998) 4855--4858,
  \href{https://arxiv.org/abs/hep-th/9802183}{{\tt hep-th/9802183}}.

\bibitem{Lawrence:1998ja}
A.~E. Lawrence, N.~Nekrasov, and C.~Vafa, ``{On conformal field theories in
  four-dimensions},'' {\em Nucl. Phys. B} {\bf 533} (1998) 199--209,
  \href{https://arxiv.org/abs/hep-th/9803015}{{\tt hep-th/9803015}}.

\bibitem{Bershadsky:1998mb}
M.~Bershadsky, Z.~Kakushadze, and C.~Vafa, ``{String expansion as large N
  expansion of gauge theories},'' {\em Nucl. Phys. B} {\bf 523} (1998) 59--72,
  \href{https://arxiv.org/abs/hep-th/9803076}{{\tt hep-th/9803076}}.

\bibitem{Bershadsky:1998cb}
M.~Bershadsky and A.~Johansen, ``{Large N limit of orbifold field theories},''
  {\em Nucl. Phys. B} {\bf 536} (1998) 141--148,
  \href{https://arxiv.org/abs/hep-th/9803249}{{\tt hep-th/9803249}}.

\bibitem{Armoni:2003gp}
A.~Armoni, M.~Shifman, and G.~Veneziano, ``{Exact results in non-supersymmetric
  large N orientifold field theories},'' {\em Nucl. Phys. B} {\bf 667} (2003)
  170--182, \href{https://arxiv.org/abs/hep-th/0302163}{{\tt hep-th/0302163}}.

\bibitem{Kovtun:2004bz}
P.~Kovtun, M.~Unsal, and L.~G. Yaffe, ``{Necessary and sufficient conditions
  for non-perturbative equivalences of large N(c) orbifold gauge theories},''
  {\em JHEP} {\bf 07} (2005) 008,
  \href{https://arxiv.org/abs/hep-th/0411177}{{\tt hep-th/0411177}}.

\bibitem{Unsal:2006pj}
M.~Unsal and L.~G. Yaffe, ``{(In)validity of large N orientifold
  equivalence},'' {\em Phys. Rev. D} {\bf 74} (2006) 105019,
  \href{https://arxiv.org/abs/hep-th/0608180}{{\tt hep-th/0608180}}.

\bibitem{Douglas:1996sw}
M.~R. Douglas and G.~W. Moore, ``{D-branes, quivers, and ALE instantons},''
  \href{https://arxiv.org/abs/hep-th/9603167}{{\tt hep-th/9603167}}.

\bibitem{Tseytlin:1999ii}
A.~A. Tseytlin and K.~Zarembo, ``{Effective potential in nonsupersymmetric
  SU(N) x SU(N) gauge theory and interactions of type 0 D3-branes},'' {\em
  Phys. Lett. B} {\bf 457} (1999) 77--86,
  \href{https://arxiv.org/abs/hep-th/9902095}{{\tt hep-th/9902095}}.

\bibitem{Dymarsky:2005uh}
A.~Dymarsky, I.~Klebanov, and R.~Roiban, ``{Perturbative search for fixed lines
  in large N gauge theories},'' {\em JHEP} {\bf 08} (2005) 011,
  \href{https://arxiv.org/abs/hep-th/0505099}{{\tt hep-th/0505099}}.

\bibitem{Dymarsky:2005nc}
A.~Dymarsky, I.~Klebanov, and R.~Roiban, ``{Perturbative gauge theory and
  closed string tachyons},'' {\em JHEP} {\bf 11} (2005) 038,
  \href{https://arxiv.org/abs/hep-th/0509132}{{\tt hep-th/0509132}}.

\bibitem{Pomoni:2008de}
E.~Pomoni and L.~Rastelli, ``{Large N Field Theory and AdS Tachyons},'' {\em
  JHEP} {\bf 04} (2009) 020, \href{https://arxiv.org/abs/0805.2261}{{\tt
  0805.2261}}.

\bibitem{Gorbenko:2018ncu}
V.~Gorbenko, S.~Rychkov, and B.~Zan, ``{Walking, Weak first-order transitions,
  and Complex CFTs},'' {\em JHEP} {\bf 10} (2018) 108,
  \href{https://arxiv.org/abs/1807.11512}{{\tt 1807.11512}}.

\bibitem{Gorbenko:2018dtm}
V.~Gorbenko, S.~Rychkov, and B.~Zan, ``{Walking, Weak first-order transitions,
  and Complex CFTs II. Two-dimensional Potts model at $Q>4$},'' {\em SciPost
  Phys.} {\bf 5} (2018), no.~5 050,
  \href{https://arxiv.org/abs/1808.04380}{{\tt 1808.04380}}.

\bibitem{Kaplan:2009kr}
D.~B. Kaplan, J.-W. Lee, D.~T. Son, and M.~A. Stephanov, ``{Conformality
  Lost},'' {\em Phys. Rev. D} {\bf 80} (2009) 125005,
  \href{https://arxiv.org/abs/0905.4752}{{\tt 0905.4752}}.

\bibitem{Osborn:2017ucf}
H.~Osborn and A.~Stergiou, ``{Seeking fixed points in multiple coupling scalar
  theories in the $\epsilon$ expansion},'' {\em JHEP} {\bf 05} (2018) 051,
  \href{https://arxiv.org/abs/1707.06165}{{\tt 1707.06165}}.

\bibitem{Hager:2002uq}
J.~Hager, ``{Six-loop renormalization group functions of O(n)-symmetric
  phi**6-theory and epsilon-expansions of tricritical exponents up to
  epsilon**3},'' {\em J. Phys. A} {\bf 35} (2002) 2703--2711.

\bibitem{Giombi:2017dtl}
S.~Giombi, I.~R. Klebanov, and G.~Tarnopolsky, ``{Bosonic tensor models at
  large $N$ and small $\epsilon$},'' {\em Phys. Rev. D} {\bf 96} (2017), no.~10
  106014, \href{https://arxiv.org/abs/1707.03866}{{\tt 1707.03866}}.

\bibitem{Giombi:2018qgp}
S.~Giombi, I.~R. Klebanov, F.~Popov, S.~Prakash, and G.~Tarnopolsky,
  ``{Prismatic Large $N$ Models for Bosonic Tensors},'' {\em Phys. Rev. D} {\bf
  98} (2018), no.~10 105005, \href{https://arxiv.org/abs/1808.04344}{{\tt
  1808.04344}}.

\bibitem{hartman1960lemma}
P.~Hartman, ``A lemma in the theory of structural stability of differential
  equations,'' {\em Proceedings of the American Mathematical Society} {\bf 11}
  (1960), no.~4 610--620.

\bibitem{grobman1959homeomorphism}
D.~M. Grobman, ``Homeomorphism of systems of differential equations,'' {\em
  Doklady Akademii Nauk SSSR} {\bf 128} (1959), no.~5 880--881.

\bibitem{guckenheimer2013nonlinear}
J.~Guckenheimer and P.~Holmes, {\em Nonlinear oscillations, dynamical systems,
  and bifurcations of vector fields}, vol.~42.
\newblock Springer Science \& Business Media, 2013.

\bibitem{chow1977integral}
S.-N. Chow and J.~Mallet-Paret, ``Integral averaging and bifurcation,'' {\em
  Journal of differential equations} {\bf 26} (1977), no.~1 112--159.

\end{thebibliography}\endgroup

\end{document}